%% file: 0-main.tex
\documentclass[sigconf]{acmart}

\AtBeginDocument{%
  \providecommand\BibTeX{{%
    \normalfont B\kern-0.5em{\scshape i\kern-0.25em b}\kern-0.8em\TeX}}}

\newcommand{\nameOurs}{CoSLight}
\newcommand{\abbr}{CoS}

\usepackage{algorithm}
\usepackage{algorithmic}
\renewcommand{\arraystretch}{1.5}

\usepackage{multirow}
\usepackage{multicol}
\usepackage{subfigure} 

\usepackage{natbib} 

\usepackage[toc,page]{appendix}
\usepackage{longtable}

\usepackage{diagbox}
\usepackage{graphicx}
\usepackage{bm} 
\usepackage{soul}
\usepackage{graphics}
\usepackage{xcolor} 

\definecolor{steelblue}{RGB}{70,130,180}
\definecolor{cornflowerblue}{RGB}{100,149,237}
\definecolor{royalblue}{RGB}{65,105,225}
\definecolor{tealblue}{RGB}{0,128,128}
\definecolor{steelblue2}{RGB}{0,90,158}

\newcommand{\revise}[1]{{\color{black}#1}}
\newcommand{\re}[1]{{\color{black}#1}}

\setcopyright{acmlicensed}
\copyrightyear{2024}
\acmYear{2024}
\setcopyright{rightsretained}
\acmConference[KDD '24]{Proceedings of the 30th ACM SIGKDD Conference on Knowledge Discovery and Data Mining}{August 25--29, 2024}{Barcelona, Spain}
\acmBooktitle{Proceedings of the 30th ACM SIGKDD Conference on Knowledge Discovery and Data Mining (KDD '24), August 25--29, 2024, Barcelona, Spain}\acmDOI{10.1145/3637528.3671998}
\acmISBN{979-8-4007-0490-1/24/08}

\begin{document}

\title{\nameOurs: Co-optimizing Collaborator Selection and Decision-making to Enhance Traffic Signal Control}

\author{Jingqing Ruan}
\email{ruanjingqing2019@ia.ac.cn}
\orcid{0000-0002-4857-9053}
\affiliation{%
  \institution{Institute of Automation, Chinese Academy of Sciences}
  \city{Beijing}
  \country{China}
}

\author{Ziyue Li$^{\ast}$}
\orcid{0000-0003-4983-9352}
\email{zlibn@wiso.uni-koeln.de}
\affiliation{%
  \institution{University of Cologne \\ EWI gGmbH}
  \city{Cologne}
  \country{Germany}
}

\author{Hua Wei}
\orcid{0000-0003-4983-9352}
\email{hua.wei@asu.edu}
\affiliation{%
  \institution{Arizona State University}
  \city{Arizona}
  \country{U.S.A}
}

\author{Haoyuan	Jiang}
\orcid{0000-0002-3768-865X}
\email{jianghaoyuan@zju.edu.cn}
\affiliation{
  \institution{Baidu Inc.}
  \city{Shenzhen}
  \country{China}}

\author{Jiaming Lu}
\email{lujia_ming@126.com}
\affiliation{
  \institution{Fudan University}
  \city{Shanghai}
  \country{China}}

\author{Xuantang Xiong}
\email{xiongxuantang2021@ia.ac.cn}
\affiliation{%
  \institution{Institute of Automation, Chinese Academy of Sciences}
  \city{Beijing}
  \country{China}
}

\author{Hangyu Mao$^{\ast}$}
\orcid{0000-0002-4499-7581}
\email{hy.mao@pku.edu.cn}
\affiliation{
  \institution{Peking University}
  \city{Beijing}
  \country{China}}

\author{Rui	Zhao}
\orcid{0000-0001-5874-131X}
\affiliation{
  \institution{Qing Yuan Research Institute of Shanghai Jiao Tong University}
  \city{Shanghai}
  \country{China}}

\thanks{$\ast$ The corresponding author.}

\renewcommand{\shortauthors}{Trovato and Tobin, et al.}


\begin{abstract}
Effective multi-intersection collaboration is pivotal \re{for reinforcement-learning-based traffic signal control to alleviate congestion. Existing work mainly chooses neighboring intersections as collaborators. However, quite an amount of congestion, even some wide-range congestion, is caused by non-neighbors failing to collaborate. }
To address these issues, we propose to \re{separate the collaborator selection as a second policy to be learned, concurrently being updated with the original signal-controlling policy. Specifically, the selection policy in real-time adaptively selects the best teammates according to phase- and intersection-level features.} 
Empirical results on both synthetic and real-world datasets provide robust validation for the superiority of our approach, offering significant improvements over existing state-of-the-art methods. Code is available at \url{https://github.com/bonaldli/CoSLight}.
\end{abstract}






\maketitle

\input{1-intro-new-hua}


\input{2-related}


\input{2-pro_setup}

\input{3-method}

\input{4-exp}


\input{5-con}

\clearpage

\bibliographystyle{acm}
\bibliography{mybib}

\clearpage
\onecolumn
\renewcommand*\appendixpagename{\centering Appendices}
\begin{appendices}
\setcounter{figure}{0}
\setcounter{section}{0}
\renewcommand{\thefigure}{A\arabic{figure}}
\setcounter{table}{0}
\renewcommand{\thetable}{A\arabic{table}}
\input{6-app}
\end{appendices}

\end{document}

%% file: 1-intro-new-hua.tex
\section{Introduction}


Traffic Signal Control (TSC) plays a critical role in managing urban traffic flow and alleviating congestion. Over the past decade, multi-agent reinforcement learning (MARL) has emerged as a powerful tool for optimizing TSC~\citep{wei2019survey,jiang2024general,mei2023libsignal,du2024felight}. 
However, effectively promoting multi-intersection collaboration remains a persistent challenge in applying MARL to TSC.

Traditionally, researchers have treated geographically adjacent intersections as natural collaborators, and combined MARL methods with graph neural networks (GNNs)~\citep{wei2019colight,wu2021dynstgat,lou2022meta,devailly2021ig,yu2021macar,ruan2022gcs,bohmer2020deep} and approaches multi-level embeddings~\citep{liang2022oam,oroojlooy2020attendlight,zhang2021attentionlight,hao2023gat,jiang2018graph,ruan2024made} to model multi-intersection collaboration. 
\revise{However, the collaboration among intersections might be far beyond topological proximity in the real world~\citep{wu2021inductive,zhong2023contrastive}.}
\revise{
For instance, during morning rush hours, signals from residential to business areas must be strategically coordinated to facilitate driving in town. As shown in Figure~\ref{fig:intro}(a), intersections in upper streams should regulate incoming traffic to prevent downstream congestion, requiring them to synchronize with signals closer to business districts, which direct traffic towards parking and alternate routes. This coordination goes beyond mere proximity, emphasizing the need for dynamic and non-adjacent collaboration across the network. Likewise, during evening rush hours in Figure~\ref{fig:intro}(b), signals from business to residential areas also need to cooperate for better driving out of town. The coordination between areas of intersections in the morning and evening peaks mostly differ due to different origin-destination flows and other factors such as weather and trip purpose.
}

In this paper, we propose \nameOurs~to investigate the collaboration among intersections beyond topological neighbors, which comes along with two questions:

\begin{figure}[t]
	\centering
	\includegraphics[width=0.9\linewidth]{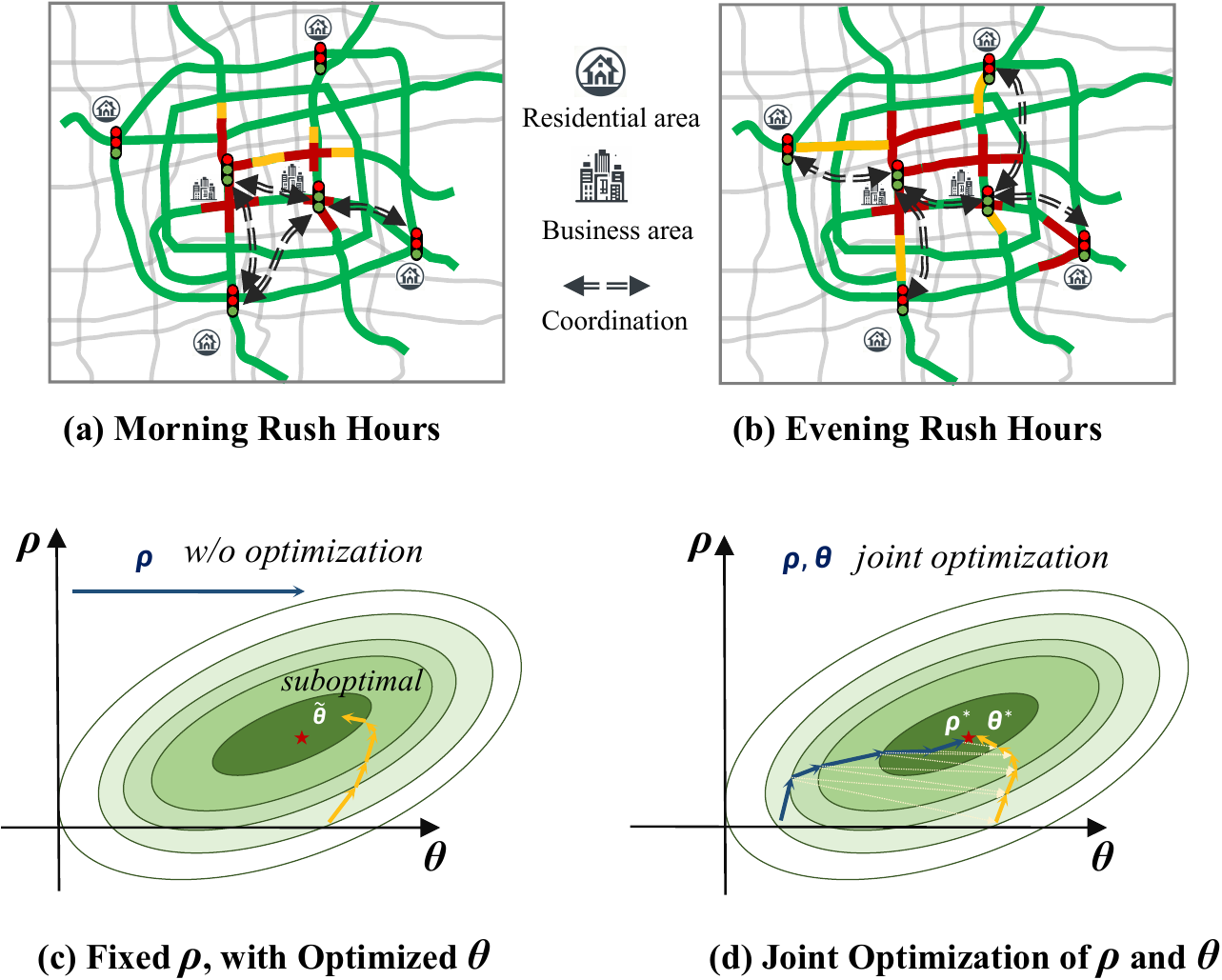}
	\caption{{(a)-(b): The coordination between areas of intersections during rush hours; (c)-(d): The collaboration policy $\boldsymbol{\rho}$ and decision policy $\boldsymbol{\theta}$ should be jointly optimized to prevent suboptimal.}
 }
    \label{fig:intro}
    \vspace{-10pt}
\end{figure}

\noindent $\bullet$~\textbf{Whom to collaborate for better decisions?}
While existing methods relying on heavy GNNs to learn whom to collaborate require information propagation within the whole network, in real-time, the collaborator selection must be light and agile. In this paper, we utilize a simple two-layer MLP structure for top-$k$ collaborator selection, which not only reduces computational complexity~\citep{han2022mlpinit, mlpmixer2021, wang2023gfs} but also achieves better performance in experiments. Instead of counting on GNNs to learn the relationships by themselves,  we incorporate two golden rules in the MLP collaboration matrix: that ``\textit{you are your biggest collaborator}'' and ``\textit{mutual reciprocity}'', which penalize the diagonal to be the largest and the matrix to be symmetric. These rules significantly improve collaboration benefits.

\noindent $\bullet$~\textbf{How to collaborate and optimize decisions?}
The goodness of collaborator selection largely influences the goodness of the decision policy. For example, the traditional practice is to first select collaborators without training and then to decide the signal policy, as shown in Figure~\ref{fig:intro}(c), which comes with two drawbacks: (1) the decision policy for one intersection might need a longer time to adjust to its collaborators,  (2) the decision policy is optimized towards maximizing the cumulative reward, while the collaborators are selected separately without acknowledgment of the performance of decision policy. To address this challenge, in this paper, we design a joint optimization scheme that simultaneously trains the collaboration selection policy and decision policy to maximize the cumulative return through a joint policy gradient, reinforcing the strong coupling between collaboration policy and decision policy. 

We conduct comprehensive experiments using both synthetic- and real data with different traffic flow and network structures. Our method consistently outperforms state-of-the-art RL methods, which shows that the effectiveness of collaborator selection and joint optimization with decision policy. We further showcase that the selected collaborators are not necessarily geographic neighbors, and visualize several interesting collaboration strategies learned by our method to show that our collaborator selection is effective and generalizable to different road networks.

In summary, our contribution is the following:
\begin{itemize}
\vspace{-5pt}
\item \nameOurs~is the first work to decouple and co-optimize the collaborator selection policy and signal planning policy. 
\item Specifically, \nameOurs~combines a Dual-Feature Extractor, capturing both phase- and intersection-level features, with a Multi-Intersection Collaboration module designed to strategically choose cooperating intersections. Moreover, a joint optimization strategy is derived to co-train the collaborator selection and decision-making policies, which uses a joint policy gradient to enhance the cumulative return, emphasizing the interdependence between collaboration selection and decision-making.
\item Extensive evaluations and multi-dimensional analysis using synthetic and real-world datasets demonstrate the effectiveness and superiority of~\nameOurs.
\end{itemize}

%% file: 2-related.tex
\vspace{-10pt}

\section{Related Work}
In this section, we review existing approaches 
based on how they collaborate: implicit and 
explicit collaboration.


\textbf{Implicit collaboration} only accesses information from other agents during the update phase to assist gradient backpropagation. MPLight repurposes FRAP~\citep{zheng2019learning,vlachogiannis2024humanlight} for a multi-agent context. IDQN~\cite{da2023uncertainty,du2023safelight,mei2023reinforcement,da2024prompt}, IPPO~\citep{ault2019learning} and MAPPO~\citep{yu2022surprising} tackle TSC problems directly from the MARL perspective. 
Works like ~\citep{kuyer2008multiagent,van2016coordinated,wei2019deep} extend the single-agent DQN solution to multi-agent scenarios using the max-plus coordination algorithm. 
Other methods such as IntelliLight~\citep{wei2018intellilight} and PressLight~\citep{wei2019presslight} enrich the state space by leveraging different types of additional information, like image frames or max pressure, respectively. 
FMA2C~\citep{ma2020feudal} and FedLight~\citep{ye2021fedlight} consider collaborative optimization from federated RL.
However, these methods focus on multi-intersection information from an update mechanism or gradient design perspective, while the decentralized execution leads to weak attention and may hinder efficient multi-intersection collaboration.

\textbf{Explicit collaboration} allows accessing other agents' information during the decision-making process to enhance collaboration. To overcome the shortcomings of weak collaboration in implicit strategies, explicit collaboration methods can be further categorized into two sub-classes. 
One line of research focuses on multi-intersection representation extraction, such as CoLight~\citep{wei2019colight,mei2023libsignal}, DynSTGAT~\citep{wu2021dynstgat}, IG-RL~\citep{devailly2021ig}, MaCAR~\citep{yu2021macar}, and MetaGAT~\citep{lou2022meta}, which utilize GNNs as the feature extractor to model representations of the current intersection and its neighbors.  X-Light \cite{jiang2024x} instead feeds the neighbors' MDP information into a Transformer. However, these methods risk introducing noise, such as unrelated intersection features, into collaboration; {it also suffers from the computational complexity of matrix multiplication}.
Another line of research leverages group-based cooperation~\cite{wang2020roma,wang2020rode,ruan2024made,ruan2023learning,mao2020neighborhood}. MT-GAD~\citep{jiang2021dynamic}, and JointLight~\citep{labres2021improving} uses heuristic grouping, which requires manual design, while CGB-MATSC~\citep{wang2021adaptive} applies KNN grouping directly based on state observation, a non-parametric method. GeneraLight~\citep{zhang2020generalight} utilizes $k$-means to execute the flow clustering to learn diverse metaparameters.
GPLight~\cite{liu2023gplight} utilizes mutual information and gathering constraints to derive latent representations, which are subsequently clustered into groups. 
However, their grouping strategies cannot be directly co-optimized with the decision-making process using the reward signal in RL, leading to a potentially suboptimal signal policy.

In contrast to these existing methods, we introduce a dual-feature extractor to derive collaborator representations beneficial for collaboration. Furthermore, we propose~\abbr~policy co-learned with decision policy with MIRL. This allows for the selection of optimal collaborators and primitive action for each intersection, guiding intense collaboration among multiple intersections.

%% file: 2-pro_setup.tex
\section{Preliminary and Problem Statement}

\subsection{Preliminary}
\begin{figure}[ht!]
	\centering
	\includegraphics[width=0.9\linewidth]{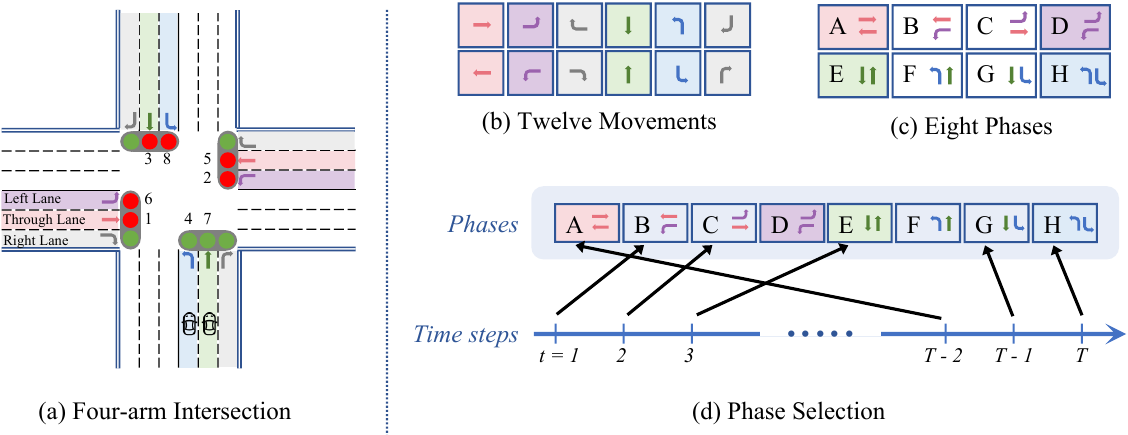}
	\caption{\re{(a) The illustration of intersection. (b) There are 12 movements: [North, South, West, East (four approaches)] $\times$ [Left, Go-through, Right (three directions)]. Usually, turning right isn't signal-controlled, so only 8 movements (index from 1-8 in (a)) are signal-controlled. (c) A phase is two non-conflicting movements that can be released together. There are 8 phases, e.g., phase-A combines movements 1 and 5. (d) The signal-control policy is to select one phase for the next time step according to the traffic condition.}}
    \label{fig:pre_}
\end{figure}

Firstly, we introduce some fundamental concepts related to traffic signal control, including traffic movement, signal phase, traffic intersection, multi-intersection traffic signal control, queue length, and pressure.
In Figure~\ref{fig:pre_}, we give a visual representation of these concepts to further aid understanding.

 \textit{\textbf{Traffic Intersection.}}
A traffic intersection, where multiple roads intersect, uses traffic signals to control vehicle flow. Figure~\ref{fig:pre_} depicts a four-arm intersection, equipped with lanes designated for left turns, straight travel, and right turns. It features four roads and twelve lanes for both entering and exiting traffic. We denote the incoming and outgoing lanes at intersection $i$ as $\mathcal{L}^i_{in}$ and $\mathcal{L}^i_{out}$.

 \textit{\textbf{Traffic Movement.}}
It refers to the progression of vehicles across an intersection in a specific direction, namely, turning left, going straight, or turning right. 
As depicted in Figure~\ref{fig:pre_}, twelve distinct traffic movements can be identified. 
Right-turning vehicles usually proceed regardless of the signal status.

 \textit{\textbf{Signal Phase.}}
It is a set of two traffic movements that can be released together without conflicts.
As illustrated in Figure~\ref{fig:pre_}, the intersection comprises eight distinct phases A-H.

 \textit{\textbf{Multi-Intersection Traffic Signal Control (TSC).}}
At each intersection, an RL agent is deployed to manage the traffic signal control. During each time unit denoted as $t_{duration}$,  the RL agent $i$ observes the state of the environment, represented as $o^i_t$. It then determines the action $a^i_t$, which dictates the signal phase for intersection $i$. The objective of the agent $i$ is to choose an optimal action (i.e., determining the most appropriate signal phase ) to maximize the cumulative reward.

 \textit{\textbf{Queue Length.}}
The queue length at each intersection $i$ is defined as the aggregate length of vehicle queues in all incoming lanes toward the intersection, denoted as:
\begin{equation}
    Q^i_{len} = \sum{q(l)}, l \in \mathcal{L}^i_{in},
\end{equation}
where $q(l)$ is the queue length in the lane $l$.

 \textit{\textbf{Pressure.}}
Pressure can be divided into intersection-wise and phase-wise categories. 
Intersection-wise pressure measures the imbalance between incoming and outgoing vehicle queues at an intersection, indicating traffic load discrepancies. 
Phase-wise pressure concerns a specific signal phase $p$. 
Each signal phase permits several traffic movements, each marked by $(l, m)$. 
Let $x(l, m)$ signify the vehicle count difference between lane $l$ and lane $m$ for a movement $(l, m)$, 
and the phase-wise pressure $P(p)$ denotes the cumulative sum of the pressures of all permissible movements within that phase, i.e., $\sum _{(l,m)} x(l, m)$.

\begin{figure}[ht]
\vspace{-15pt}
    \setcounter{subfigure}{0}
	\centering
	\subfigure[Grid 4 $\times$ 4]{
		\begin{minipage}[t]{0.22\textwidth}
			\centering
			\includegraphics[width=\textwidth]{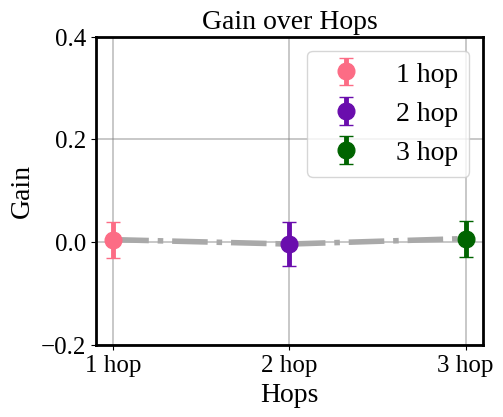}
		\end{minipage}
	}
	\subfigure[Grid 5 $\times$ 5]{
		\begin{minipage}[t]{0.22\textwidth}
			\centering
			\includegraphics[width=\textwidth]{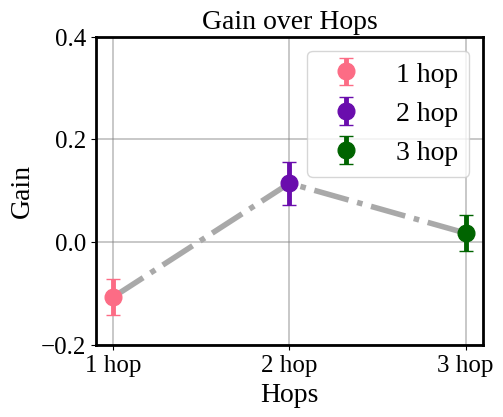}
		\end{minipage}
	}
\vspace{-10pt}	
 \caption{Gains over hops: choosing the right collaborators is important.}
	\label{exp:app_hops}
 \vspace{-15pt}
\end{figure}

 \textit{\textbf{Collaborator Matters.}}
As shown in Figure~\ref{exp:app_hops}, we conducted experiments on Grid 4 $\times$ 4 and Grid 5 $\times$ 5 to further substantiate our assertion: the optimal number and range of collaborating intersections vary across different scenarios. For instance, in Grid 4 $\times$ 4, the impact of collaboration remains largely consistent regardless of increasing distances, suggesting that the benefits of collaboration might be distance-independent in this setting. Alternatively, it could indicate that merely selecting topologically adjacent intersections might not enhance the collaborative outcomes. In contrast, Grid 5 $\times$ 5 displays a negative impact with one-hop collaboration, whereas two-hop collaboration produces the greatest benefits. This underscores the significance of precisely and judiciously selecting collaborators, highlighting that not just the presence, but the quality and context of collaboration matters.

\subsection{Problem Statement}

Then, we formulate $N$-intersection problem based on multiagent Markov decision processes (MMDPs)~\citep{boutilier1996planning} as collaborator-based MMDPs, which can be expressed as $<\{\mathcal{I}\}, \mathcal{S}, \mathcal{O},\{\mathcal{C}^i\}_{i=1}^N, \{\mathcal{A}^i\}_{i=1}^N, $ $\mathbb{P},r, \gamma>$.
$i \in \mathcal{I}$ is the $i^{th}$ intersection, $N$ is the number of intersections, and $\mathcal{S}, \mathcal{O}$ are the global state space and local observation space.
$\mathcal{C}^i$ and $\mathcal{A}^i$ denote the selected collaborator and action space for the $i^{th}$ intersection.
We label $\bm{ids}:=(\bm{ids}^1,...,\bm{ids}^N) \in \mathcal{C}$ and $\bm{a}:=(a^1,...,a^N) \in \mathcal{A}$ the joint collaborator identifiers and actions for all intersections. 
$\mathbb{P}(\cdot |s,\bm{ids},\bm{a})$ is the transition dynamics. 
All intersections share the same reward function $r(s,\bm{ids},\bm{a}):\mathcal{S} \times \mathcal{C}  \times \mathcal{A}  \to \mathbb{R}$. 
$\gamma \in (0,1)$ denotes a discount factor.
Here, we can denote $\tau  = ({s_0},{\bm{ids}_0},{\bm{a}_0},{s_1},...)$ 
as the trajectory induced by the policy $\bm{\pi}^{all}=\{\{{\rho^i}\cdot\pi^i\}_{i=1}^N$\}, where ${\rho^i}$ denotes the collaborator selection policy, and ${\pi^i}$ is the decision policy. 
All the intersections coordinate together to maximize the cumulative discounted return ${\mathbb{E}_{\bm{\tau}  \sim \bm{\pi}^{all} }}\left[ {\sum\nolimits_{t = 0}^\infty  {{\gamma ^t}r({{s}_t},\bm{ids}_t,{\bm{a}_t})} } \right]$.
We can define the overall joint policy as \re{the product of selection policy $\rho$ and decision policy $\pi$ based on Bayesian chain rules}. 
\begin{equation}
    {\bm{\pi} ^{all}}( \bm{ids}, \bm{a} |s) = \prod\limits_{i = 1}^N {{\rho^i}(\bm{ids}^i|{o^i}) \times {\pi^i}({a^i}|{o^i},\bm{ids}^i)} .
\end{equation}


%% file: 3-method.tex
\section{Methodology}

\begin{figure*}[ht!]
	\centering
	\includegraphics[width=0.9\linewidth]{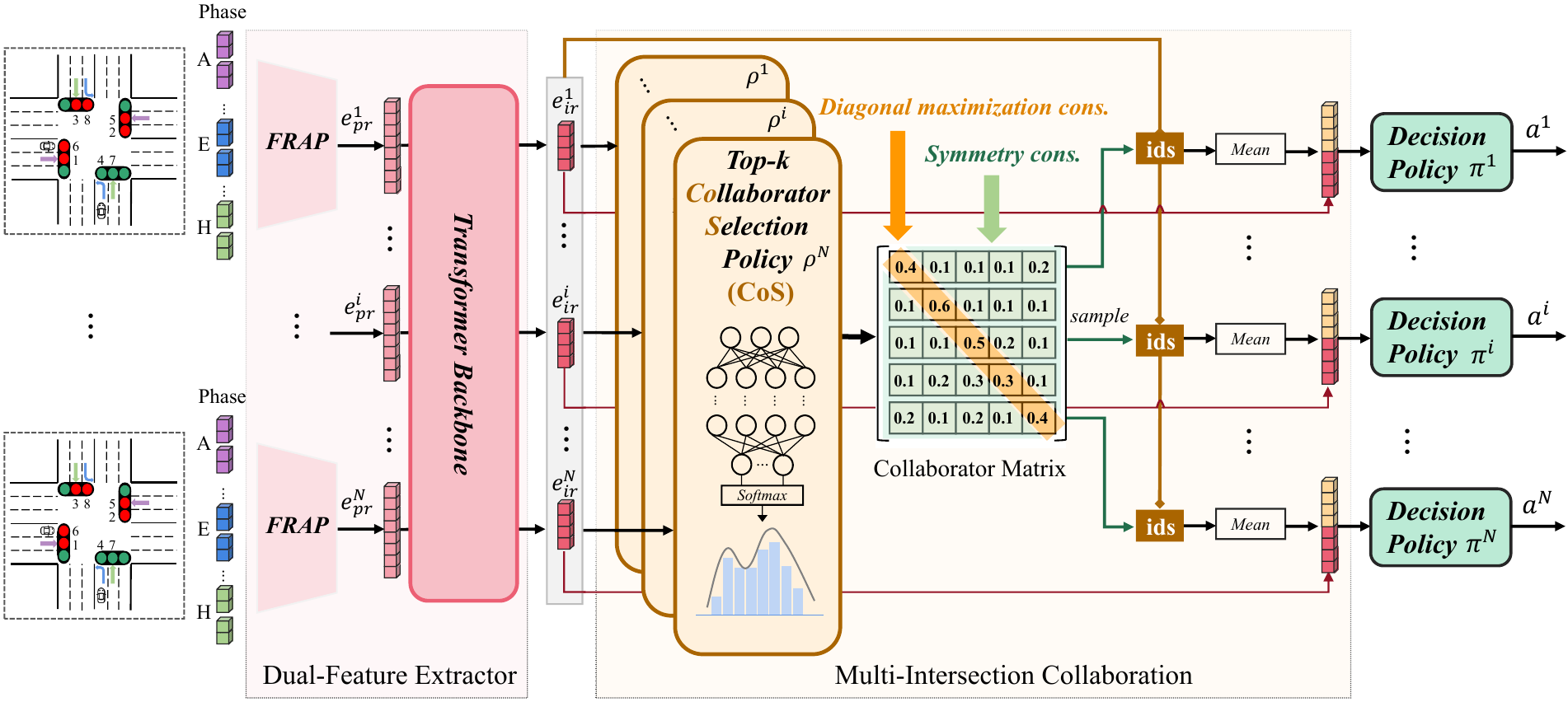}
	\caption{\revise{Overview of our proposed~\nameOurs: integrating Dual-Feature Extractor for phase- and intersection-level features with the module of Multi-Intersection Collaboration to select teammates for cooperation.}}
    \label{fig:framework}
    \vspace{-10pt}
\end{figure*}

\subsection{Main Modules}
As shown in Figure~\ref{fig:framework}, the dual-feature extractor pre-processes each intersection by obtaining phase- and intersection-level representations. Subsequently, the~\abbr~identifies the most appropriate collaborators and collects their information to assist with the decision-making process. 

\subsubsection{Dual-Feature Extractor.}

To optimize policy-making in TSC, a comprehensive representation of traffic situations is vital. While phase-level features inform individual intersections, they lack broader coordination insights from correlated intersections. Thus, we complement with intersection-level features to better capture network-wide 
traffic dynamics based on Transformer~\citep{vaswani2017attention}.

At the \textbf{phase} level, we adopt FRAP~\citep{zheng2019learning} to obtain the phase-level representation. 
The raw observations $o$ from the simulator include $K$ features, such as the number of vehicles, queue length, the current phase, the flow, etc. For any traffic movement $m \in \{1,...,8\}$ in an intersection $i$, the $k$-th feature in the raw observation can be denoted as $o_{m,k}^i$. For brevity, the superscript $i$ is omitted hereinafter.
The process can be summarized as follows:
$\small
\bm{e}_{m}  =  ||_{k=1}^K \sigma (\text{MLP}_k(o_{m,k})), \bm{e}_{pcr}  =  \text{FRAP}(\bm{e}_{m_1},\dots,\bm{e}_{m_8}),
\bm{e}^i_{pr}  =  \text{MLP}(\bm{e}_{pcr}),
$
where $||$ is concatenation, \re{$\sigma$} is the activation function, and $\bm{e}_{m}$ is the embedding of traffic movement.
Then $\text{FRAP}(\cdot)$ module {(details in Appendix~\ref{app:frap})} is applied to extract the phase competition representation $\bm{e}_{pcr}$, and we reshape $\bm{e}_{pcr}$ as a vector through \textit{flatten} operation.
Finally, $\bm{e}^i_{pr}$ is the phase representation in the intersection $i$. 

   

At the \textbf{intersection} level, we adopt the Transformer encoder~\cite{mao2022transformer,mao2023pdit,zhang2023stackelberg} as the backbone to model the relationship across multiple intersections: it takes in the phase representations $\bm{e}_{pr} = \{\bm{e}^i_{pr}\}_{i=1}^N, \bm{e}_{pr} \in \mathbb{R}^{N \times d}$, $d$ is the input dimension.
Attention can be calculated to obtain the intersection-level representation $\bm{e}_{ir}$,  
\re{attending to information from different intersections' representations}.
\begin{equation}\small
   \bm{e}_{ir} = \text{TransformerEnc}(\bm{e}_{pr}) , \ \ \bm{e}_{ir} \in \mathbb{R}^{N \times d},
    \label{eq: transformer}
\end{equation}
\re{The $\text{TransformerEnc}$ is the standard Transformer encoder with multi-head, followed by a feed-forward network.}

In summary, by extracting features from both the phase- and intersection-level simultaneously, we are able to obtain a more comprehensive insight to provide sufficient representation for subsequent modeling.

\subsubsection{Multi-Intersection Collaboration.}
The module facilitates cooperation among multiple intersections. By leveraging the rich representation $\bm{e}_{ir}$, we design the~\abbr~to select suitable collaborators for each intersection.


{As mentioned before, since the collaboration policy (\abbr) will be co-optimized with the decision policy, a lite and fast module is preferred to avoid becoming the computation bottleneck. Thus, we pursue minimalism: to use two-stacked MLPs. In computer vision, MLP-Mixer \citep{mlpmixer2021} is exclusively based on MLP and attains competitive scores as CNN-based methods or even ViT. In the spatiotemporal domain, MLPInit \citep{han2022mlpinit} also proves that MLPs could be 600$\times$ faster than GNN and achieves comparable results since MLP is free from matrix multiplication. GFS \citep{wang2023gfs} gives an insightful and theoretically-rigid explanation: \textbf{MLPs with general layer normalization provide similar functionality and effectiveness of aggregating nodes}.}
Moreover, an ablation study by replacing MLPs with GNNs in Section~\ref{sec:ablation} empirically confirms the effectiveness of MLPs.

Thus, {the top-$k$ Collaborator Selection (\abbr) policy is implemented by two-stacked MLP layers,}  
$\small
\bm{\alpha}^i = f_{{\tiny{\abbr}}}(\bm{e}_{ir})=\text{MLP}(\bm{e}_{ir}),$
where $\bm{\alpha}^i$ represents the logits from the MLP.
Then the probability distribution can be constructed:
\begin{equation}\small
\mathcal{P}^i := \text{Categorical}\left( {\frac{{\exp (\alpha _j^i)}}{{\sum\nolimits_{j' \in \mathcal{I} } {\exp (\alpha _{j'}^i)} }},j \in \mathcal{I} } \right),
\end{equation}
from which the indices of the selected collaborators for the intersection $i$ can be sampled. 
Here, we set the hyper-parameter $k$ as 5. We can sample the top-$k$ collaborator identifiers without replacement from $\mathcal{P}^i$, which means sampling the first element, then renormalizing the remaining probabilities to sample the next element. 
Let $ID^{*}_1,...,ID^{*}_k$ be an (ordered) sample without replacement from the $\mathcal{P}^i$, then the joint probability can be formulated as follows.
\begin{equation}\small
\label{eq:topk}
    P(ID_1^* = id_1^*,...,ID_k^* = id_K^*) = \prod\limits_{i = 1}^k {\frac{{\exp ({\alpha^i_{id_i^*}})}}{{\sum\nolimits_{\iota  \in \mathcal{I}_i^*} {\exp ({\alpha^i_{\iota }})} }}}, 
\end{equation}
where $id_1^*,...,id_k^* = \text{arg top} $-$ k({\mathcal{P}^i })$
is the sampled top-$K$ realization of variables $ID_1^*,...,ID_k^*$, and $\mathcal{I}_i^* = \mathcal{I}\backslash \{ id_1^*,...,id_{i - 1}^*\} $ is the domain (without replacement) for the $i$-th sampled element.
Summarized as $\bm{ids}^i \sim \rho^i(\bm{e}_{ir})$,
where $\bm{ids}^i \in \mathbb{Z}^{1 \times k}$ refers to the indices of the selected collaborators for intersection $i$, and $\rho$ represents the~\abbr, whose parameters are shared among all intersections.

With $\bm{ids}^i$ and $\bm{e}_{ir}$, we obtain the $k$ collaborators' representations: $\bm{e}_{team}^{i} = \text{lookup}(\bm{e}_{ir}, \bm{ids}^i)$,
where $\text{lookup}$ operator extracts the vector at index $\bm{ids}^i \in \mathbb{Z}^{1 \times k}$ from matrix $\bm{e}_{ir} \in \mathbb{R}^{N \times d}$, thus $\bm{e}_{team}^{i} \in \mathbb{R}^{k \times d}$.
After that, $\textit{mean-pooling}$ is used to obtain the final collaboration representation.
$\small
    \bm{\bar e}_{team}^{i} = \textit{mean-pooling}(\bm{e}_{team}^{i}), \ \ \bm{\bar e}_{team}^{i} \in \mathbb{R}^{1 \times d}.$

For the decision policy $\pi^i$, it receives the intersection-level representation and collaborator representation and outputs the action $a^i$ for the intersection $i$:
$\small
    {a^i} = \pi^i ( \cdot |[\bm{e}_{ir}^i|| \bm{\bar e}_{team}^i]).$


\subsection{Optimization Scheme}
We have introduced the inference process of the main modules. In this section, we propose an end-to-end joint optimization scheme to obtain the optimal~\abbr~$\rho$ and decision policy $\pi$.

\subsubsection{Overall Optimization Objective.}
The overall objective $\eta$ is to maximize the cumulative return, formulated as follows:
\begin{equation}\small
\label{eq:overall}
\mathop {\max }\limits_{\varphi ,\theta } \mathcal{J}(\varphi ,\theta ) = \mathop {\max }\limits_{\varphi ,\theta } {\mathbb{E}_{\scriptstyle \bm{ids} \sim {\bm{\rho} _\varphi }\hfill\atop
\scriptstyle \bm{a} \sim {\bm{\pi} _\theta }( \cdot |s, \bm{ids})\hfill}}\left[ {\sum\limits_{t = 0}^\infty  {{\gamma ^t}r({s_t},\bm{a_t})} } \right],
\end{equation}
where $\bm{\rho}=\{\rho^1,...,\rho^N\}$ and $\bm{\pi}=\{\pi^1,...,\pi^N\}$ are joint~\abbr~policy and decision policy, parameterized by $\{\varphi^1,...,\varphi^N\}$ and $\{\theta^1,...,\theta^N\}$, respectively.

According to Bellman Equation~\citep{bellman1966dynamic,sutton1998introduction}, the Q-function provides an estimate to guide the agent toward the optimal policy that maximizes cumulative rewards. Thus, the objective can be written:
\revise{
\begin{equation}\small
\label{eq:objective}
    \begin{array}{l}
\mathcal{J}(\varphi ,\theta ) = {\mathbb{E}_{s\sim{p^{{\pi ^{all}}}},(\bm{ids},\bm{a})\sim{\pi ^{all}}}}\left[ {{\pi ^{all}}(\bm{ids},\bm{a}|s){Q^{\pi^{all}} }(s,\bm{a})} \right]\\
 = {\mathbb{E}_{{p^{{\pi ^{all}}}},{\pi ^{all}}}}\left[ {\prod\limits_{i = 1}^N {{\rho_{\varphi} ^i}(\bm{ids}^i|{o^i}){\pi_{\theta} ^i}({a^i}|{o^i},\bm{ids}^i)} {Q^{\pi^{i}} }(o^i,{a}^i)} \right],
\end{array}
\end{equation}
}
where $p^{{\pi ^{all}}}$ denotes the stationary distribution induced by policy $\pi^{all}$, $Q^{\pi^{all}}$ is the joint action-value function, \revise{and $Q^{\pi^{i}}$ is the individual action-value function. Here, $Q^{\pi^{i}}$ simplifies the learning process by focusing on local state-action pairs rather than the global state, effectively mitigating the scalability issue in large-scale environments}.

\subsubsection{Optimizing~\abbr~Policy.}

\revise{The ~\abbr~policy $\boldsymbol{\rho}_{t}$} aims to choose optimal collaborators for intersections.
We follow two rules: 
(a) Intersections should primarily focus on their own decisions. 
(b) They should also account for and collaborate with each other's decisions. 
Thus, two key constraints are introduced to ensure these rules in the collaborator selection process.

Firstly, we can denote the distributions of $\boldsymbol{\rho}_{t}$ for all intersections as a collaborator matrix $\bm{M}^\rho \in \mathbb{R}^{N \times N}$, where each element $M^\rho_{ij}$ in the matrix signifies the probability of collaborator selection between the intersections $i$ and $j$.

\textbf{\textit{{Rule 1: ``You are your biggest collaborator''.} Diagonal maximization constraint}} is imposed.
To prioritize its decision-making for each intersection, we enforce the diagonal maximization constraint on the matrix $\bm{M}^\rho$, ensuring it remains a valid probability distribution matrix. The main objective is to maximize the sum of the diagonal elements while complying with specific constraints that require the elements to be non-negative, and the row sums must equal to one:

\begin{equation}
\small
    \max \sum\limits_{i = 1}^N {{M}_{ii}^\rho } \ \ s.t.\ 
 \ {M}_{ij}^\rho  \ge 0,\forall i,j; \ \ \sum\limits_{j = 1}^N {{M}_{ij}^\rho }  = 1,\forall i .
 \label{eq:diag}
\end{equation}

\textbf{\textit{{Rule 2: ``Collaboration should be mutually reciprocal''.} Symmetry constraint}} is enforced.
To encourage collaboration and mutual consideration between intersections, we incorporate a symmetry constraint into the training. 
The symmetry constraint, calculated as the mean squared difference between the matrix $M^\rho$ and its transpose ${{[M^\rho]}^T}$, guides the neural network to learn symmetric collaborator selection.
Mathematically, the symmetry constraint is formulated as follows:
\begin{equation}\small
\min \frac{1}{{{N^2}}}\sum\limits_{i = 1}^N {\sum\limits_{j = 1}^N {{{({M}_{ij}^\rho  - {{[{M}_{ij}^\rho ]}^{\rm{T}} })}^2}} } .
\label{eq:symm}
\end{equation}

In summary, we can define the loss function for optimizing the~\abbr~policy:
\begin{equation}\small
    \label{eq:loss_assign}
    \mathcal{L}(\rho_{\varphi}) = -\mathcal{J}(\varphi ,\theta ) - \sum\limits_{i = 1}^N {{M}_{ii}^\rho } + \frac{1}{{{N^2}}}\sum\limits_{i = 1}^N {\sum\limits_{j = 1}^N {{{({M}_{ij}^\rho  - {{[{M}_{ij}^\rho ]}^{\rm{T}} })}^2}} } .
\end{equation}

Following the multi-agent policy gradient optimization~\citep{zhang2018fully,foerster2018counterfactual}, we can derive the gradient for $\rho_{\varphi}$:
\revise{
\begin{equation}\small
\begin{array}{l}
{\nabla _{{\varphi ^i}}}\mathcal{J}(\varphi ,\theta ) = \int_S {{p^{{\pi ^{all}}}}(s)} \sum\nolimits_{\bm{ids}} {{\rho _\varphi }(\bm{ids}|s)}  \cdot \\
\left. {\left[ {{\nabla _\varphi }\sum\nolimits_i {\log \rho _{{\varphi ^i}}^i(\bm{ids}^i|s) \cdot } } \right.\sum\nolimits_{\bm{a}} {\pi (\bm{a}|s,\bm{ids})} } \right] \cdot {Q_{{\pi ^{i}}}}(o^i,a^i)ds \\
- \frac{1}{{\rm{N}}}\nabla \varphi {{\bm{M}}^\rho }
+ \frac{2}{{\rm{N}}}\left( {{{\bm{M}}^\rho } - {{\left[ {{{\bm{M}}^\rho }} \right]}^{\rm{T}}}} \right)\nabla \varphi {{\bm{M}}^\rho }
\end{array}
\label{eq:team_policy}
\end{equation}
}



\begin{algorithm}[ht!]
\caption{The optimization process.}
\label{alg:algo}
\begin{algorithmic}[1] 
\STATE \textbf{Ensure}: Collaborator assignment policy $\{\rho^i\}_{i=1}^N$, collaborator-based multi-agent actor $\{\pi^i\}_{i=1}^N$, and critics \revise{$\{Q^i\}_{i=1}^N$}; \\
\textbf{Initialize}: $\gamma, \mathcal{D}\leftarrow \emptyset, L_1, L_2, B$; // $L_1, L_2$ are intervals; $B$ is the mini-batch size. \\
\STATE \textbf{Initialize}: the parameters $\theta^i,\varphi^i,\phi^i$ 
for $\rho^i$, $\pi^i$, and \revise{$Q^i$};\\
\FOR {each episode}
\STATE Reset state $\leftarrow \{o_0^i\}_{i=1}^N$; // drop $i$ as $\bm{o_0}$ for brevity;\\
\FOR {each timestep $t$}
\STATE Obtain the dual-features $\bm{e}_{ir}$;
\FOR {each intersection $i$}
\STATE Get the collaborator indices $\bm{ids}^i=\rho^i(\cdot|\bm{e}_{ir})$ for each intersection; \\
\STATE Query the collaborator representation ${\bar e}_{team}^{i}$ with $\bm{e}_{ir}$ and $\bm{ids}^i$;\\
Get the action $a^i =  \pi^i ( \cdot |[e_{ir}^i||\bar e_{team}^i])$;\\
\ENDFOR \\ 
\STATE Take joint actions $\bm{a}_t$; \\
\STATE Receive reward {${\bm{r}_t}$} and observe the next state ${\bf{o}}_{t+1}$;\\
\STATE Add transition $\{{\bm{o}}_t, \bm{ids}_t, {\bm{a}}_t, {\bm{r}}_t, {\bm{o}}_{t+1}\}$ into $\mathcal{D}$;\\
\ENDFOR

\IF{episodes $\ge L_1$}
\STATE Sample batch {$\{{\bm{o}}_j, \bm{ids}_j, \bm{{a}}_j, \bm{{r}}_j, {\bm{o}}_{j+1}\}_{j=0}^B\sim \mathcal{D}$}; \\
\STATE Update the~\abbr~policy $\rho^i$ using~(\ref{eq:team_policy});\\
\STATE Update the decision policy $\pi^i$ using (\ref{eq:actor});\\
\STATE Update the critic \revise{$\phi^i$} using~(\ref{eq:critic});\\
\ENDIF

Updating the target networks every $L_2$ episodes ;\\

\ENDFOR

\end{algorithmic}
\end{algorithm}

Thus, {incorporating these constraints into the~\abbr~policy gradient optimization enables the policy $\rho_{\varphi}$ to learn the optimal collaborator composition progressively.} As a result, the policy can promote cooperative decision-making among intersections, resulting in enhanced traffic signal control and improved traffic flow efficiency.


\subsubsection{Optimizing Decision Policy.}

Assumed we have $\bm{ids} \sim \boldsymbol{\rho}_{t}$ to identify the most appropriate collaborators and then to optimize the multi-agent decision policy $\boldsymbol{\pi}_t$ as follows. For the decision policy, we can derive its gradient with the Eq.~(\ref{eq:objective}).
\revise{
\begin{equation}\small
\label{eq:actor}
    \begin{array}{l}
{\nabla _{{\theta ^i}}}\mathcal{J}(\varphi ,\theta ) = \int_S {{p^{{\pi ^{all}}}}(s)} \sum\nolimits_{\bm{ids}} {\rho (\bm{ids}|s)} \sum\nolimits_{\bm{a}} {{\pi _\theta }(\bm{a}|s,\bm{ids})}  \cdot \\
{\nabla _\theta }\sum\nolimits_i {\log \pi _{{\theta ^i}}^i({a^i}|{o^i},\bm{ids}^i)}  \cdot {Q_{{\pi ^{i}}}}(o^i, a^i)ds \\
\approx {\mathbb{E}_{(s,\bm{a},\bm{ids})\sim \mathcal{D}}}\left[ {{\nabla _\theta }\sum\nolimits_i {\log \pi _{{\theta ^i}}^i({a^i}|{o^i},\bm{ids}^i)}  \cdot {Q_{{\pi ^{i}}}}(o^i, a^i)} \right]\\
 = {\mathbb{E}_{(s,\bm{a},\bm{ids})\sim D}}\left[ {\sum\nolimits_i {\frac{{{\nabla _{{\theta ^i}}}\pi _{{\theta ^i}}^i({a^i}|{o^i},\bm{ids}^i)}}{{\pi _{{{\overline \theta  }^i}}^i({a^i}|{o^i},\bm{ids}^i)}}}  \cdot {Q_{{\pi ^{i}}}}(o^i, a^i)} \right] , 
\end{array}
\end{equation}
}
where ${\pi _{{{\overline \theta  }^i}}^i}$ is the old decision policy used for sampling.

\revise{
The critic is updated to minimize the difference between the predicted and actual returns, which resembles the action-value TD-learning~\citep{sutton1988learning}.
The loss function for $Q^{\pi^{i}}$ is formulated as follows.
\begin{equation}\small
\label{eq:critic}
    {\mathcal{L}_{{Q^{{\pi ^{i}}}}}}(\phi ) = {\mathbb{E}_{(o^i,a^i,r^i,o^{i'})\sim \mathcal{D}}}\left[ {{{\left( {y^i - {Q^{{\pi ^{i}}}}(o^i,a^i;\phi )} \right)}^2}} \right],
\end{equation}
where $y^i = {r^i} + \gamma {\max _{a^{i'}}}{\widetilde Q^{{\pi ^{i}}}}(o^{i'}, a^{i'})$ is the learning target, $\gamma$ is the discounted factor, and $\widetilde Q^{{\pi ^{i}}}$ is the target network for intersection $i$.
}


The joint optimization scheme ensures the~\abbr~policy and the collaborator-based decision policy converge in the same direction by optimizing the same objective function, which is to maximize the cumulative discounted return in an end-to-end manner. 
The proposed method facilitates effective decision-making and collaboration among intersections, allowing the two policies to work together harmoniously toward achieving the overall goal. 



\subsection{Algorithmic Framework}
\label{app:alg}

Here we provide a detailed pseudocode to elaborate the overall inference and training process. 
In the inference process, we first select suitable teammate IDs $\bm{ids}^i$ for each intersection $i$ based on its observations $o^i$. Then we calculate the teammate vectors $\bar e^i_{team}$ according to the teammate IDs and concatenate them to the self vector $e^i_{ir}$ to obtain the action for decision making, and finally interact with the SUMO environment. In training, we use the derived loss for backpropagation training. The detailed process is presented in Algorithm~\ref{alg:algo}.

%% file: 4-exp.tex
\section{Experiments}


\subsection{Environments}
The evaluation scenarios come from the Simulation of Urban Mobility (SUMO)\footnote{https://www.eclipse.org/sumo/}, which contains three synthetic scenarios and two real road network maps of different scales, including Grid $4 \times 4$, Avenue $4 \times 4$, Grid $5 \times 5$, Cologne8, and Nanshan.
In Table~\ref{tab:scenarios}, we present detailed statistics including the total number of intersections (\#Total Int.), along with the quantity of 2-arm, 3-arm, and 4-arm intersections in each scenario.

\begin{table}[t]
\centering

\resizebox{0.99\columnwidth}{!}{%
\begin{tabular}{cccccccccc}
\hline
Dataset & Country  & \#Total Int. & \#2-arm & \#3-arm & \#4-arm & Flow Type & min Flow (/hour) & max Flow & mean Flow \\ \hline
Grid $4 \times 4$   & synthetic  & 16   & 0  & 0  & 16 & multi-modal Gaussian (m.m.G)	& 66 & 136 & 94.5 \\
Avenue $4 \times 4$   & synthetic  & 16 & 0 & 0 & 16 & m.m.G &  94.5 & 666 & 364.6 \\
Grid $5 \times 5$     & synthetic  & 25 & 0 & 0 & 25 & m.m.G  & 120 & 1363 & 269.8 \\
Cologne8     & Germany & 8 & 1  & 3  & 4  & real flow, morning peak (8AM-9AM) & 134 & 212 &  139.0 \\
Nanshan      & China  & 29 & 1 & 6    & 22 & real flow, evening peak (6PM-7PM)  & 160 & 1440 & 172.6 \\ \hline
\end{tabular}
}
\caption{The detailed statistics about evaluation scenarios}
\vspace{-20pt}
\label{tab:scenarios}
\end{table}

As for each scenario, each individual episode lasts for a time span of 3600 seconds, during which the action interval is $\Delta t$ = 15 seconds. The network design and hyper-parameter settings are provided in Appendix~\ref{app:network_para}.

\begin{table*}[t]
\centering  
\resizebox{\textwidth}{!}{%
\begin{tabular}{c|lllll|lllll}
\hline
\multirow{2}{*}{\textbf{Methods}} & \multicolumn{5}{c|}{\textbf{Average Trip Time (seconds)}}  & \multicolumn{5}{c}{\textbf{Average Delay Time (seconds)}}   
\\
  & \multicolumn{1}{c}{\textbf{Grid 4$\times$4}} & \multicolumn{1}{c}{\textbf{Avenue 4$\times$4}} & \multicolumn{1}{c}
  {\textbf{Grid 5$\times$5}} & \multicolumn{1}{c}
  {\textbf{Cologne8}} & \multicolumn{1}{c|}{\textbf{Nanshan}} & \multicolumn{1}{c}{\textbf{Grid 4$\times$4}} & \multicolumn{1}{c}{\textbf{Avenue 4$\times$4}} & \multicolumn{1}{c}{\textbf{Grid 5$\times$5}} & \multicolumn{1}{c}{\textbf{Cologne8}} & \multicolumn{1}{c}{\textbf{Nanshan}} \\ \hline

\textbf{FTC}                               
& 206.68 \fontsize{8pt}{8pt}\selectfont{±} 0.54               
& 828.38 \fontsize{8pt}{8pt}\selectfont{± 8.17}                 
& 550.38 \fontsize{8pt}{8pt}\selectfont{± 8.31}                   
& 124.4 \fontsize{8pt}{8pt}\selectfont{± 1.99}                 
& 729.02 \fontsize{8pt}{8pt}\selectfont{± 37.03}                  
& 94.64 \fontsize{8pt}{8pt}\selectfont{± 0.43}                
& 1234.30 \fontsize{8pt}{8pt}\selectfont{± 6.50}               
& 790.18 \fontsize{8pt}{8pt}\selectfont{± 7.96}                  
& 62.38 \fontsize{8pt}{8pt}\selectfont{± 2.95}                   
& 561.69 \fontsize{8pt}{8pt}\selectfont{± 37.09}                 \\

\textbf{MaxPressure}                       
& 175.97 \fontsize{8pt}{8pt}\selectfont{± 0.70}               
& 686.12 \fontsize{8pt}{8pt}\selectfont{± 9.57}             
& 274.15 \fontsize{8pt}{8pt}\selectfont{± 15.23}                    
& 95.96 \fontsize{8pt}{8pt}\selectfont{± 1.11}                  
& 720.89 \fontsize{8pt}{8pt}\selectfont{± 29.94}                  
& 64.01 \fontsize{8pt}{8pt}\selectfont{± 0.71}                 
& 952.53 \fontsize{8pt}{8pt}\selectfont{± 12.48}             
& 240.00 \fontsize{8pt}{8pt}\selectfont{± 18.43}   
& 31.93 \fontsize{8pt}{8pt}\selectfont{± 1.07}
& 553.94 \fontsize{8pt}{8pt}\selectfont{± 32.61}                   
      \\

\textbf{IPPO}                              
& 167.62 \fontsize{8pt}{8pt}\selectfont{± 2.42}               
& 431.31 \fontsize{8pt}{8pt}\selectfont{± 28.55}              
& 259.28 \fontsize{8pt}{8pt}\selectfont{± 9.55}                      
& 90.87 \fontsize{8pt}{8pt}\selectfont{± 0.40}                
& 743.69 \fontsize{8pt}{8pt}\selectfont{± 38.9}     
& 56.38 \fontsize{8pt}{8pt}\selectfont{± 1.46}             
& 914.58 \fontsize{8pt}{8pt}\selectfont{± 36.90}               
& 243.58 \fontsize{8pt}{8pt}\selectfont{± 9.29}   
& 26.82 \fontsize{8pt}{8pt}\selectfont{± 0.43}
& 577.99 \fontsize{8pt}{8pt}\selectfont{± 42.22}                   
    \\          

\textbf{MAPPO}                            
& 164.96 \fontsize{8pt}{8pt}\selectfont{± 1.87}                
& 565.67 \fontsize{8pt}{8pt}\selectfont{± 44.8}              
& 300.90 \fontsize{8pt}{8pt}\selectfont{± 8.31}                 
& 97.68 \fontsize{8pt}{8pt}\selectfont{± 2.03}                  
& 744.47 \fontsize{8pt}{8pt}\selectfont{± 30.07}                    
& 53.65 \fontsize{8pt}{8pt}\selectfont{± 1.00}            
& 1185.2 \fontsize{8pt}{8pt}\selectfont{± 167.48}                
& 346.78 \fontsize{8pt}{8pt}\selectfont{± 28.25}     
& 33.37 \fontsize{8pt}{8pt}\selectfont{± 1.97} 
& 580.49 \fontsize{8pt}{8pt}\selectfont{± 33.6}                    
\\

\textbf{MAT}                            
& 246.13 \fontsize{8pt}{8pt}\selectfont{± 24.23}                
& 421.85 \fontsize{8pt}{8pt}\selectfont{± 73.13 }              
& 356.81 \fontsize{8pt}{8pt}\selectfont{± 11.05 }                 
& 111.59 \fontsize{8pt}{8pt}\selectfont{± 18.82}                  
& 754.28 \fontsize{8pt}{8pt}\selectfont{± 58.70   }  

& 106.70 \fontsize{8pt}{8pt}\selectfont{±  14.07}            
& \textbf{565.42 \fontsize{8pt}{8pt}\selectfont{± 91.35}   }             
& 217.93 \fontsize{8pt}{8pt}\selectfont{± 40.64}     
& 25.23  \fontsize{8pt}{8pt}\selectfont{± 8.69 } 
& \textbf{415.84 \fontsize{8pt}{8pt}\selectfont{± 75.59} } 
\\

\textbf{FRAP}       
& {161.58 \fontsize{8pt}{8pt}\selectfont{± 1.9} }               
& 383.71 \fontsize{8pt}{8pt}\selectfont{± 4.42}             
& {238.41 \fontsize{8pt}{8pt}\selectfont{± 10.66} }                 
& \textbf{88.61 \fontsize{8pt}{8pt}\selectfont{± 0.33}}                   
& 709.18 \fontsize{8pt}{8pt}\selectfont{± 21.46} 

& {50.02 \fontsize{8pt}{8pt}\selectfont{± 0.93} }              
& 794.13 \fontsize{8pt}{8pt}\selectfont{± 42.52}             
& \underline{203.95 \fontsize{8pt}{8pt}\selectfont{± 8.92}  }                
& 27.5 \fontsize{8pt}{8pt}\selectfont{± 0.24}                    
& 542.43 \fontsize{8pt}{8pt}\selectfont{± 21.51}  
\\

\textbf{MPLight}                           
& 179.51 \fontsize{8pt}{8pt}\selectfont{± 0.95}                   
& 541.29 \fontsize{8pt}{8pt}\selectfont{± 45.24}             
& 261.76 \fontsize{8pt}{8pt}\selectfont{± 6.60}                  
& 98.44 \fontsize{8pt}{8pt}\selectfont{± 0.62}                  
& 668.81 \fontsize{8pt}{8pt}\selectfont{± 7.92}                  
& 67.52 \fontsize{8pt}{8pt}\selectfont{± 0.97}                 
& 1083.18 \fontsize{8pt}{8pt}\selectfont{± 63.38}             
& 213.78 \fontsize{8pt}{8pt}\selectfont{± 14.44}   
& 34.38 \fontsize{8pt}{8pt}\selectfont{± 0.63} 
& 494.05 \fontsize{8pt}{8pt}\selectfont{± 7.52}                   
                \\

\textbf{CoLight}                          
& 163.52 \fontsize{8pt}{8pt}\selectfont{± 0.00}               
& 409.93 \fontsize{8pt}{8pt}\selectfont{± 0.00}               
&  {242.37 \fontsize{8pt}{8pt}\selectfont{± 0.00}}                   
&  \underline{89.72 \fontsize{8pt}{8pt}\selectfont{± 0.00}}                
& \textbf{608.01 \fontsize{8pt}{8pt}\selectfont{± 0.00}  } 

& 51.58 \fontsize{8pt}{8pt}\selectfont{± 0.00}              
& 776.61 \fontsize{8pt}{8pt}\selectfont{± 0.00}               
& 248.32 \fontsize{8pt}{8pt}\selectfont{± 0.00}  
&  {25.56 \fontsize{8pt}{8pt}\selectfont{± 0.00}}
& \underline{428.95 \fontsize{8pt}{8pt}\selectfont{± 0.00}}                    
  \\


\textbf{Advanced-CoLight}                            
& 171.63 \fontsize{8pt}{8pt}\selectfont{± 1.71}                
& 421.44 \fontsize{8pt}{8pt}\selectfont{± 5.61}              
& 237.67 \fontsize{8pt}{8pt}\selectfont{± 3.02 }                 
& 91.22 \fontsize{8pt}{8pt}\selectfont{± 1.01 }                  
& 612.34 \fontsize{8pt}{8pt}\selectfont{± 9.79  }  

& 52.31 \fontsize{8pt}{8pt}\selectfont{± 0.01}            
& 763.78 \fontsize{8pt}{8pt}\selectfont{± 14.01}                
& 242.50 \fontsize{8pt}{8pt}\selectfont{± 5.06 }     
& \underline{25.12 \fontsize{8pt}{8pt}\selectfont{± 1.08 } }
& 512.45 \fontsize{8pt}{8pt}\selectfont{± 6.98  }  
\\     

\textbf{MetaGAT}                           
& 165.23 \fontsize{8pt}{8pt}\selectfont{± 0.00}                       
&  \underline{374.80 \fontsize{8pt}{8pt}\selectfont{± 0.87}}                 
& 266.60 \fontsize{8pt}{8pt}\selectfont{± 0.00}                      
& 90.74 \fontsize{8pt}{8pt}\selectfont{± 0.00}                  
&   {676.42 \selectfont{± 0.00}}                                            
& 53.20 \fontsize{8pt}{8pt}\selectfont{± 0.00}                    
& {772.36 \fontsize{8pt}{8pt}\selectfont{± 0.00}}  &  234.80 \fontsize{8pt}{8pt}\selectfont{± 0.00}                 
&  26.85 \fontsize{8pt}{8pt}\selectfont{± 0.00}                        
&  {503.42 \fontsize{8pt}{8pt}\selectfont{±0.00}}                           \\

\textbf{DuaLight}                            
& \underline{161.0 \fontsize{8pt}{8pt}\selectfont{± 0.00}  }              
& 396.65 \fontsize{8pt}{8pt}\selectfont{± 0.00}              
& \underline{221.83 \fontsize{8pt}{8pt}\selectfont{± 0.00}   }              
& 89.74 \fontsize{8pt}{8pt}\selectfont{± 0.00}                  
& \underline{609.89 \fontsize{8pt}{8pt}\selectfont{± 0.00}}  

& \underline{49.32 \fontsize{8pt}{8pt}\selectfont{± 0.00}}            
& {756.99 \fontsize{8pt}{8pt}\selectfont{± 69.44} }               
& 237.71 \fontsize{8pt}{8pt}\selectfont{± 0.00}     
& 25.35 \fontsize{8pt}{8pt}\selectfont{± 0.00} 
& 429.49 \fontsize{8pt}{8pt}\selectfont{± 0.00}  
\\



\textbf{\nameOurs}                            
&  \textbf{159.1 \fontsize{8pt}{8pt}\selectfont{± 3.12}}                
&  \textbf{364.21 \fontsize{8pt}{8pt}\selectfont{± 4.78}}  
&  \textbf{220.32\fontsize{8pt}{8pt}\selectfont{ ± 1.71}}          
&  { {90.46 \fontsize{8pt}{8pt}\selectfont{± 0.55}}}          
&  { {621.23 \fontsize{8pt}{8pt}\selectfont{± 7.17}}}      

&  \textbf{46.11 \fontsize{8pt}{8pt}\selectfont{± 1.21}}                 
&   \underline{744.98 \fontsize{8pt}{8pt}\selectfont{± 16.49}}       
&   \textbf{178.54\fontsize{8pt}{8pt} \selectfont{± 4.34}}          
&   \textbf{24.79 \fontsize{8pt}{8pt}\selectfont{± 1.23}}        
&  { {476.57 \fontsize{8pt}{8pt}\selectfont{± 17.12}}}         \\ 
\hline
\end{tabular}%
}
\caption{
Scenario-wise evaluation. \revise{Our method achieves either the best (boldface) or the second-best (underlined) performance.}}
\label{tab:main-table1}
\vspace{-10pt}
\end{table*}

\begin{table}[t]
\centering  
\resizebox{\columnwidth}{!}{%
\begin{tabular}{c|lllll}
\hline
\multirow{2}{*}{\textbf{Methods}} & \multicolumn{5}{c}{\textbf{Average Rewards}}     
\\
  & \multicolumn{1}{c}{\textbf{Grid 4$\times$4}} & \multicolumn{1}{c}{\textbf{Avenue 4$\times$4}} & \multicolumn{1}{c}
  {\textbf{Grid 5$\times$5}} & \multicolumn{1}{c}
  {\textbf{Cologne8}} & \multicolumn{1}{c}{\textbf{Nanshan}} \\ \hline

\textbf{FTC}     

& -0.614  \scriptsize{$\pm$ 0.015} 
& -4.503  \scriptsize{$\pm$ 0.025} 
& -2.346  \scriptsize{$\pm$ 0.052} 
& -2.114  \scriptsize{$\pm$ 0.021} 
& -3.479  \scriptsize{$\pm$ 0.186}
\\

\textbf{MaxPressure}  
& -0.393  \scriptsize{$\pm$ 0.003} 
& -4.032  \scriptsize{$\pm$ 0.040} 
& -1.132  \scriptsize{$\pm$ 0.013} 
& -0.756  \scriptsize{$\pm$ 0.012} 
& -3.055  \scriptsize{$\pm$ 0.162} 
\\

\textbf{IPPO}      
& -0.336  \scriptsize{$\pm$ 0.004} 
& {-2.558  \scriptsize{$\pm$ 0.213}} 
& -0.943  \scriptsize{$\pm$ 0.037} 
& {-0.646  \scriptsize{$\pm$ 0.015}} 
& -3.555  \scriptsize{$\pm$ 0.097} 
\\ 

\textbf{MAPPO}   
& -0.308  \scriptsize{$\pm$ 0.006} 
& -2.744  \scriptsize{$\pm$ 0.238} 
& -1.273  \scriptsize{$\pm$ 0.107} 
& -1.697  \scriptsize{$\pm$ 0.132} 
& -4.161  \scriptsize{$\pm$ 0.195} 
\\

\textbf{MAT}   
& -0.328  \scriptsize{$\pm$ 0.001} 
& -3.100  \scriptsize{$\pm$ 0.279} 
& -1.109  \scriptsize{$\pm$ 0.132} 
& -1.811  \scriptsize{$\pm$ 0.117} 
& -5.002  \scriptsize{$\pm$ 0.755} 
\\

\textbf{FRAP}    
& \underline{-0.274  \scriptsize{$\pm$ 0.000}} 
& -2.573  \scriptsize{$\pm$ 0.012} 
& -1.064  \scriptsize{$\pm$ 0.002} 
& -0.705  \scriptsize{$\pm$ 0.020} 
& -3.191  \scriptsize{$\pm$ 0.170} 
\\ 

\textbf{MPLight} 
& -0.414  \scriptsize{$\pm$ 0.012} 
& -4.079  \scriptsize{$\pm$ 0.049} 
& -1.087  \scriptsize{$\pm$ 0.041} 
& -0.842  \scriptsize{$\pm$ 0.026} 
& -3.117  \scriptsize{$\pm$ 0.116} 
\\ 

\textbf{CoLight}  
& -0.309  \scriptsize{$\pm$ 0.006} 
& -2.326  \scriptsize{$\pm$ 0.057} 
& \underline{-0.918  \scriptsize{$\pm$ 0.042}} 
& -0.695  \scriptsize{$\pm$ 0.008} 
& {-2.939  \scriptsize{$\pm$ 0.092}} 
\\ 

\textbf{Advanced-CoLight}  
& -0.291  \scriptsize{$\pm$ 0.011} 
& \underline{-2.317  \scriptsize{$\pm$ 0.018}} 
& {-1.112  \scriptsize{$\pm$ 0.018}} 
& \underline{-0.607  \scriptsize{$\pm$ 0.001}} 
& \underline{-2.904  \scriptsize{$\pm$ 0.016}} 
\\

\textbf{MetaGAT}   
& -0.468  \scriptsize{$\pm$ 0.126} 
& -2.538  \scriptsize{$\pm$ 0.077} 
& -1.326  \scriptsize{$\pm$ 0.311} 
& -0.805  \scriptsize{$\pm$ 0.168} 
& -3.289  \scriptsize{$\pm$ 0.261} 
\\

\textbf{DuaLight}   
& -0.331  \scriptsize{$\pm$ 0.112} 
& -2.711  \scriptsize{$\pm$ 0.005} 
& -1.007  \scriptsize{$\pm$ 0.622} 
& -0.724  \scriptsize{$\pm$ 0.338} 
& -4.330  \scriptsize{$\pm$ 0.306} 
\\

\textbf{\nameOurs}     
& \textbf{-0.251  \scriptsize{$\pm$ 0.000}} 
& \textbf{-2.309  \scriptsize{$\pm$ 0.068}} 
& \textbf{-0.890  \scriptsize{$\pm$ 0.012}} 
& \textbf{-0.538  \scriptsize{$\pm$ 0.007}}
& \textbf{-2.899  \scriptsize{$\pm$ 0.014}} \\

\hline
\end{tabular}%
}
\caption{
Intersection-wise evaluation. \revise{Our method constantly achieves the best performance.}}
\label{tab:main-table22}
\vspace{-30pt}
\end{table}

\subsection{Baselines}

We analyze the performance of our method by comparing it with two conventional transportation techniques and six state-of-the-art (SOTA) RL/MARL algorithms.

\noindent\textbf{Conventional Methods:}
\\\noindent$\bullet$~\textbf{Fixed Time Control (FTC)}~\citep{roess2004traffic}  with random offset executes each phase within a loop, utilizing a pre-defined phase duration.
\\\noindent$\bullet$~\textbf{MaxPressure}~\citep{varaiya2013max,kouvelas2014maximum} greedily chooses the phase with the maximum pressure, which is a SOTA transportation control method.

\noindent\textbf{RL-based Methods:}
\\\noindent$\bullet$~\textbf{IPPO}~\citep{ault2019learning,ault2021reinforcement} controls each intersection with an independent PPO agent, which is trained with the data from the current intersection.
\\\noindent$\bullet$~\textbf{MAPPO}~\citep{schulman2017proximal,yu2022surprising} executes with an independent PPO agent and is trained collectively using the data from all intersections, enabling an optimized coordinated traffic flow.
\\\noindent$\bullet$~\textbf{MAT}~\citep{wen2022multi} is a strong baseline in MARL with centralized training with centralized execution paradigm, modeling the TSC as a sequential problem.
\\\noindent$\bullet$~\textbf{FRAP}~\citep{zheng2019learning} models phase competition and employs deep Q-network agent for each intersection to optimize traffic phase operation.
\\\noindent$\bullet$~\textbf{MPLight}~\citep{chen2020toward} utilizes the concept of pressure as both state and reward to coordinate multiple intersections, which is based on FRAP.
\\\noindent$\bullet$~\textbf{CoLight}~\citep{wei2019colight} leverages a GAT to extract neighboring information, thereby assisting the agent in optimizing queue length.
\\\noindent$\bullet$~\textbf{Advanced-CoLight}~\citep{zhang2022expression} combines advanced traffic state for the traffic movement representation with a pressure of queuing and demand of running for vehicles with CoLight, to enhance the decision-making process. 
\\\noindent$\bullet$~\textbf{MetaGAT}~\citep{lou2022meta} leverages GAT-based context to boost cooperation among intersections.
\\\noindent$\bullet$~\textbf{DuaLight}~\citep{lu2023dualight} introduces a scenario-specific experiential weight module and a scenario-shared co-train module to facilitate the information extraction of scenarios and intersections.

\subsection{Evaluation Metrics}

We utilize two evaluation metrics in our study. 
Firstly, at the scenario level~\citep{ault2021reinforcement}, we compute the average delay, and average trip time by tracking vehicles in the scenario. Specifically, delay signifies the holdup caused by signalized intersections (either stop or approach delay) for a vehicle, and trip time denotes the complete duration of a vehicle's journey from its origin to its destination.

Secondly, at the intersection level, we employ the external reward of the environment as an evaluation criterion, including the average delay time, average wait time, average queue length, and average pressure for each intersection. These metrics are calculated at each individual intersection by averaging the values across all vehicles.


\subsection{Main Results}

\paragraph{\textbf{Scenario-wise evaluation}}
As illustrated in Table~\ref{tab:main-table1}, the performances marked in boldface and underlined represent the best and second-best results, respectively. 
\nameOurs~consistently achieves substantial performance improvements, reducing the average delay time by 7.68\%, the average trip time by 1.98\%, 
which not only validates the effectiveness of~\nameOurs~but also highlights its potential to efficiently manage and enhance multi-intersection collaboration in various traffic scenarios. 

\paragraph{\textbf{Intersection-wise evaluation}}
Table~\ref{tab:main-table22} shows~\nameOurs~achieves the best results in all scenarios.
Compared to the second-best result, ours achieved a 7.71 \% improvement on average.
This consistent performance enhancement across intersection-wise evaluation metrics underlines the robustness of our proposed method.

Notably, these findings provide strong evidence that our algorithm performs well not only in terms of global cooperation (scenario-wise evaluation) but also from the perspective of benefits at individual intersections (intersection-wise evaluation). This dual-level efficacy showcases the effectiveness of our approach, signifying its ability to foster overall road network performance while simultaneously optimizing individual intersection operations. 

\subsection{Ablation Analysis of Three Settings}

\label{sec:ablation}
\begin{figure}[ht!]
	\centering
	\includegraphics[width=\linewidth]{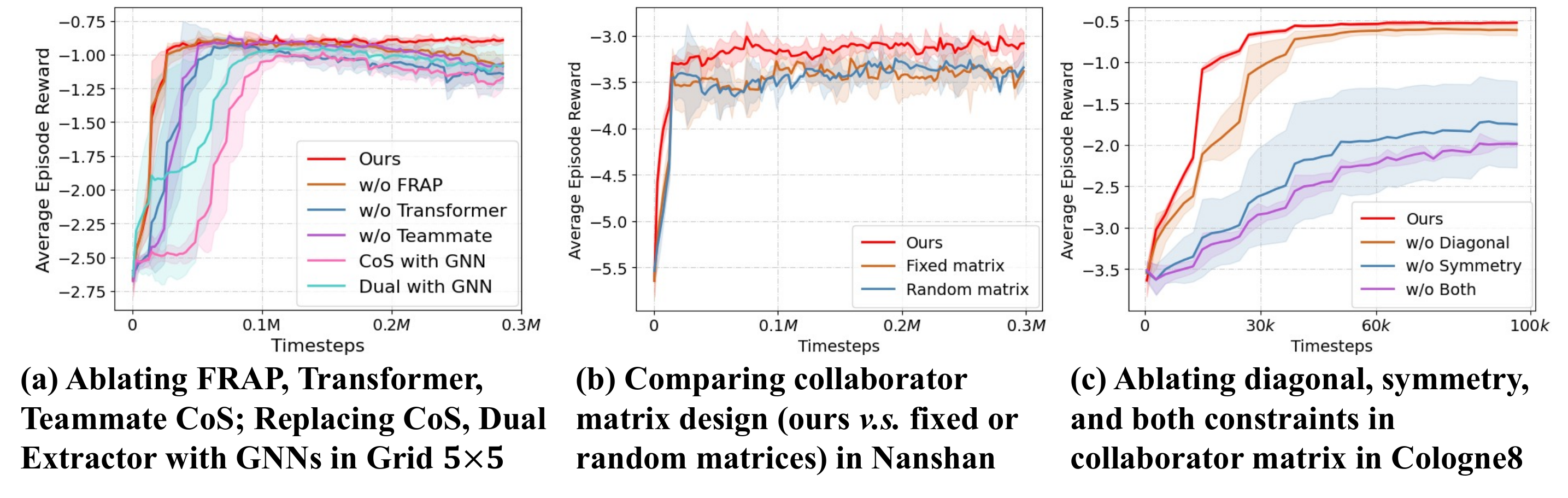}
    \vspace{-20pt}
	\caption{Ablation Studies}
    \label{fig:ablation}
    \vspace{-5pt}
\end{figure}


\paragraph{\textbf{Ablation of Components}}
Firstly, we selectively remove different modules, specifically the FRAP, Transformer, and Collaborator modules, to validate the necessity of each component in~\nameOurs.
In Figure~\ref{fig:ablation}(a), (1) it is evident that the Transformer module that aggregates other intersections' information and the Teammate CoS module that adaptively selects collaborators are the most important. 
(2) Moreover, to justify the MLP design for CoS, we replace it in~\abbr~ with GNNs, where we could observe a significant slowdown of convergence and performance drop in \textit{\abbr~with GNN}, due to GNN's computational complexity as the bottleneck. (3)  \revise{
To justify Transformer as a better design to aggregate other intersections' information, we replaced the transformer module in the dual extractor with GNNs (\textit{Dual with GNN}). We see similar performance drops. Such a feature extractor encourages learning better embedding for each agent and understanding them better, shown in Figure \ref{fig:vis_dual_fea}.
}

\paragraph{\textbf{Ablation of the Collaborator Matrix}} 
To further assess the impact of the co-learned collaborator matrix, we conduct experiments where the matrix in the proposed~\abbr~is replaced with both fixed (as topological adjacency matrix) and random (freezing the collaborator selection after randomly initializing top-$k$ selection) matrices. 
Figure~\ref{fig:ablation}(b) shows that using co-learned collaborator matrices in CoS can boost performance, highlighting the critical role of the dynamical collaboration matrix in achieving effective coordination. 
Thus, the joint optimization of the collaborator matrix with decision policies is key for optimizing cumulative rewards.


\paragraph{\textbf{Ablation of  Constraints on Collaborator Matrix}}
In Figure \ref{fig:ablation}(c), we further evaluate the contributions of the constraints in Eq~(\ref{eq:diag}) and~(\ref{eq:symm}) by removing the Diagonal constraint, the Symmetry constraint, or both. 
Removing the Symmetry constraint significantly degrades performance, which underlines the symmetric interplay between each other is essential.
Conversely, the Diagonal constraint has a marginal impact, primarily enhancing the convergence speed. 
These insights highlight the value of the Symmetry constraint for optimality and the Diagonal constraint for efficiency.

In summary, the ablations provide empirical evidence that each dimension of~\abbr~is vital. The co-learning of the collaborator matrix, the adherence to specific constraints, and the integration of crucial components such as the FRAP, Transformer, and Collaborator modules all contribute to the robustness and effectiveness of the system. 
Through validation, we demonstrate that our model leads to enhanced performance
in complex environments.

\subsection{Visualization Analysis of Dual-feature}

\begin{figure}[ht!]
\vspace{-15pt}
    \setcounter{subfigure}{0}
	\centering
	\subfigure[Avenue 4 $\times$ 4]{
		\begin{minipage}[t]{0.22\textwidth}
			\centering
			\includegraphics[width=\textwidth]{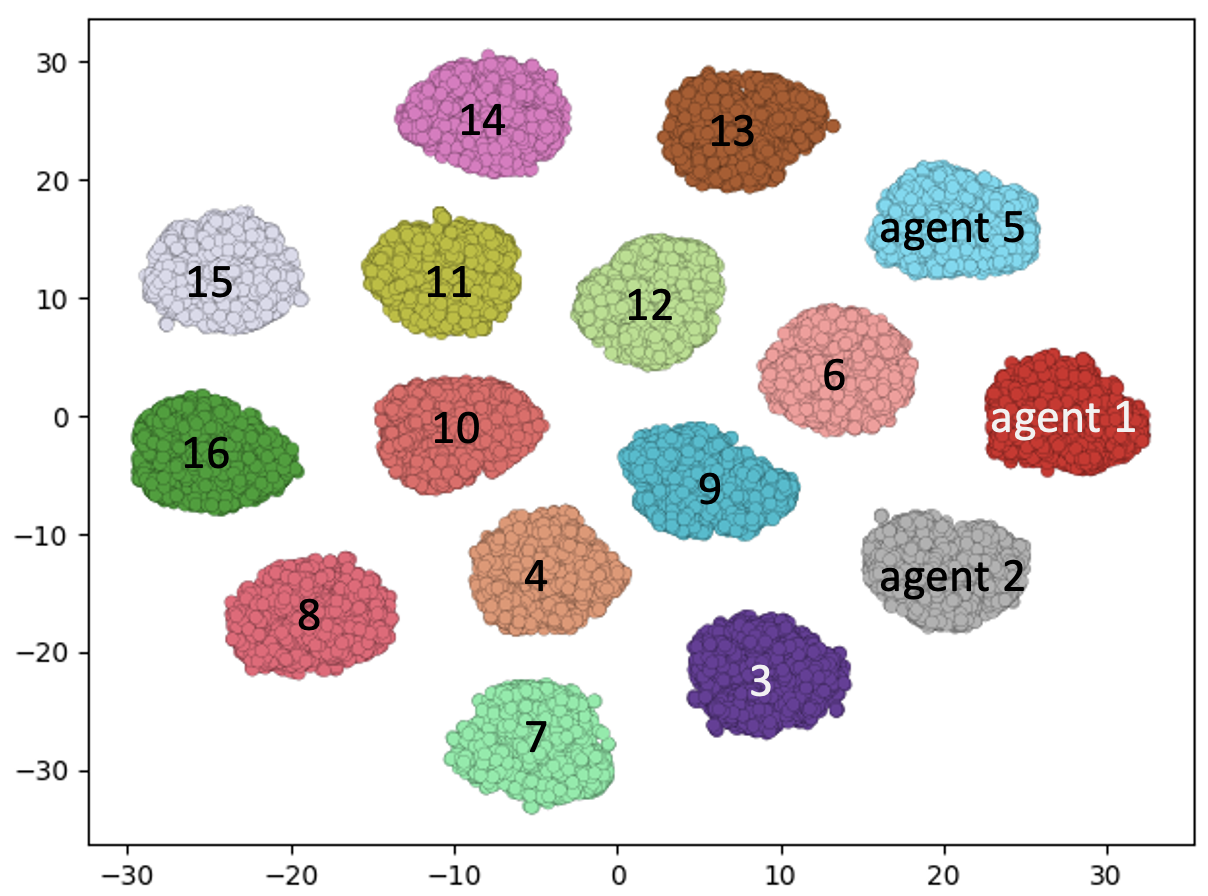}
		\end{minipage}
	}
	\subfigure[Nanshan]{
		\begin{minipage}[t]{0.22\textwidth}
			\centering
			\includegraphics[width=\textwidth]{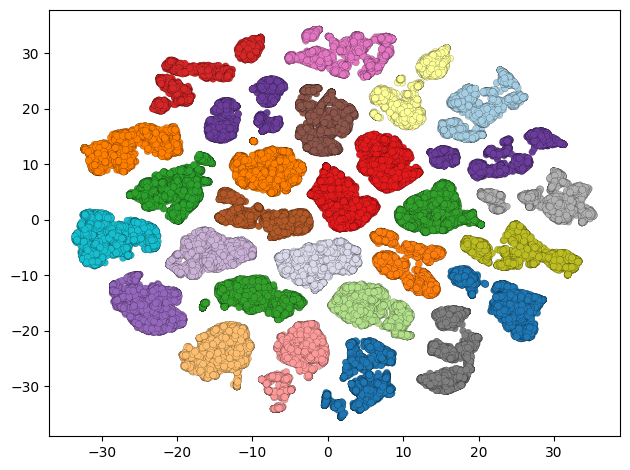}
		\end{minipage}
	}
 \vspace{-10pt}
 \caption{Visualization of Dual-feature. Each color represents a specific intersection. \revise{CoSLight has captured unique features at each intersection with distinct clustering patterns.}}
	\label{fig:vis_dual_fea}
\end{figure}

We analyze the dual-feature embeddings from our Dual-Feature Extractor. As shown in Figure~\ref{fig:vis_dual_fea}, we test 10 episodes for each intersection in each scenario, resulting in 2400 dual-feature embeddings visualized using the t-SNE technique~\citep{van2008visualizing}. 

These embeddings demonstrate a distinct clustering pattern, suggesting that our model captures unique features at each intersection effectively. This allows the model to group similar states together, adapting to variations in traffic conditions and intersection-level characteristics. This adaptability is a crucial advantage of our approach, contributing significantly to its performance improvement.

More results of all five scenarios under the above settings are in Appendix~\ref{app:res_inter} - \ref{app:vis_teammate_assign}.
In conclusion, through rigorous experiments and insightful analysis, our study confirms that our method, which integrates dual-feature extraction and multi-intersection collaboration, provides an effective and efficient solution for the TSC.

\subsection{Analysis of Collaborator Number $k$}


\begin{figure}[ht]
\vspace{-15pt}
    \setcounter{subfigure}{0}
	\centering
	\subfigure[Avenue 4 $\times$ 4]{
		\begin{minipage}[t]{0.22\textwidth}
			\centering
            \label{fig:ave44_k1}
			\includegraphics[width=\textwidth]{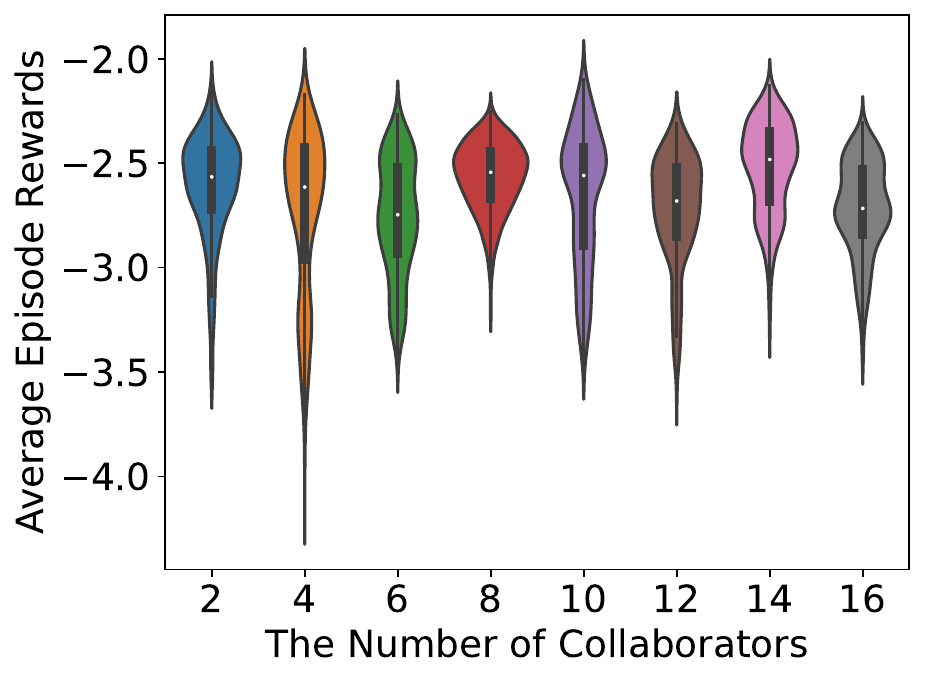}
		\end{minipage}
	}
	\subfigure[Grid 5 $\times$ 5]{
		\begin{minipage}[t]{0.22\textwidth}
			\centering
            \label{fig:grid55_k1}
			\includegraphics[width=\textwidth]{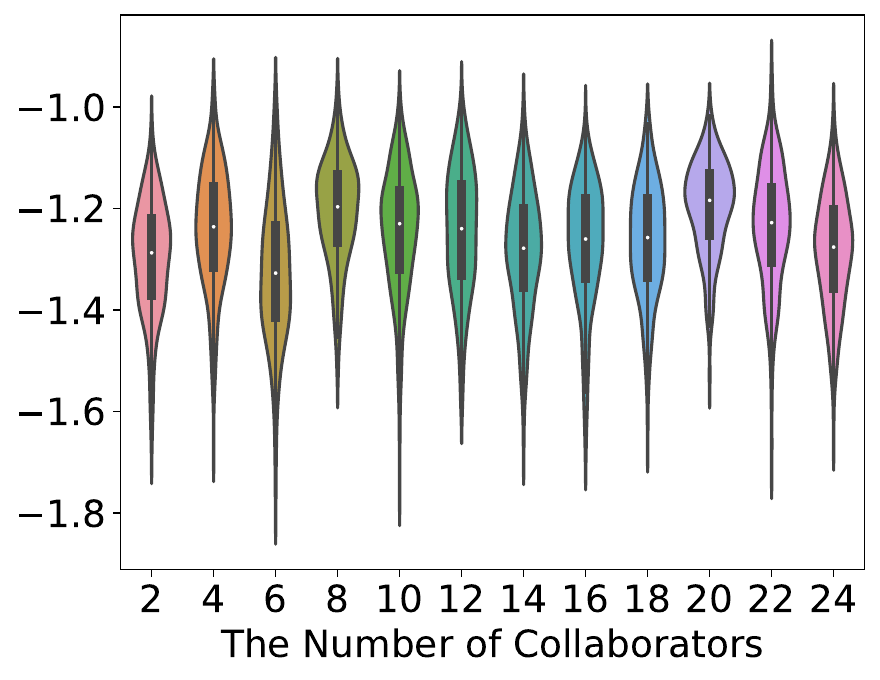}
		\end{minipage}
	}


    \vspace{-15pt}
	\caption{\revise{Violin plots display the performance trade-offs at varying numbers of collaborators.}}
	\label{fig:abs_k}
\end{figure}

Figure~\ref{fig:abs_k} shows the trade-off between the number of collaborators and performance. 
For example, in Avenue 4$\times$4 (Figure~\ref{fig:ave44_k1}), $k=8$ yields the best results, suggesting an optimal balance between useful information and performance gains.
Information from more collaborators beyond this point does not guarantee improved results and may lead to higher resource usage. 
This finding poses a direction for future work to dynamically determine the ideal number of collaborators, potentially enhancing the algorithm's efficiency.

\subsection{Visualization of Collaborator Matrix}

\begin{figure}[ht!]
\vspace{-10pt}
	\centering
	\includegraphics[width=\linewidth]{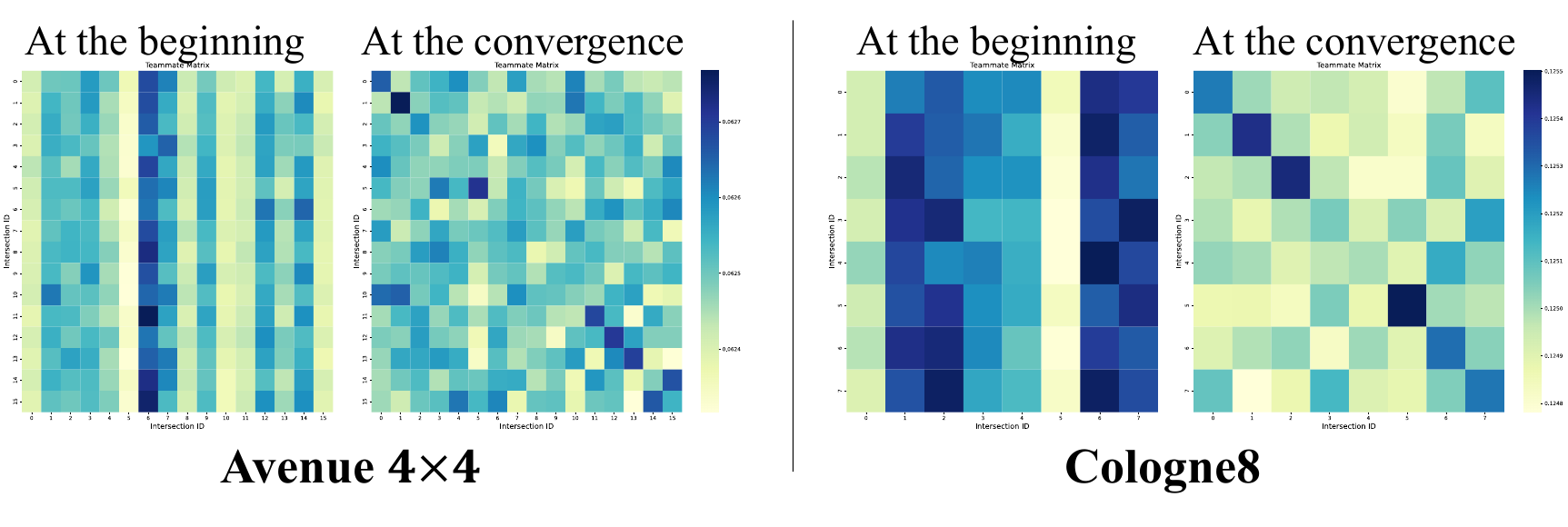}
    \vspace{-20pt}
	\caption{Saliency maps of collaborator matrix. The deeper the color, the stronger the correlation. \revise{CoSLight has learned diagonal maximization and symmetry constraints.}}
    \label{fig:vis_team_matrix}
    \vspace{-5pt}
\end{figure}

In this section, we visually analyze the collaborator matrix to offer an intuitive understanding of attention distribution among intersections during training.
The results are shown at the start and end of training in the Cologne8 (8 intersections) and Avenue 4 $\times$ 4 (16 intersections) scenarios, respectively.
The saliency maps in Figure~\ref{fig:vis_team_matrix} depict the state of the collaborator matrix at the beginning and the end of the training, respectively. 

Upon the conclusion of the training, we notice a deepening of the color along the diagonal elements of the saliency map. This signifies an increased self-attention, indicating that the intersections have adapted to pay more heed to their own states. Additionally, the symmetry apparent in the saliency map suggests mutual awareness among intersections. As the training progresses, intersections not only learn to focus on themselves but also pay attention to their peers, signifying a learned mutual collaboration.

These observations validate the effectiveness of our approach in creating a collaborative environment among intersections, thus leading to enhanced performance.


\subsection{Visualization of Collaborator Selection}

\begin{figure}[ht!]
	\centering
	\includegraphics[width=\columnwidth]{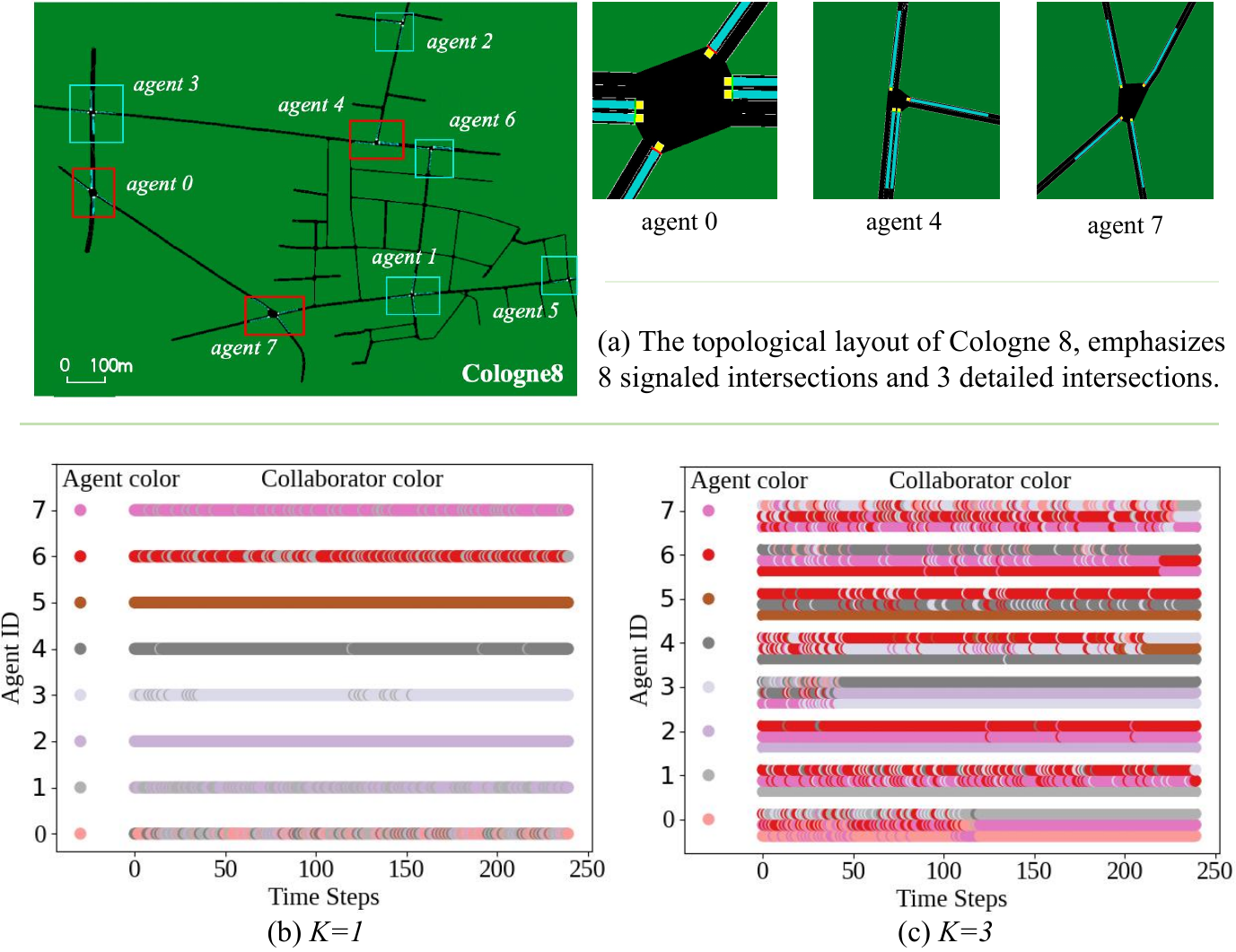}
\caption{Collaborator selection on Cologne8\revise{, showing self-selection and efficient cross-collaboration at (b) K = 1 with predominantly solo dynamics, and (c) K = 3 with inter-agent collaboration 
 with not just the topological selection.}}
\label{fig:vis_cteammate_assign}
\end{figure}


Figure~\ref{fig:vis_cteammate_assign} depicts the collaborator selection process. For $K=1$ in Figure~\ref{fig:vis_cteammate_assign} (b), self-selection is prevalent; however, agent 0 displays varied collaboration patterns, likely due to the intricacy of its signal control tasks (refer to Figure~\ref{fig:vis_cteammate_assign} (a)), which require engaging with multiple collaborators for optimal traffic management. 
When $K$ increases to 3, as shown in Figure~\ref{fig:vis_cteammate_assign} (c), agents exhibit both self-selection and mutual collaboration, forming complex interaction networks. For example, agent 3 largely collaborates with agents 2 (non-neighbor) and 4 (neighbor); Similarly, agent 4 with agents 3 (neighbor) and 6 (neighbor); Also, agent 2 with agents 6 (non-neighbor) and 7 (non-neighbor). Specifically, agent 2 is quite far from 6 and 7, but they form a strong collaboration since agent 2 is in the office building and agents 6 and 7 are in the community residential region. 

Overall, such learned patterns suggest strategic selection beyond topological neighbors.
Collectively, these results affirm that the collaborator selection mechanism is adaptively responsive to both the complexity of traffic tasks and the benefits of strategic collaboration to optimize traffic flow.


\subsection{Average Inference and Training Time}
We collected 100 episodes over 100 training epochs to obtain the average inference time for CoS and Decision policies per episode and average training time per epoch. We experimented on 4 NVIDIA TITAN Xp GPUs(12G).
In Table~\ref{tab:time_}, across five scenarios, CoS inference and training times average 2.83\% and 31.29\%, respectively, while Decision policy averages 2.66\% for inference and 63.21\% for training. The CoS strategy adds a 34.42\% time overhead, justifiable by its performance benefits.

\begin{table}[htbp]
\centering
\resizebox{0.99\columnwidth}{!}{%
\begin{tabular}{lcccccc}
\toprule
 & CoS & Decision policy & CoS Training & Decision Training & Total Time \\
\midrule
Grid 4x4 & 0.153±0.008 & 0.143±0.008 & 1.522±0.112 & 2.940±0.163 & 4.758±0.179 \\
Avenue 4x4 & 0.161±0.006 & 0.152±0.006 & 1.423±0.176 & 3.050±0.203 & 4.785±0.276 \\
Grid 5x5 & 0.168±0.011 & 0.153±0.011 & 2.531±0.093 & 6.148±0.123 & 9.000±0.123 \\
Cologne8 & 0.111±0.007 & 0.105±0.007 & 2.001±0.096 & 2.336±0.192 & 4.553±0.186 \\
Nanshan & 0.190±0.008 & 0.181±0.008 & 1.321±0.124 & 4.130±0.511 & 5.822±0.507 \\
Avg (Percent) & 0.156(2.83\%) & 0.147(2.66\%) & 1.760(31.29\%) & 3.721(63.21\%) & 5.778 \\
\bottomrule
\end{tabular}
}
\caption{Average inference and training time (s).}
\label{tab:time_}
\end{table}


%% file: 5-con.tex
\section{Conclusion}

In this paper, we introduce an innovative approach to traffic signal control, employing a top-$k$ collaborator selection policy with a dual-feature extractor. 
This unique strategy allows for the effective extraction of phase- and intersection-level representations while adaptively selecting collaborators for enhanced multi-intersection collaboration.
Moreover, we are the first to propose a joint optimization regime to train the~\abbr~and decision policies simultaneously for maximizing the cumulative discounted return.
Comprehensive experiments on both synthetic and real-world datasets validate our approach's superiority. 
The extensive analysis further reinforces the effectiveness and efficacy of~\nameOurs.

\textbf{Future research} could potentially explore an adaptive mechanism to efficiently determine the optimal number of collaborators, thereby enhancing the performance and effectiveness of traffic signal control. 
Moreover, enhancing the explainability of collaborator selection processes could provide valuable insights, potentially enabling more intuitive and transparent decision-making to promote cooperation.

%% file: 6-app.tex
\setcounter{secnumdepth}{2}

\section{Detailed Network Architecture and Hyper-parameters Descriptions}
\label{app:network_para}
There is a summary of all the neural networks used in our framework about the network structure, layers, and activation functions.
\begin{table}[h!]
    \centering
    
    \begin{tabular}{ccccc}
       \hline
       ~ & Network Structure &  Layers & Hidden Size & Activation Functions  \\
       \hline
       FRAP & FRAP\footnotemark[1] & - &  -  & - \\
        \hline
        \multirow{3}{*}{Transformer Backbone} 
        & Positional Embedding & 1 & 64 & None \\
        & Encoder Layer &  2 & 64 & ReLu \\
        & Output Layer (MLP) 3 & 1 & 32 & ReLu \\
        \hline
      Top-k Collaborator Assignment & MLP & 2 &  64  & ReLu   \\
        \hline
       Actor &   MLP+RNN+MLP  & 3 & [64]+[64]+[32]  &  ReLu \\
        \hline
       Critic & MLP & 2 & 64 & ReLu\\
       \hline
    \end{tabular}
   \caption{The Summary for Network Architecture}
   \label{tab:net}
\end{table}
\footnotetext[1]{Refer to the implementation: \url{https://github.com/Chacha-Chen/MPLight}}


There are our hyper-parameter settings for the training, shown in Table~\ref{tab:parameter}.



\begin{table}[ht!]
\centering
\begin{tabular}{cc}
\hline
\textbf{Description} &  \\textbf{Value}  \\
\hline
    optimizer & $AdamW$  \\
    learning rate & $5*10^{-4}$  \\
    group embedding size  &   $32$ \\
    attention head &    $8$ \\
    transformer layer &    $2$ \\
    actor embedding size &    $32$ \\
    state key  & $['current\_phase', 'car\_num',  'queue\_length',$ \\ & $'occupancy', 'flow', 'stop\_car\_num', 'pressure']$\\
    number of actions & 8 \\
    interaction steps & 300000 \\
    $\alpha$        &    $10^{-4}$  \\
    $\beta_1$       &    $0.9$  \\
    $\beta_2$       &    $0.999$  \\
    $\varepsilon$-greedy $\varepsilon$   &    $10^{-5}$  \\
    clipping $\epsilon$       &   $0.2$ \\
    seed             &   $[0,10)$  \\
    number of process &  $64$  \\
    eval interval  &     $4000$  \\
    eval episodes  &     $100$   \\
\hline
\end{tabular}
\caption{The hyper-parameter settings.}
\label{tab:parameter}
\end{table}

\vspace{10pt}

By the way, 
we refer readers to the source code in the supplementary to check the detailed hyper-parameters.

\section{The Details about FRAP}
\label{app:frap}
At the \textbf{phase} level, we adopt FRAP~\citep{zheng2019learning} to obtain the phase-wise representation. 
The raw observations $o$ from the simulator include $K$ features, such as the number of vehicles, queue length, the current phase, the flow, etc. For any traffic movement $m, m \in \{1,...,8\}$ in an intersection $i$, the $k$-th feature in the raw observation can be denoted as $o_{m,k}^i$. For brevity, the superscript $i$ is omitted hereinafter.
First, the embedding of traffic movement $m$ is obtained by Multi-Layer Perceptron (MLP):
\begin{equation} 
\bm{e}_{m} = ||_{k=1}^K Sigmoid(MLP_k(o_{m,k})),
\end{equation} 
where $||$ denotes concatenation, and $Sigmoid$ is the activation function.
Then FRAP module is applied to extract the phase competition representation, denoted as follows.
\begin{equation}
    \bm{e}_{pcr} = FRAP(\bm{e}_{m_1},...,\bm{e}_{m_8}).
\end{equation}
The process can be summarized as follows.
\begin{enumerate}
    \item \textbf{Phase embedding}: Each phase $p$ consists of two movements $m_1, m_2$, and we get the phase embedding $\bm{e}_p = \bm{e}_{m_1} + \bm{e}_{m_2}$.
    \item \textbf{Phase pair representation}: For any pair $p_k, p_l$ from different phases, the pairwise relation vector is $\bm{e}_{p_k, p_l} = \bm{e}_{p_k} || \bm{e}_{p_l}$. Gathering the vectors of all phase pairs can obtain the pair demand embedding volume $\bm{E}$. Then the phase pair representation can be denoted as $\bm{e}_{ppr} = Conv_{1 \times 1}(\bm{E})$, where $Conv_{1 \times 1}$ is the convolutional layer with 1 × 1 filters. 
    \item \textbf{Phase competition}: Let $\bm{M}$ be phase competition mask, and the phase competition representation can be obtained by: $\bm{e}_{pcr} = Conv_{1 \times 1}(\bm{e}_{ppr} \otimes \bm{M})$, where $\otimes$ is the element-wise multiplication. Here, we reshape $\bm{e}_{pcr}$ as a vector through \textit{flatten} operation.
\end{enumerate}

Finally, an MLP is utilized to mine the phase representation in the intersection $i$ as follows.
\begin{equation}
    \bm{e}^i_{pr} = MLP(\bm{e}_{pcr})
\end{equation}

\section{The Details about Evaluation Settings}
\label{app:eva_detail}

In Table~\ref{tab:scenarios}, we present detailed statistics including the total number of intersections (\#Total Int.), along with the quantity of 2-arm, 3-arm, and 4-arm intersections in each scenario.

In real-world traffic scenarios, many intersections don't conform to a standard four-arm structure, potentially having varied lane counts and orientations. To ensure the broad application of our method across diverse scenarios, we intentionally conducted experiments in two settings that feature non-standard intersections: Cologne8 and Nanshan.

These scenarios incorporate not just the typical four-arm intersections but also irregular ones where the number and direction of lanes deviate from the standard. By running experiments in such environments, we demonstrate that our approach is effective not only in handling standard intersections but also excels in managing these irregularities. This further underscores the robustness and potential wide-scale applicability of our methodology.

\section{Additional Main Results of Intersection-wise Evaluations}

\label{app:res_inter}

In the section, we provide the additional results with the intersection-level evaluation in Table~\ref{tab:main-table2}.
Our algorithm consistently yielded favorable results, even achieving superior evaluation improvements in many cases. This consistent performance enhancement across intersection-wise evaluation metrics underlines the robustness of our proposed method.

\begin{table*}[ht!]
    \centering
    
    \small 
    \renewcommand{\arraystretch}{1.2} 
    \begin{tabular}{ccllllll}
    \hline
        Model & Intersection-wise Metrics & Grid 4 $\times$ 4 & Avenue 4 $\times$ 4 & Grid 5 $\times$ 5 & Cologne 8 & Nanshan \\ \hline
        \multirow{5}{*}{FTC} & Total & -0.614  \scriptsize{$\pm$ 0.015} & -4.503  \scriptsize{$\pm$ 0.025} & -2.346  \scriptsize{$\pm$ 0.052} & -2.114  \scriptsize{$\pm$ 0.021} & -3.479  \scriptsize{$\pm$ 0.186} \\ 
        ~ & Delay time & -0.065  \scriptsize{$\pm$ 0.001} & -0.660  \scriptsize{$\pm$ 0.004} & -0.213  \scriptsize{$\pm$ 0.010} & -0.214  \scriptsize{$\pm$ 0.000} & -0.603  \scriptsize{$\pm$ 0.050} \\ 
        ~ & Wait time & -0.279  \scriptsize{$\pm$ 0.007} & -1.175  \scriptsize{$\pm$ 0.010} & -0.605  \scriptsize{$\pm$ 0.006} & -1.210  \scriptsize{$\pm$ 0.013} & -1.566  \scriptsize{$\pm$ 0.043} \\ 
        ~ & Queue length & -0.203  \scriptsize{$\pm$ 0.005} & -2.294  \scriptsize{$\pm$ 0.018} & -1.154  \scriptsize{$\pm$ 0.028} & -0.564  \scriptsize{$\pm$ 0.007} & -1.079  \scriptsize{$\pm$ 0.095} \\ 
        ~ & Pressure & -0.066  \scriptsize{$\pm$ 0.001} & -0.372  \scriptsize{$\pm$ 0.005} & -0.373  \scriptsize{$\pm$ 0.015} & -0.125  \scriptsize{$\pm$ 0.000} & -0.229  \scriptsize{$\pm$ 0.052} \\ \hline
        
        \multirow{5}{*}{MaxPressure} & Total & -0.393  \scriptsize{$\pm$ 0.003} & -4.032  \scriptsize{$\pm$ 0.040} & -1.132  \scriptsize{$\pm$ 0.013} & -0.756  \scriptsize{$\pm$ 0.012} & -3.055  \scriptsize{$\pm$ 0.162} \\ 
        ~ & Delay time & -0.055  \scriptsize{$\pm$ 0.000} & -0.600  \scriptsize{$\pm$ 0.009} & -0.123  \scriptsize{$\pm$ 0.002} & -0.143  \scriptsize{$\pm$ 0.002} & \textcolor{steelblue2}{-0.547  \scriptsize{$\pm$ 0.062}} \\ 
        ~ & Wait time & -0.173  \scriptsize{$\pm$ 0.002} & -1.168  \scriptsize{$\pm$ 0.012} & -0.379  \scriptsize{$\pm$ 0.008} & -0.373  \scriptsize{$\pm$ 0.006} & -1.398  \scriptsize{$\pm$ 0.032} \\ 
        ~ & Queue length & -0.115  \scriptsize{$\pm$ 0.000} & -1.811  \scriptsize{$\pm$ 0.023} & -0.447  \scriptsize{$\pm$ 0.009} & -0.198  \scriptsize{$\pm$ 0.005} & -0.958  \scriptsize{$\pm$ 0.066} \\ 
        ~ & Pressure & -0.050  \scriptsize{$\pm$ 0.001} & -0.453  \scriptsize{$\pm$ 0.006} & -0.184  \scriptsize{$\pm$ 0.003} & -0.042  \scriptsize{$\pm$ 0.001} & \textcolor{steelblue2}{\textbf{-0.153  \scriptsize{$\pm$ 0.020}}} \\ \hline
        
        \multirow{5}{*}{Frap} & Total & \textcolor{steelblue2}{-0.274  \scriptsize{$\pm$ 0.000}} & -2.573  \scriptsize{$\pm$ 0.012} & -1.064  \scriptsize{$\pm$ 0.002} & -0.705  \scriptsize{$\pm$ 0.020} & -3.191  \scriptsize{$\pm$ 0.170} \\ 
        ~ & Delay time & \textcolor{steelblue2}{-0.051  \scriptsize{$\pm$ 0.001}} & -0.403  \scriptsize{$\pm$ 0.003} & -0.120  \scriptsize{$\pm$ 0.001} & -0.146  \scriptsize{$\pm$ 0.001} & -0.600  \scriptsize{$\pm$ 0.012 }\\ 
        ~ & Wait time & \textcolor{steelblue2}{-0.107  \scriptsize{$\pm$ 0.001}} & -0.525  \scriptsize{$\pm$ 0.004} & {-0.352  \scriptsize{$\pm$ 0.002}} & {-0.313  \scriptsize{$\pm$ 0.003}} & -1.391  \scriptsize{$\pm$ 0.021} \\ 
        ~ & Queue length & \textcolor{steelblue2}{-0.079  \scriptsize{$\pm$ 0.000}} & -1.287  \scriptsize{$\pm$ 0.010} & -0.392  \scriptsize{$\pm$ 0.002} & -0.196  \scriptsize{$\pm$ 0.001} & -0.985  \scriptsize{$\pm$ 0.090} \\ 
        ~ & Pressure & \textcolor{steelblue2}{\textbf{-0.035  \scriptsize{$\pm$ 0.000}}} & -0.436  \scriptsize{$\pm$ 0.002} & -0.193  \scriptsize{$\pm$ 0.001} & -0.042  \scriptsize{$\pm$ 0.000} & -0.215  \scriptsize{$\pm$ 0.020} \\ \hline
        
        \multirow{5}{*}{IPPO} & Total & -0.336  \scriptsize{$\pm$ 0.004} & {-2.558  \scriptsize{$\pm$ 0.213}} & -0.943  \scriptsize{$\pm$ 0.037} & \textcolor{steelblue2}{-0.646  \scriptsize{$\pm$ 0.015}} & -3.555  \scriptsize{$\pm$ 0.097} \\ 
        ~ & Delay time & {-0.053  \scriptsize{$\pm$ 0.000}} & \textcolor{steelblue2}{\textbf{-0.314  \scriptsize{$\pm$ 0.011}}} & -0.104  \scriptsize{$\pm$ 0.005} & {-0.133  \scriptsize{$\pm$ 0.001}} & -0.586  \scriptsize{$\pm$ 0.027} \\ 
        ~ & Wait time & -0.145  \scriptsize{$\pm$ 0.002} & \textcolor{steelblue2}{\textbf{-0.395  \scriptsize{$\pm$ 0.064}}} & -0.325  \scriptsize{$\pm$ 0.015} & \textcolor{steelblue2}{-0.314  \scriptsize{$\pm$ 0.001}} & -1.572  \scriptsize{$\pm$ 0.056} \\ 
        ~ & Queue length & -0.095  \scriptsize{$\pm$ 0.001} & \textcolor{steelblue2}{-1.103  \scriptsize{$\pm$ 0.029}} & \textcolor{steelblue2}{-0.352  \scriptsize{$\pm$ 0.011}} & \textcolor{steelblue2}{-0.164  \scriptsize{$\pm$ 0.004}} & -1.170  \scriptsize{$\pm$ 0.026 }\\ 
        ~ & Pressure & -0.043  \scriptsize{$\pm$ 0.001} & \textcolor{steelblue2}{\textbf{-0.346  \scriptsize{$\pm$ 0.021}}} & \textcolor{steelblue2}{-0.162  \scriptsize{$\pm$ 0.009}} & \textcolor{steelblue2}{-0.037  \scriptsize{$\pm$ 0.001}} & -0.227  \scriptsize{$\pm$ 0.017} \\ \hline
        
        \multirow{5}{*}{MAPPO} & Total & -0.308  \scriptsize{$\pm$ 0.006} & -2.744  \scriptsize{$\pm$ 0.238} & -1.273  \scriptsize{$\pm$ 0.107} & -1.697  \scriptsize{$\pm$ 0.132} & -4.161  \scriptsize{$\pm$ 0.195} \\ 
        ~ & Delay time & \textcolor{steelblue2}{-0.051  \scriptsize{$\pm$ 0.001}} & \textcolor{steelblue2}{-0.390  \scriptsize{$\pm$ 0.039}} & -0.126  \scriptsize{$\pm$ 0.008} & -0.207  \scriptsize{$\pm$ 0.013} & -0.682  \scriptsize{$\pm$ 0.020} \\ 
        ~ & Wait time & -0.132  \scriptsize{$\pm$ 0.003} & -0.497  \scriptsize{$\pm$ 0.060} & -0.334  \scriptsize{$\pm$ 0.019} & -0.899  \scriptsize{$\pm$ 0.064} & -1.710  \scriptsize{$\pm$ 0.094} \\ 
        ~ & Queue length & -0.085  \scriptsize{$\pm$ 0.002} & -1.433  \scriptsize{$\pm$ 0.139} & -0.552  \scriptsize{$\pm$ 0.054} & -0.443  \scriptsize{$\pm$ 0.042} & -1.446  \scriptsize{$\pm$ 0.096} \\ 
        ~ & Pressure & -0.040  \scriptsize{$\pm$ 0.001} & -0.423  \scriptsize{$\pm$ 0.037} & -0.260  \scriptsize{$\pm$ 0.031} & -0.147  \scriptsize{$\pm$ 0.022} & -0.323  \scriptsize{$\pm$ 0.010 }\\ \hline
        
        \multirow{5}{*}{MPlight} & Total & -0.414  \scriptsize{$\pm$ 0.012} & -4.079  \scriptsize{$\pm$ 0.049} & -1.087  \scriptsize{$\pm$ 0.041} & -0.842  \scriptsize{$\pm$ 0.026} & -3.117  \scriptsize{$\pm$ 0.116} \\ 
        ~ & Delay time & -0.056  \scriptsize{$\pm$ 0.001} & -0.656  \scriptsize{$\pm$ 0.022} & -0.115  \scriptsize{$\pm$ 0.006} & -0.150  \scriptsize{$\pm$ 0.002} & -0.561  \scriptsize{$\pm$ 0.019 }\\ 
        ~ & Wait time & -0.182  \scriptsize{$\pm$ 0.006} & -0.907  \scriptsize{$\pm$ 0.112} & -0.362  \scriptsize{$\pm$ 0.008} & -0.424  \scriptsize{$\pm$ 0.015} & -1.430  \scriptsize{$\pm$ 0.053} \\ 
        ~ & Queue length & -0.124  \scriptsize{$\pm$ 0.003} & -2.057  \scriptsize{$\pm$ 0.120} & -0.429  \scriptsize{$\pm$ 0.018} & -0.225  \scriptsize{$\pm$ 0.008} & -0.964  \scriptsize{$\pm$ 0.043} \\ 
        ~ & Pressure & -0.052  \scriptsize{$\pm$ 0.001} & -0.459  \scriptsize{$\pm$ 0.013} & -0.181  \scriptsize{$\pm$ 0.013} & -0.045  \scriptsize{$\pm$ 0.002} & \textcolor{steelblue2}{-0.162  \scriptsize{$\pm$ 0.006}} \\ \hline
        
        \multirow{5}{*}{CoLight} & Total & -0.309  \scriptsize{$\pm$ 0.006} & -2.326  \scriptsize{$\pm$ 0.057} & \textcolor{steelblue2}{-0.918  \scriptsize{$\pm$ 0.042}} & -0.695  \scriptsize{$\pm$ 0.008} & \textcolor{steelblue2}{-2.939  \scriptsize{$\pm$ 0.092}} \\ 
        ~ & Delay time & \textcolor{steelblue2}{-0.051  \scriptsize{$\pm$ 0.000}} & {-0.395  \scriptsize{$\pm$ 0.010}} & \textcolor{steelblue2}{-0.099  \scriptsize{$\pm$ 0.004}} & \textcolor{steelblue2}{-0.132  \scriptsize{$\pm$ 0.001}} & -0.559  \scriptsize{$\pm$ 0.041} \\ 
        ~ & Wait time & -0.135  \scriptsize{$\pm$ 0.003} & \textcolor{steelblue2}{-0.440  \scriptsize{$\pm$ 0.015}} & \textcolor{steelblue2}{-0.283  \scriptsize{$\pm$ 0.011}} & -0.351  \scriptsize{$\pm$ 0.005} & \textcolor{steelblue2}{-1.321  \scriptsize{$\pm$ 0.049 }}\\ 
        ~ & Queue length & -0.083  \scriptsize{$\pm$ 0.002} & -1.114  \scriptsize{$\pm$ 0.032} & {-0.356  \scriptsize{$\pm$ 0.019}} & -0.171  \scriptsize{$\pm$ 0.003} & \textcolor{steelblue2}{-0.880  \scriptsize{$\pm$ 0.012}} \\ 
        ~ & Pressure & -0.039  \scriptsize{$\pm$ 0.001} & \textcolor{steelblue2}{-0.376  \scriptsize{$\pm$ 0.007}} & -0.180  \scriptsize{$\pm$ 0.017} & -0.041  \scriptsize{$\pm$ 0.002} & -0.179  \scriptsize{$\pm$ 0.006} \\ \hline

        \multirow{5}{*}{MetaGAT} 
        & Total 
        & -0.468  \scriptsize{$\pm$ 0.126} 
        & \textcolor{steelblue2}{-2.538  \scriptsize{$\pm$ 0.077} }
        & -1.326  \scriptsize{$\pm$ 0.311} 
        & -0.805  \scriptsize{$\pm$ 0.168} 
        & -3.289  \scriptsize{$\pm$ 0.261} \\ 
        ~ & Delay time 
        & -0.059  \scriptsize{$\pm$ 0.008} 
        & -0.399  \scriptsize{$\pm$ 0.009} 
        & -0.139  \scriptsize{$\pm$ 0.029} 
        & -0.144  \scriptsize{$\pm$ 0.009} 
        & -0.627  \scriptsize{$\pm$ 0.031} \\ 
        ~ & Wait time 
        & -0.195  \scriptsize{$\pm$ 0.042} 
        & -0.531  \scriptsize{$\pm$ 0.055} 
        & -0.377  \scriptsize{$\pm$ 0.045} 
        & -0.409  \scriptsize{$\pm$ 0.101} 
        & -1.424  \scriptsize{$\pm$ 0.08} \\ 
        ~ & Queue length 
        & -0.153  \scriptsize{$\pm$ 0.06} 
        & -1.207  \scriptsize{$\pm$ 0.061} 
        & -0.55  \scriptsize{$\pm$ 0.155} 
        & -0.207  \scriptsize{$\pm$ 0.047} 
        & -1.022  \scriptsize{$\pm$ 0.13} \\ 
        ~ & Pressure 
        & -0.061  \scriptsize{$\pm$ 0.017} 
        & -0.402  \scriptsize{$\pm$ 0.014} 
        & -0.26  \scriptsize{$\pm$ 0.087} 
        & -0.046  \scriptsize{$\pm$ 0.012} 
        & -0.215  \scriptsize{$\pm$ 0.028} \\ \hline

        \multirow{5}{*}{Ours} 
        & Total 
        & \textcolor{steelblue2}{\textbf{-0.251  \scriptsize{$\pm$ 0.000}}} 
        & \textcolor{steelblue2}{\textbf{-2.309  \scriptsize{$\pm$ 0.068}}} 
        & \textcolor{steelblue2}{\textbf{-0.890  \scriptsize{$\pm$ 0.012}}} 
        & \textcolor{steelblue2}{\textbf{-0.538  \scriptsize{$\pm$ 0.007}}}
        & \textcolor{steelblue2}{\textbf{-2.899  \scriptsize{$\pm$ 0.014}}} \\ 
        
        ~ & Delay time 
        & \textcolor{steelblue2}{\textbf{-0.041  \scriptsize{$\pm$ 0.000}}} 
        & -0.551  \scriptsize{$\pm$ 0.004} 
        & \textcolor{steelblue2}{\textbf{-0.090  \scriptsize{$\pm$ 0.006}}} 
        & \textcolor{steelblue2}{\textbf{-0.117  \scriptsize{$\pm$ 0.002}}} 
        & \textcolor{steelblue2}{\textbf{-0.532  \scriptsize{$\pm$ 0.011}}} \\ 
        
        ~ & Wait time 
        & \textcolor{steelblue2}{\textbf{-0.100  \scriptsize{$\pm$ 0.002}}} 
        & -0.597  \scriptsize{$\pm$ 0.008} 
        & \textcolor{steelblue2}{\textbf{-0.281  \scriptsize{$\pm$ 0.011}}} 
        & \textcolor{steelblue2}{\textbf{-0.243  \scriptsize{$\pm$ 0.001}}} 
        & \textcolor{steelblue2}{\textbf{-1.39  \scriptsize{$\pm$ 0.037}}} \\ 
        
        ~ & Queue length 
        & \textcolor{steelblue2}{\textbf{-0.061  \scriptsize{$\pm$ 0.001}}} 
        & \textcolor{steelblue2}{\textbf{-0.231  \scriptsize{$\pm$ 0.013}}} 
        & \textcolor{steelblue2}{\textbf{-0.358  \scriptsize{$\pm$ 0.007}}} 
        & \textcolor{steelblue2}{\textbf{-0.120  \scriptsize{$\pm$ 0.002}}} 
        & \textcolor{steelblue2}{\textbf{-0.801  \scriptsize{$\pm$ 0.031}}} \\ 
        
        ~ & Pressure 
        & \textcolor{steelblue2}{{-0.039  \scriptsize{$\pm$ 0.002}}} 
        & -0.492  \scriptsize{$\pm$ 0.005} 
        & \textcolor{steelblue2}{\textbf{-0.151  \scriptsize{$\pm$ 0.003}}} 
        & \textcolor{steelblue2}{\textbf{-0.020  \scriptsize{$\pm$ 0.007}}} 
        & -0.232  \scriptsize{$\pm$ 0.011} \\ \hline

    \end{tabular}
    \caption{Performance on synthetic and real-world data using intersection-wise evaluation.}
    \label{tab:main-table2}
\end{table*}

Notably, these findings provide strong evidence that our algorithm performs well not only in terms of global cooperation but also from the perspective of benefits at individual intersections. This dual-level efficacy showcases the effectiveness of our approach, signifying its ability to foster overall road network performance while simultaneously optimizing individual intersection operations. 

Moreover, the consistent and superior performance across various intersection-wise evaluations signifies the algorithm's ability to handle different traffic scenarios and dynamics effectively. It's worth noting that traffic scenarios can be widely varied and unpredictable, thus the adaptability shown by our model underscores its real-world application potential.

\clearpage

\section{Additional Ablation of Various Settings}
\label{app:ablation}

\subsection{Additional Ablation of Various Collaborator Matrix}
\label{app:abs_coll_matrix}

We conduct a comprehensive ablation study regarding different collaborator matrices. In this context, the ``\textbf{fixed matrix}'' signifies selecting the nearest four intersections as neighbors based on the topology, centered around oneself, and keeping it constant throughout the experiment. Meanwhile, the ``\textbf{random matrix}'' refers to randomly choosing four intersections as collaborators at each instance, making it possible that different intersections to be selected at various times.

The experimental results indicate that our approach, which jointly optimizes the selection of collaborators, is superior. For instance, in scenarios like Grid 4 $\times$ 4 and Grid 5 $\times$ 5, using a fixed or random collaborator matrix leads to performance degradation, with the performance deteriorating even further in the latter stages of learning. Observing all five scenarios, both the fixed and random matrices demonstrate similar performance trends. This suggests that such non-learning-based approaches may lack adaptability to the dynamic nature of traffic scenarios and can't effectively capture the intricate interactions among intersections. In contrast, our joint optimization paradigm is better positioned to leverage the interplay and achieve optimal collaboration for boosting traffic signal control.

\begin{figure*}[ht]
    \setcounter{subfigure}{0}
	\centering
	\subfigure[Grid 4 $\times$ 4]{
		\begin{minipage}[t]{0.3\textwidth}
			\centering
			\includegraphics[width=\textwidth]{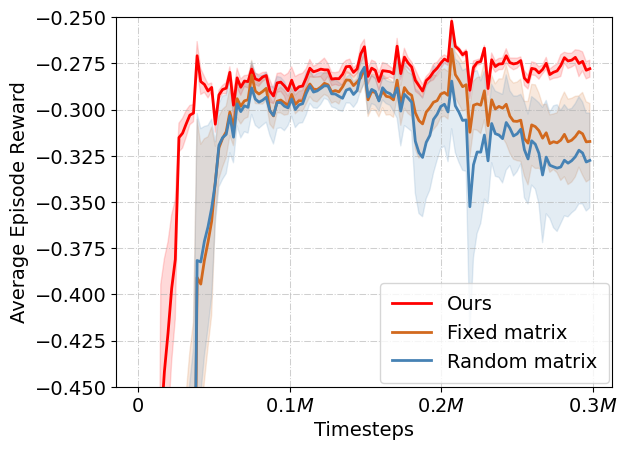}
		\end{minipage}
	}
	\subfigure[Avenue 4 $\times$ 4]{
		\begin{minipage}[t]{0.3\textwidth}
			\centering
			\includegraphics[width=\textwidth]{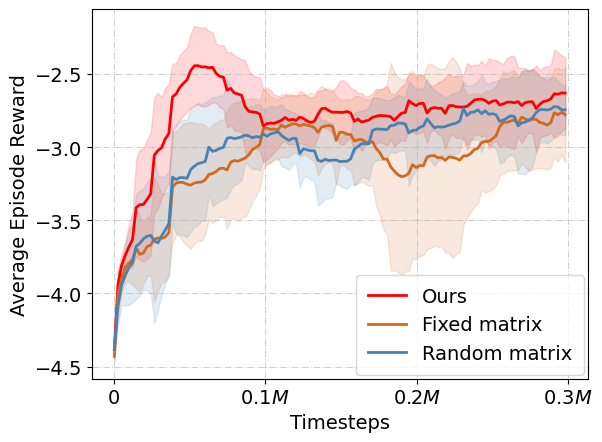}
		\end{minipage}
	}
	\subfigure[Grid 5 $\times$ 5]{
		\begin{minipage}[t]{0.3\textwidth}
			\centering
			\includegraphics[width=\textwidth]{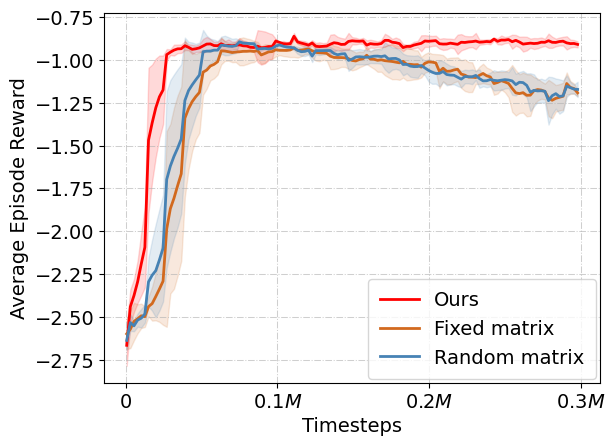}
		\end{minipage}
	}

    \subfigure[Cologne8]{
		\begin{minipage}[t]{0.3\textwidth}
			\centering
			\includegraphics[width=\textwidth]{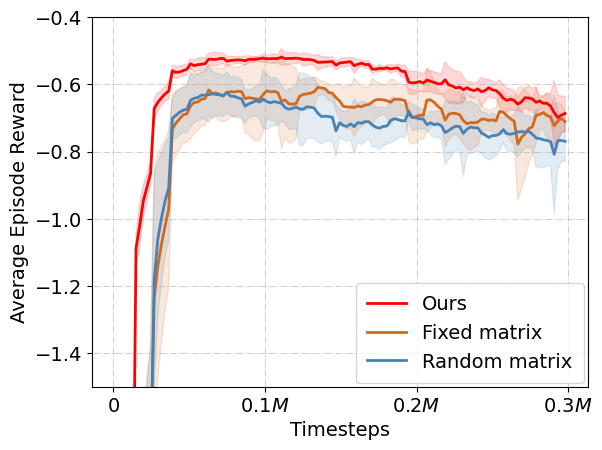}
		\end{minipage}
	}
	\subfigure[Nanshan]{
		\begin{minipage}[t]{0.3\textwidth}
			\centering
			\includegraphics[width=\textwidth]{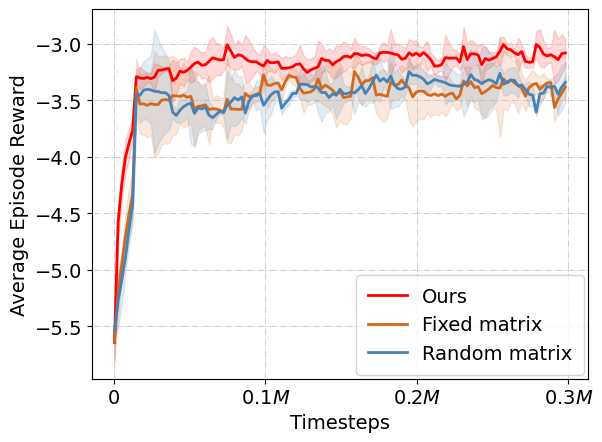}
		\end{minipage}
	}
	\caption{Learning curves of the ablation study about various collaborator matrices.}
	\label{exp:app_abs_matrix}
\end{figure*}

\subsection{Additional Ablation about the Constraints}
\label{app:abs_cons}

To assess the significance of the constraints in Equations~(\ref{eq:diag}) and~(\ref{eq:symm}), we conduct a thorough ablation study across five scenarios. Specifically:
\begin{itemize}
\item \textbf{w/o Diagonal}: The constraint for diagonal maximization is removed.
\item \textbf{w/o Symmetry}: The symmetry constraint is omitted.
\item \textbf{w/o Both}: Both the diagonal maximization and symmetry constraints are eliminated.
\end{itemize}
From this setup, we aim to understand the individual and combined roles of these constraints in the overall performance and adaptability of our model in various traffic situations.

\begin{figure*}[ht]
    \setcounter{subfigure}{0}
	\centering
	\subfigure[Grid 4 $\times$ 4]{
		\begin{minipage}[t]{0.3\textwidth}
			\centering
			\includegraphics[width=\textwidth]{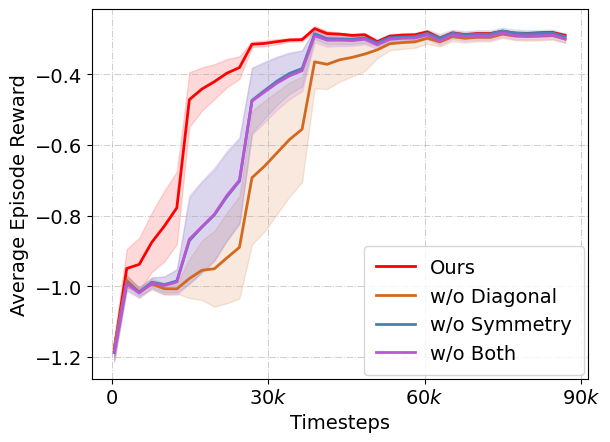}
		\end{minipage}
	}
	\subfigure[Avenue 4 $\times$ 4]{
		\begin{minipage}[t]{0.3\textwidth}
			\centering
			\includegraphics[width=\textwidth]{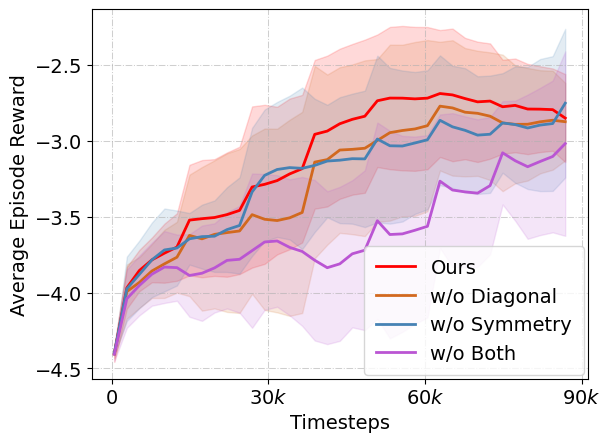}
		\end{minipage}
	}
	\subfigure[Grid 5 $\times$ 5]{
		\begin{minipage}[t]{0.3\textwidth}
			\centering
			\includegraphics[width=\textwidth]{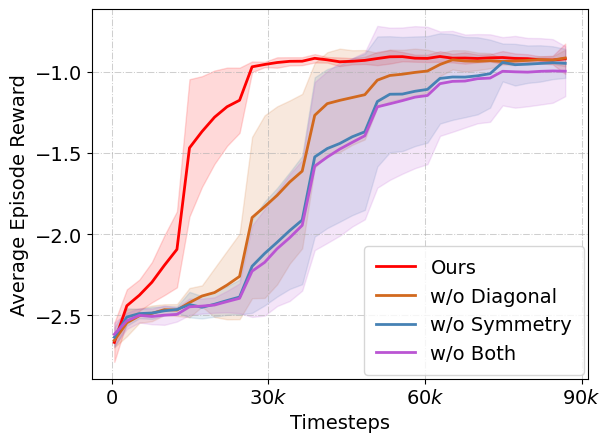}
		\end{minipage}
	}

    \subfigure[Cologne8]{
		\begin{minipage}[t]{0.3\textwidth}
			\centering
			\includegraphics[width=\textwidth]{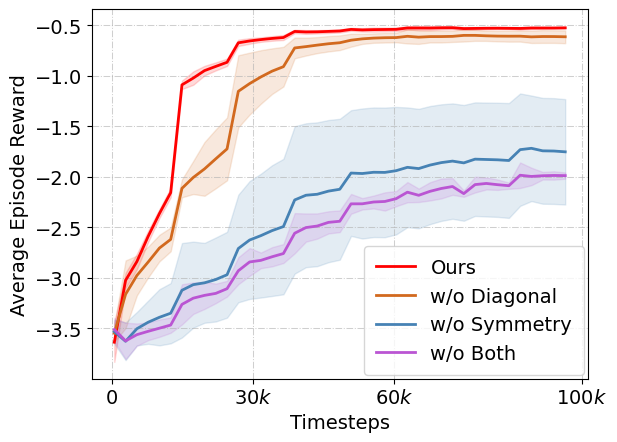}
		\end{minipage}
	}
	\subfigure[Nanshan]{
		\begin{minipage}[t]{0.3\textwidth}
			\centering
			\includegraphics[width=\textwidth]{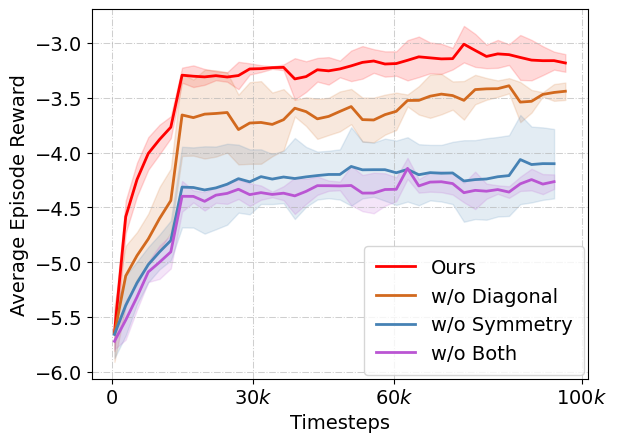}
		\end{minipage}
	}
	\caption{Learning curves of the ablation study about various constraints.}
	\label{exp:app_abs_cons}
\end{figure*}

As depicted in Figure~\ref{exp:app_abs_cons}, 
for scenarios like Grid 4 $\times$ 4 and Grid 5 $\times$ 5, all these configurations can eventually achieve similar performance levels. However, when incorporating the proposed constraints, there's a clear advantage in terms of quickly reaching peak performance, thus showcasing faster convergence. Specifically, in the Grid 4 $\times$ 4 scenario, removing the diagonal maximization constraint results in significant performance degradation in the early stages. This suggests that decisions in this scenario might rely more heavily on the features of the intersection itself. On the other hand, other scenarios display larger performance drops when the symmetry constraint is removed, indicating a greater need to consider mutual characteristics for more decisive and interoperable decision-making. Both Cologne8 and Nanshan scenarios show a trend of stagnating learning when constraints are removed. This implies that these constraints play a vital role in guiding the learning process, especially in scenarios with potentially intricate internal dynamics. By adding these constraints, the exploration space becomes more compact, making it easier to learn an effective decision-making strategy.

\subsection{Additional Ablation about the Main Components}
\label{app:abs_res}

First, let us delineate the experimental setup for this section.
\begin{itemize}
    \item \textbf{w/o FRAP}: Remove the FRAP module from the dual-feature extractor, retaining only the intersection-level features.
    \item \textbf{w/o Transformer}: Eliminate the Transformer module from the dual-feature extractor, keeping solely the phase-level features.
    \item \textbf{w/o Teammate}: Remove the~\abbr~module, resulting in no teammate coordination.
    \item \textbf{\abbr~with GNN}: Replace the MLPs in the~\abbr~module with GNNs.
\end{itemize}

In this section, we present multi-dimensional training curves for various evaluation metrics including average episode rewards, average delay time, average wait time, average queue length, and average pressure across different scenarios. These results supplement our main ablation studies and provide a more comprehensive understanding of the performance and robustness of our algorithm under various conditions.

\subsubsection{Ablation analysis on average episode rewards}

\begin{figure*}[ht]
    \setcounter{subfigure}{0}
	\centering
	\subfigure[Grid 4 $\times$ 4]{
		\begin{minipage}[t]{0.3\textwidth}
			\centering
			\includegraphics[width=\textwidth]{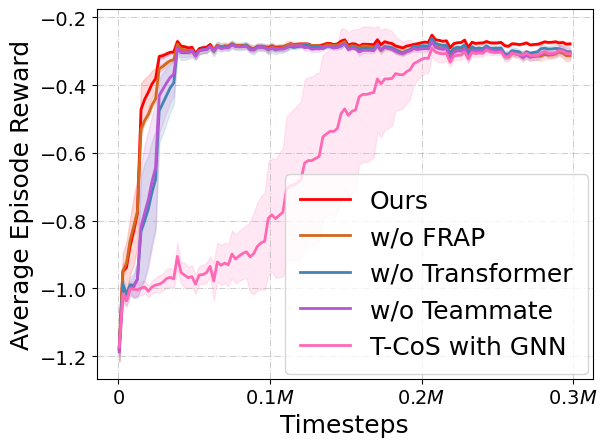}
		\end{minipage}
	}
	\subfigure[Avenue 4 $\times$ 4]{
		\begin{minipage}[t]{0.3\textwidth}
			\centering
			\includegraphics[width=\textwidth]{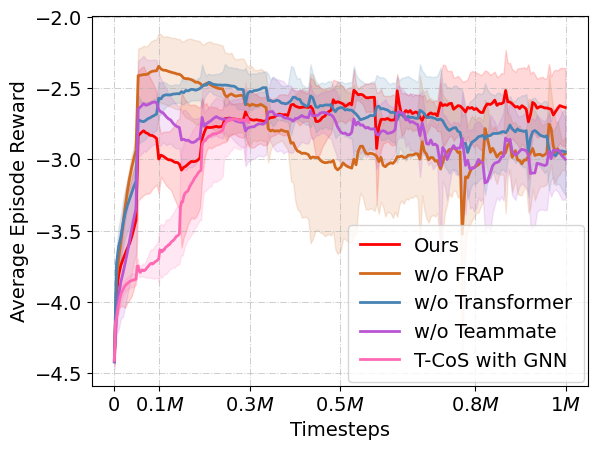}
		\end{minipage}
	}
	\subfigure[Grid 5 $\times$ 5]{
		\begin{minipage}[t]{0.3\textwidth}
			\centering
			\includegraphics[width=\textwidth]{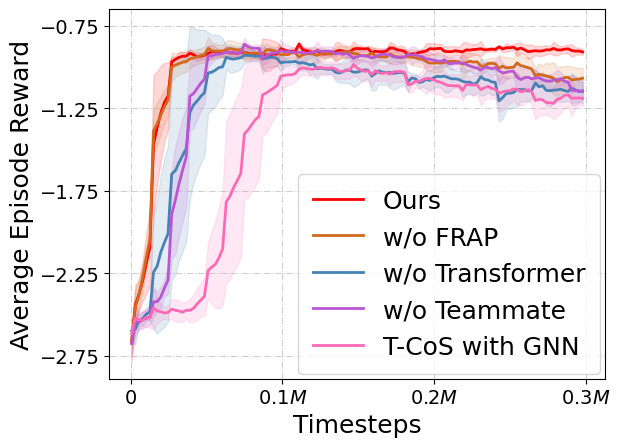}
		\end{minipage}
	}

    \subfigure[Cologne8]{
		\begin{minipage}[t]{0.3\textwidth}
			\centering
			\includegraphics[width=\textwidth]{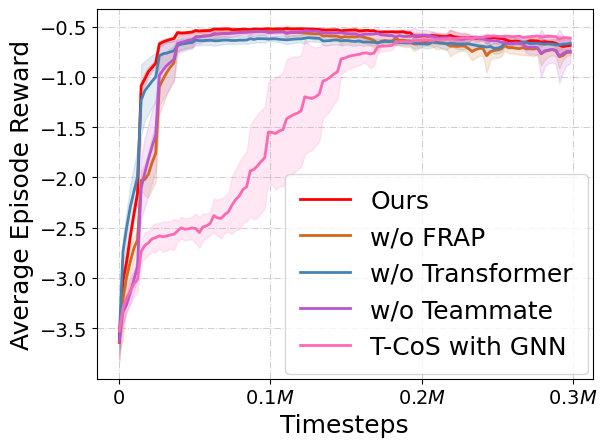}
		\end{minipage}
	}
	\subfigure[Nanshan]{
		\begin{minipage}[t]{0.3\textwidth}
			\centering
			\includegraphics[width=\textwidth]{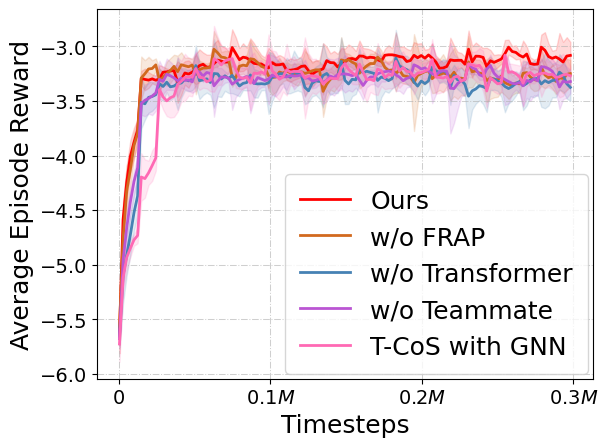}
		\end{minipage}
	}
	\caption{Learning curves of the ablation study of the various components about the average episode rewards.}
	\label{exp:app_abs_more}
\end{figure*}

As shown in Figure~\ref{exp:app_abs_more}, In experiments across Grid 4 $\times$ 4, Avenue 4 $\times$ 4, Grid 5 $\times$ 5, and Cologne8, we observed a significantly slower convergence rate for~\abbr~with GNN, taking much longer to reach a plateau compared to our algorithm. 
For the Nanshan map, which features an irregular topology with 28 intersections, the slowdown was less pronounced. We speculate that in such a complex scenario, GNNs might be better at capturing inherent dynamical information, thus resulting in a decent convergence rate and performance. 
However, for this setting, training our~\abbr~on a single seed (300k timestamps) took just 13.1 hours, whereas the GNN variant took a staggering 40.8 hours. 
In terms of GPU consumption for a single process, the MLP version of~\abbr~peaked at 1.5G, while the GNN version utilized nearly 12G. This clearly demonstrates that while achieving competitive performance,~\abbr~with MLPs significantly save on computational resources.

The above is an evaluation of the training curve concerning the average episode reward. Subsequently, the training curves for fine-grained evaluations in each scenario, including the average delay time, average wait time, average queue length, and average pressure, are provided.

\subsubsection{Ablation analysis on average delay time}

\begin{figure*}[ht]
    \setcounter{subfigure}{0}
	\centering
	\subfigure[Grid 4 $\times$ 4]{
		\begin{minipage}[t]{0.3\textwidth}
			\centering
			\includegraphics[width=\textwidth]{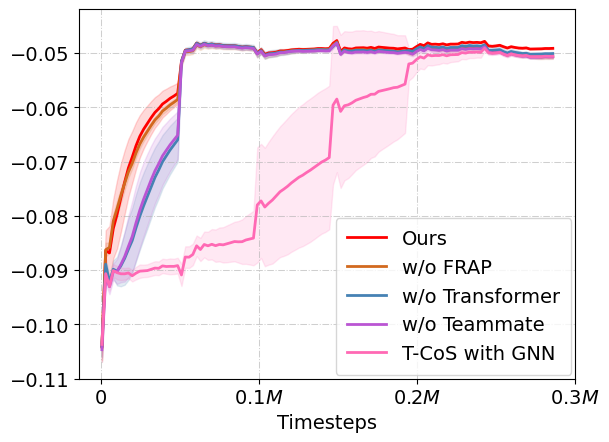}
		\end{minipage}
	}
	\subfigure[Avenue 4 $\times$ 4]{
		\begin{minipage}[t]{0.3\textwidth}
			\centering
			\includegraphics[width=\textwidth]{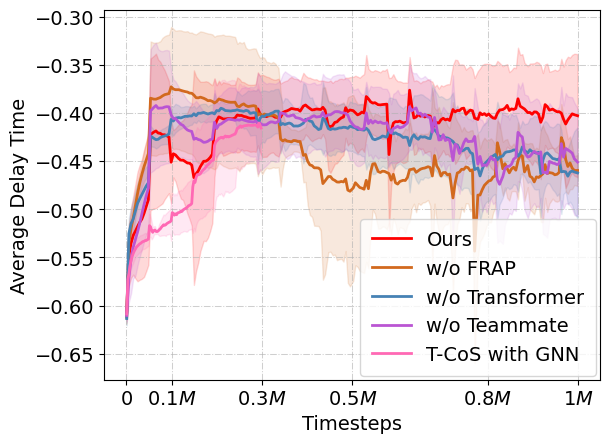}
		\end{minipage}
	}
	\subfigure[Grid 5 $\times$ 5]{
		\begin{minipage}[t]{0.3\textwidth}
			\centering
			\includegraphics[width=\textwidth]{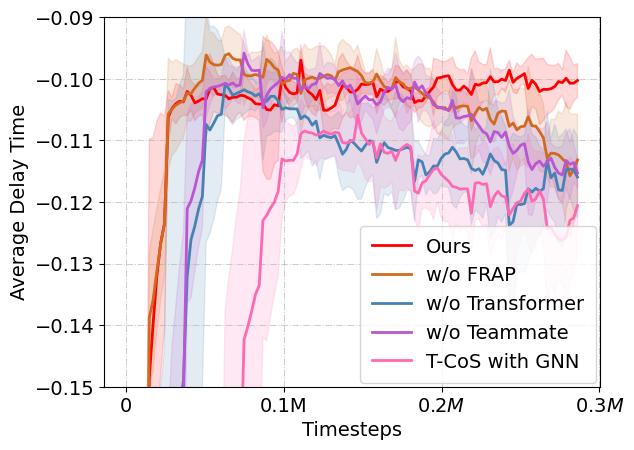}
		\end{minipage}
	}

    \subfigure[Cologne8]{
		\begin{minipage}[t]{0.3\textwidth}
			\centering
			\includegraphics[width=\textwidth]{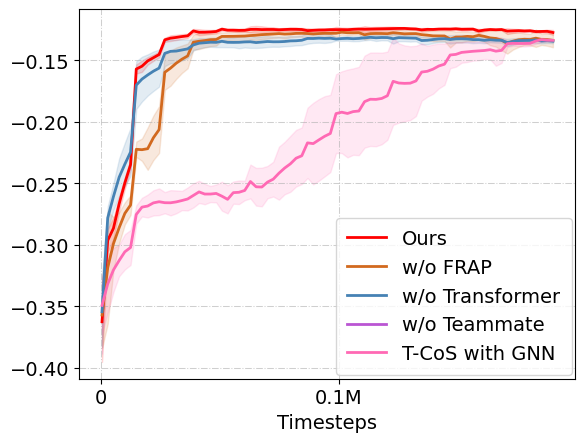}
		\end{minipage}
	}
	\subfigure[Nanshan]{
		\begin{minipage}[t]{0.3\textwidth}
			\centering
			\includegraphics[width=\textwidth]{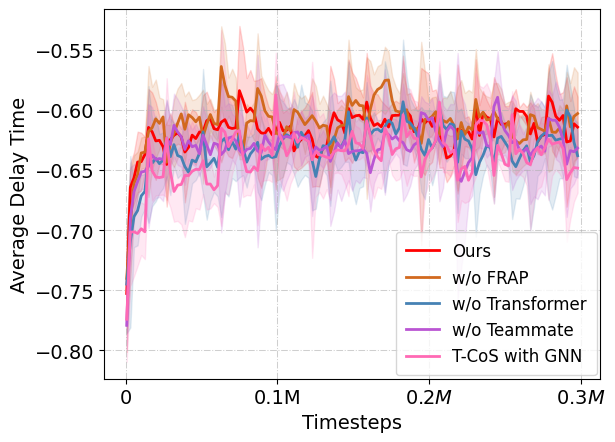}
		\end{minipage}
	}
	\caption{Learning curves of the ablation study about the average delay time.}
	\label{exp:app_abs_delay} 
\end{figure*}

As depicted in Figure~\ref{exp:app_abs_delay}, which presents the learning curves from the ablation study concerning the average delay time, our algorithm with all its modules demonstrates significant improvements in four out of the five scenarios. The one exception was Nanshan, where the impact of our finely-designed modules appeared to be less substantial.

We hypothesize that this anomaly may be due to the larger scale of the Nanshan map. The complexity inherent in larger maps might dilute the effects of our specialized modules when it comes to reducing delay times. However, the consistent improvements in the other four scenarios underscore the overall effectiveness of our algorithm and its components, proving their value in optimizing traffic signal control across a majority of scenarios.

\begin{figure*}[ht]
    \setcounter{subfigure}{0}
	\centering
	\subfigure[Grid 4 $\times$ 4]{
		\begin{minipage}[t]{0.3\textwidth}
			\centering
			\includegraphics[width=\textwidth]{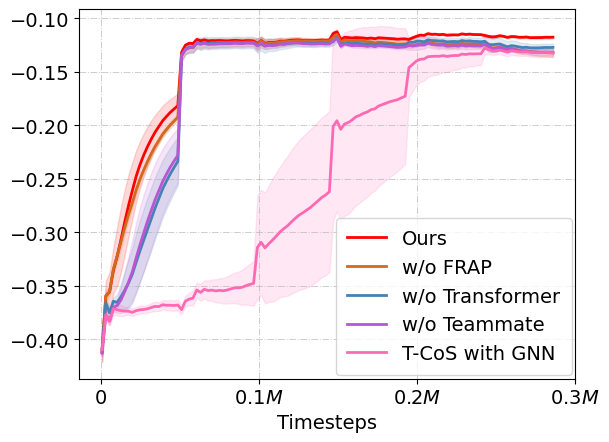}
		\end{minipage}
	}
	\subfigure[Avenue 4 $\times$ 4]{
		\begin{minipage}[t]{0.3\textwidth}
			\centering
			\includegraphics[width=\textwidth]{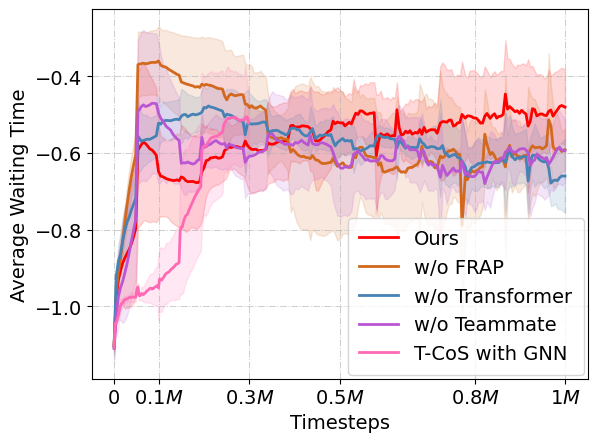}
		\end{minipage}
	}
	\subfigure[Grid 5 $\times$ 5]{
		\begin{minipage}[t]{0.3\textwidth}
			\centering
			\includegraphics[width=\textwidth]{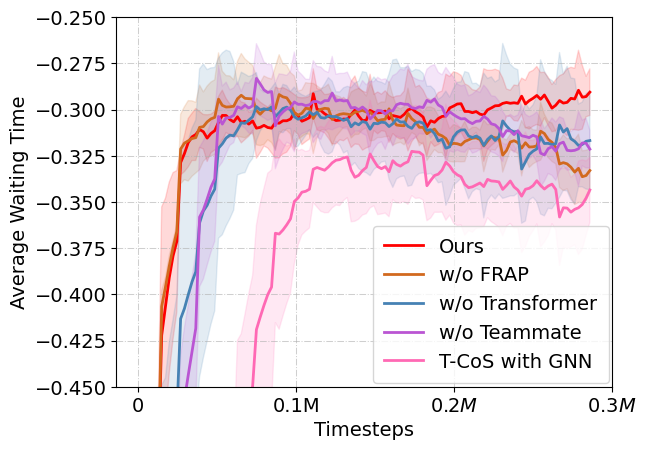}
		\end{minipage}
	}

    \subfigure[Cologne8]{
		\begin{minipage}[t]{0.3\textwidth}
			\centering
			\includegraphics[width=\textwidth]{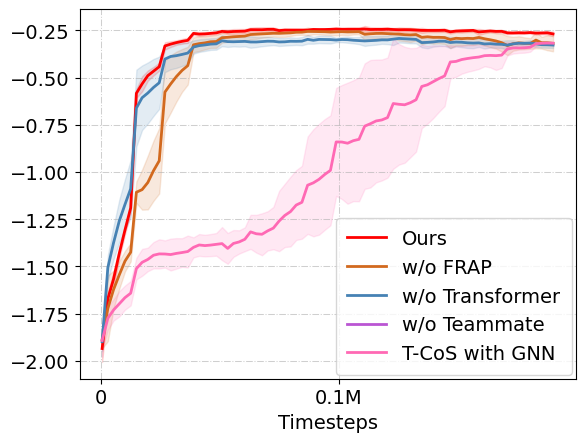}
		\end{minipage}
	}
	\subfigure[Nanshan]{
		\begin{minipage}[t]{0.3\textwidth}
			\centering
			\includegraphics[width=\textwidth]{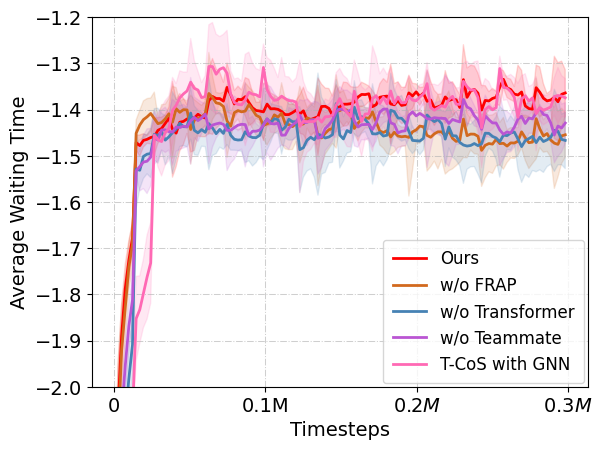}
		\end{minipage}
	}
	\caption{Learning curves of the ablation study about the average wait time.}
	\label{exp:app_abs_wait}\vspace{-10pt}
\end{figure*}


\begin{figure*}[ht]
    \setcounter{subfigure}{0}
	\centering
	\subfigure[Grid 4 $\times$ 4]{
		\begin{minipage}[t]{0.3\textwidth}
			\centering
			\includegraphics[width=\textwidth]{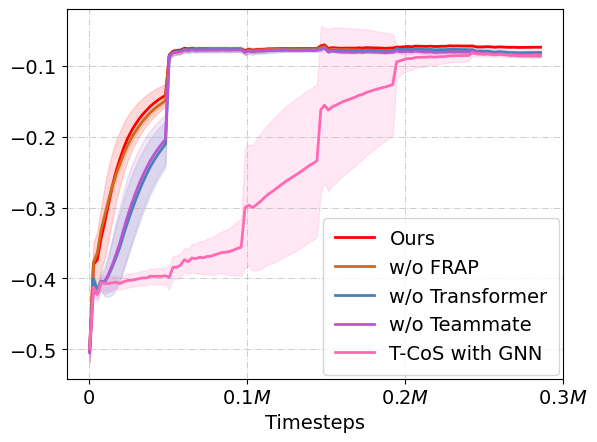}
		\end{minipage}
	}
	\subfigure[Avenue 4 $\times$ 4]{
		\begin{minipage}[t]{0.3\textwidth}
			\centering
			\includegraphics[width=\textwidth]{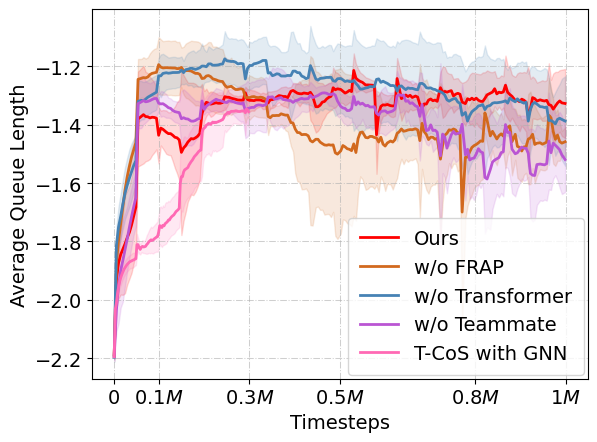}
		\end{minipage}
	}
	\subfigure[Grid 5 $\times$ 5]{
		\begin{minipage}[t]{0.3\textwidth}
			\centering
			\includegraphics[width=\textwidth]{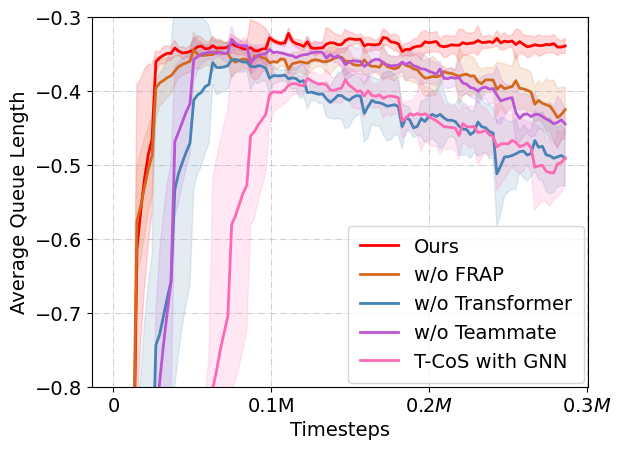}
		\end{minipage}
	}

    \subfigure[Cologne8]{
		\begin{minipage}[t]{0.3\textwidth}
			\centering
			\includegraphics[width=\textwidth]{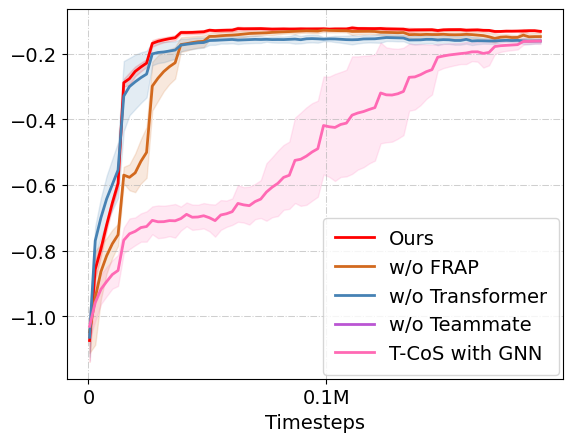}
		\end{minipage}
	}
	\subfigure[Nanshan]{
		\begin{minipage}[t]{0.3\textwidth}
			\centering
			\includegraphics[width=\textwidth]{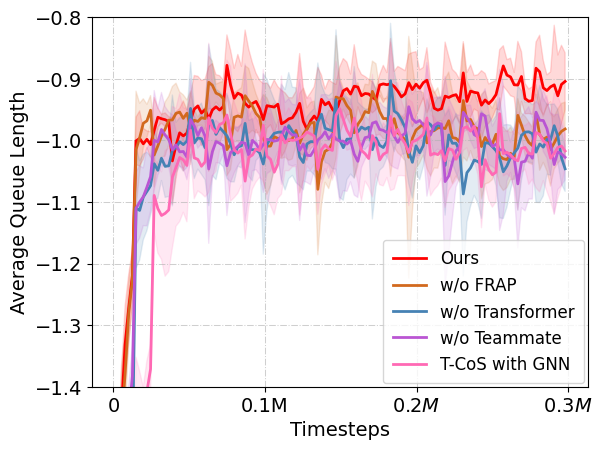}
		\end{minipage}
	}
	\caption{Learning curves of the ablation study about the average queue length.}
	\label{exp:app_abs_queue} 
\end{figure*}

\begin{figure*}[ht]
    \setcounter{subfigure}{0}
	\centering
	\subfigure[Grid 4 $\times$ 4]{
		\begin{minipage}[t]{0.3\textwidth}
			\centering
			\includegraphics[width=\textwidth]{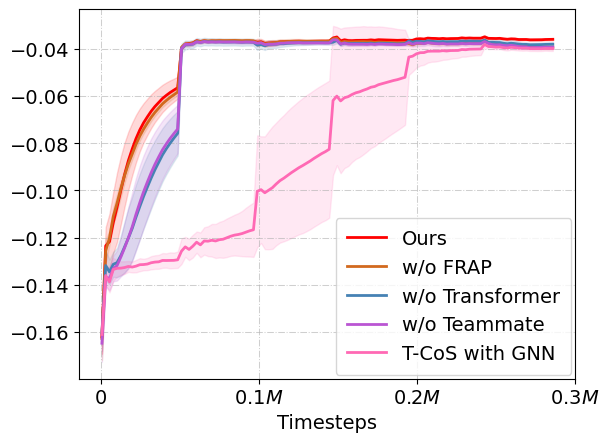}
		\end{minipage}
	}
	\subfigure[Avenue 4 $\times$ 4]{
		\begin{minipage}[t]{0.3\textwidth}
			\centering
			\includegraphics[width=\textwidth]{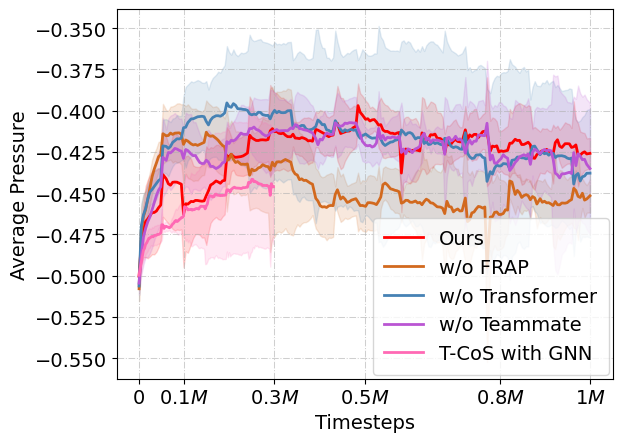}
		\end{minipage}
	}
	\subfigure[Grid 5 $\times$ 5]{
		\begin{minipage}[t]{0.3\textwidth}
			\centering
			\includegraphics[width=\textwidth]{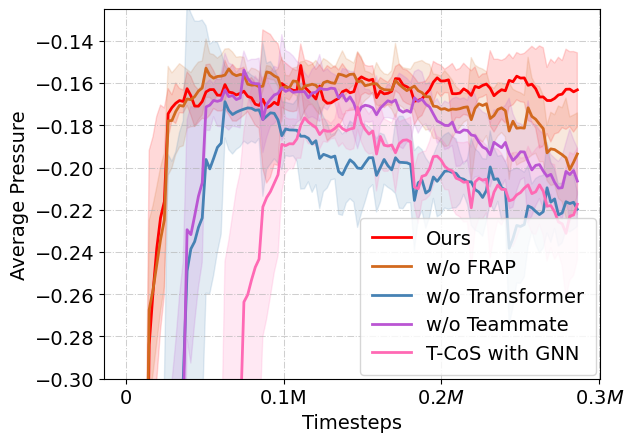}
		\end{minipage}
	}

    \subfigure[Cologne8]{
		\begin{minipage}[t]{0.3\textwidth}
			\centering
			\includegraphics[width=\textwidth]{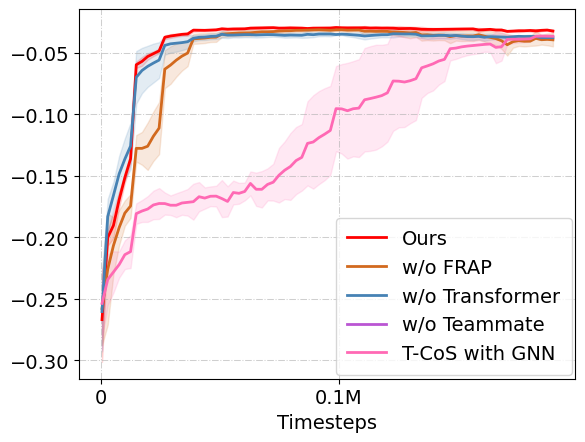}
		\end{minipage}
	}
	\subfigure[Nanshan]{
		\begin{minipage}[t]{0.3\textwidth}
			\centering
			\includegraphics[width=\textwidth]{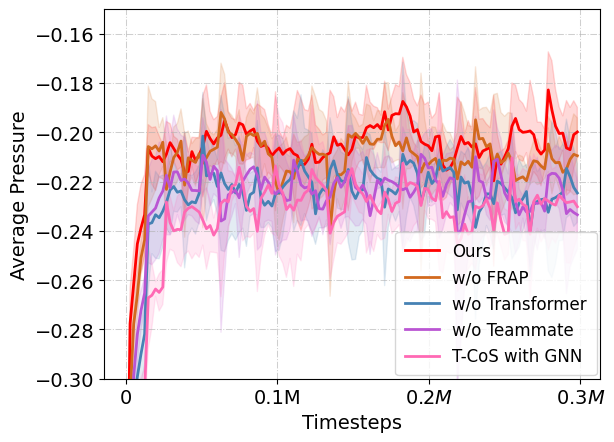}
		\end{minipage}
	}
	\caption{Learning curves of the ablation study about the average pressure.}
	\label{exp:app_abs_pressure} 
\end{figure*}


\subsubsection{Ablation analysis on average wait time}

Figure~\ref{exp:app_abs_wait} shows the learning curves of the ablation study about the average wait time.
In terms of average waiting time, our complete algorithm has shown consistent performance advantages across all maps. This indicates that the full suite of modules in our approach is highly effective in reducing wait times, a critical factor in traffic signal control. The uniform success across all scenarios reinforces the robustness of our method and its potential for broad application in diverse traffic situations.

\subsubsection{Ablation analysis on average queue length}

As illustrated in Figure~\ref{exp:app_abs_queue}, which displays the learning curves of the ablation study about the average queue length, our full algorithm showed consistent performance improvements in all scenarios, except for the Avenue 4 $\times$ 4. We believe that the complexity and high traffic flow of the Avenue 4 $\times$ 4 scenario may have resulted in the less noticeable improvement in queue length by our algorithm.

This suggests that while our method performs excellently across most scenarios, certain situations with high complexity and traffic density may present additional challenges. Despite this, the overall performance advantage in the majority of scenarios reinforces the effectiveness of our approach, demonstrating its capability to manage queue length - an important factor in traffic flow control.

\subsubsection{Ablation analysis on average pressure}

Figure~\ref{exp:app_abs_pressure} shows the learning curves of the ablation study about the average pressure. Our complete algorithm performed better in all situations, except for the Avenue 4 $\times$ 4 scenario.
We conjecture that the high complexity and significant traffic volume of the Avenue $\times$ scenario may have led to a less noticeable effect in pressure reduction from our algorithm.

This trend does shed light on a possible limitation of our method, which might yield smaller performance improvements in highly complex scenarios with dense traffic flow. However, this should not be seen as a significant shortcoming. When compared to baseline algorithms, our method still displays substantial advantages. Additionally, our complete algorithm achieved significant and consistent improvements in all other scenarios, attesting to its robustness and general effectiveness in TSC.

\clearpage

\section{Additional Analysis of Collaborator Number $k$}
\label{app:abs_K_res}
\begin{figure*}[ht]
    \setcounter{subfigure}{0}
	\centering
	\subfigure[Grid 4 $\times$ 4]{
		\begin{minipage}[t]{0.3\textwidth}
			\centering
			\includegraphics[width=\textwidth]{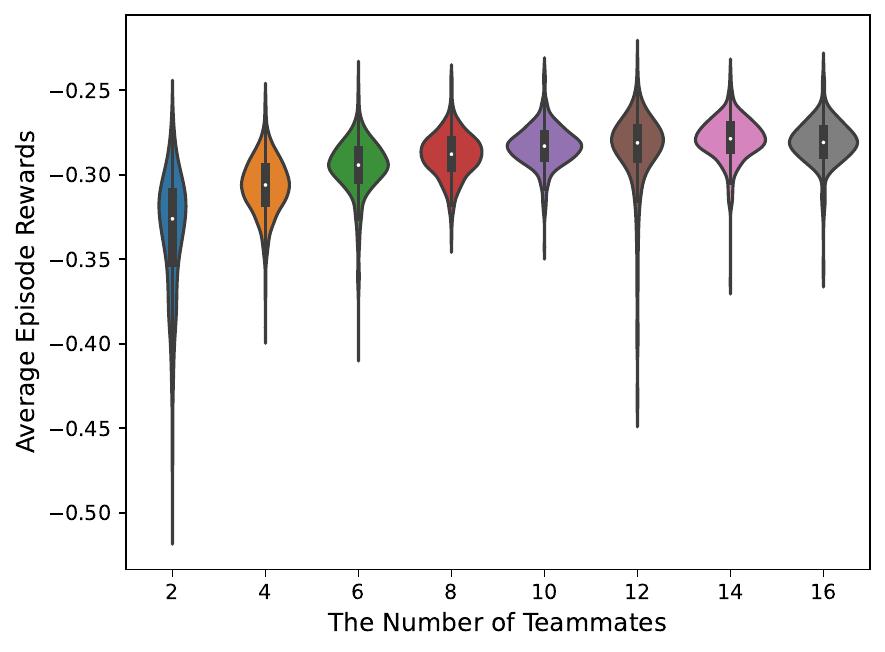}
		\end{minipage}
	}
	\subfigure[Avenue 4 $\times$ 4]{
		\begin{minipage}[t]{0.3\textwidth}
			\centering
			\includegraphics[width=\textwidth]{fig/abs-K/arterial4x4_average_episode_rewards_final_viollion.pdf}
		\end{minipage}
	}
	\subfigure[Grid 5 $\times$ 5]{
		\begin{minipage}[t]{0.3\textwidth}
			\centering
			\includegraphics[width=\textwidth]{fig/abs-K/grid5x5_average_episode_rewards_final_viollion_all.pdf}
		\end{minipage}
	}

    \subfigure[Cologne8]{
		\begin{minipage}[t]{0.3\textwidth}
			\centering
			\includegraphics[width=\textwidth]{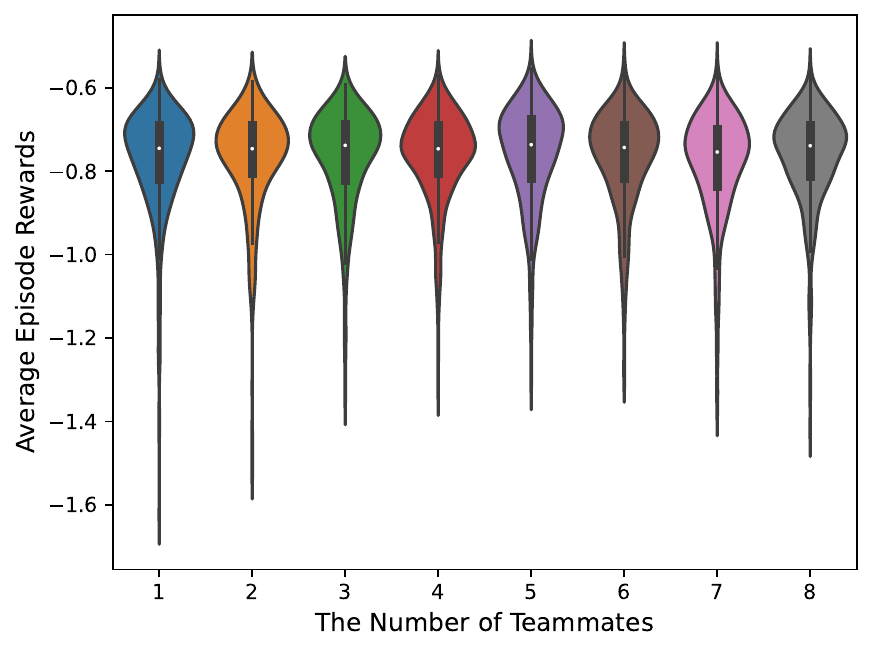}
		\end{minipage}
	}
	\subfigure[Nanshan]{
		\begin{minipage}[t]{0.3\textwidth}
			\centering
			\includegraphics[width=\textwidth]{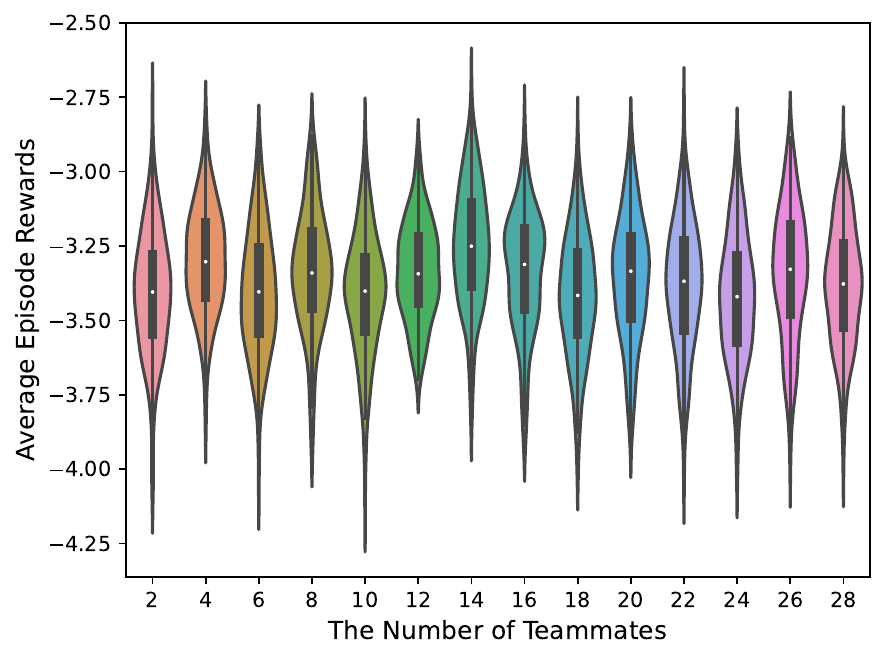}
		\end{minipage}
	}
	\caption{Violin plot of analysis about various number of collaborators.}
	\label{exp:app_change_K}\vspace{-15pt}
\end{figure*}

In this section, we present violin plots for all scenarios in Figure~\ref{exp:app_change_K}, demonstrating the effect of varying the number of collaborators. 
Notably, in the Grid 4 $\times$ 4 scenario, there is a clear trend of performance enhancement as the number of collaborators increases. This indicates that, in such a straightforward grid environment, the performance improvements brought about by targeted collaboration between collaborators may not be as significant as those from the incorporation of information from across the entire map. In this simple case, the information is either not redundant, or the decision-making is not affected by any redundant information.
In the Cologne8 scenario, the performance improvements brought about by increasing the number of collaborators are not apparent. This suggests that, in this extremely simplistic map with sparse traffic and only eight intersections, collaboration between collaborators may not be as crucial.
In other maps, a balance similar to what we mentioned earlier is displayed. It signifies the nuanced relationship between the number of collaborators and the resulting performance improvement. 

In each map, varying the number of collaborators can influence the performance differently, highlighting the importance of tailoring strategies to specific scenarios. For instance, while Grid 4 $\times$ 4 benefits more from a holistic information integration across all intersections, Cologne8 sees limited gains from extensive collaborators due to its simplicity and sparse traffic. 
The analysis also underscores the fact that the environmental complexity and specific scenario factors heavily influence this relationship. This further highlights the importance of adaptively determining the appropriate number of collaborators, a potential area for future exploration.

\section{Additional Visualization Analysis of Collaborator Matrix}
\label{app:vis_teammatrix}

\begin{figure*}[ht]
    \setcounter{subfigure}{0}
	\centering
	\subfigure[Beginning in Grid 4 $\times$ 4]{
		\begin{minipage}[t]{0.21\textwidth}
			\centering
			\includegraphics[width=\textwidth]{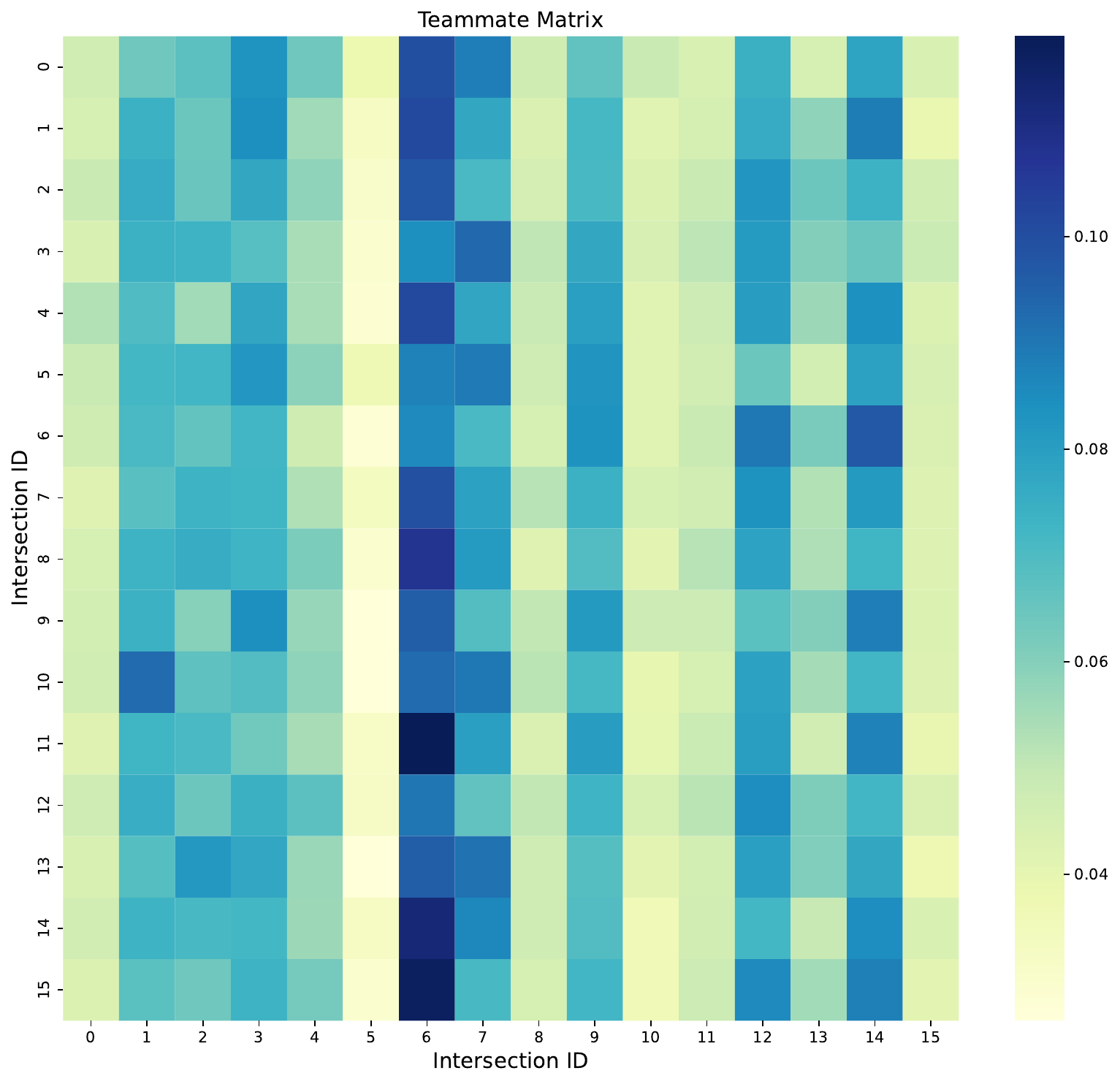}
		\end{minipage}
	}
	\subfigure[Ending in Grid 4 $\times$ 4]{
		\begin{minipage}[t]{0.21\textwidth}
			\centering
			\includegraphics[width=\textwidth]{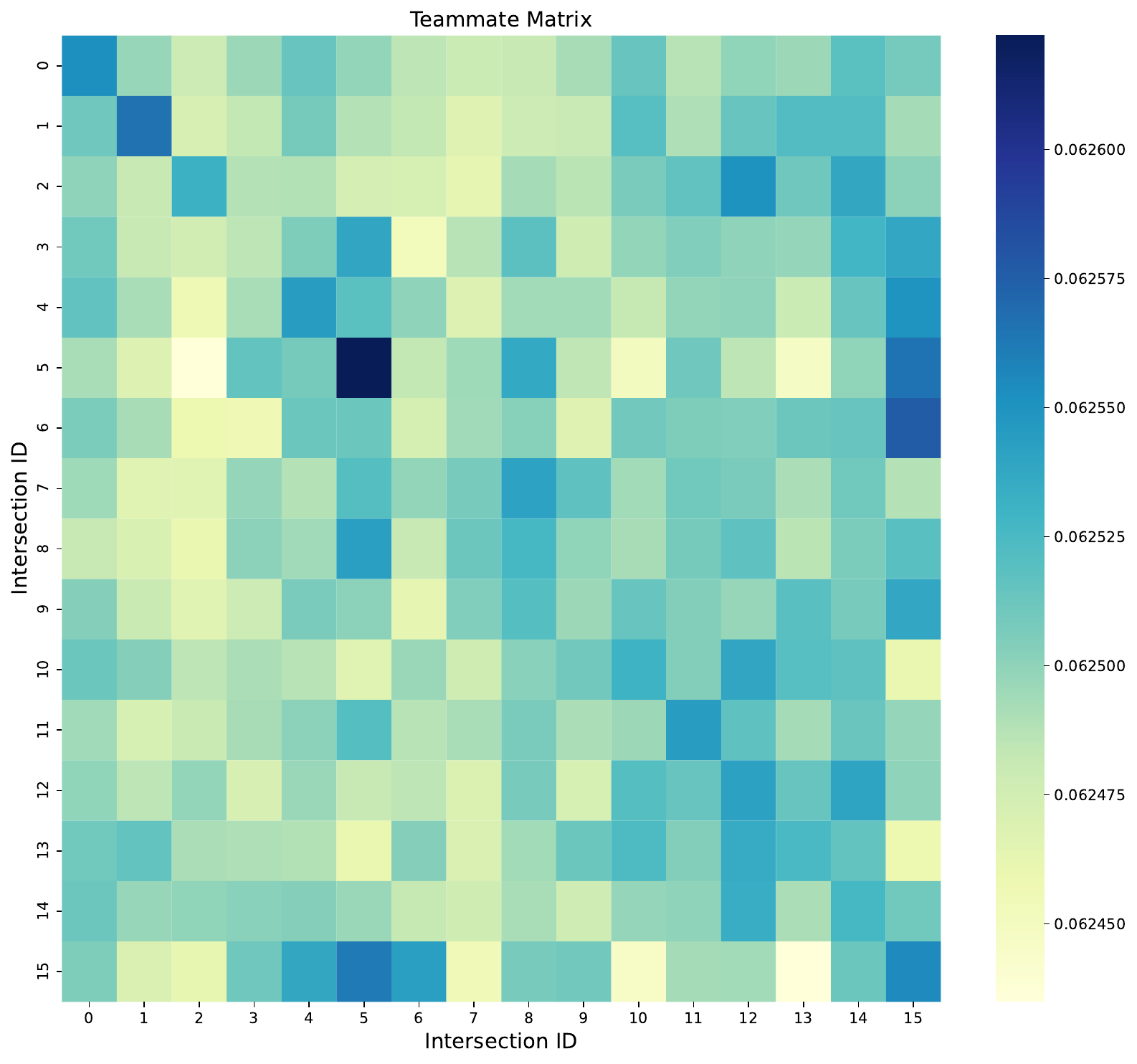}
		\end{minipage}
	}
	\subfigure[Beginning in Avenue 4 $\times$ 4]{
		\begin{minipage}[t]{0.21\textwidth}
			\centering
			\includegraphics[width=\textwidth]{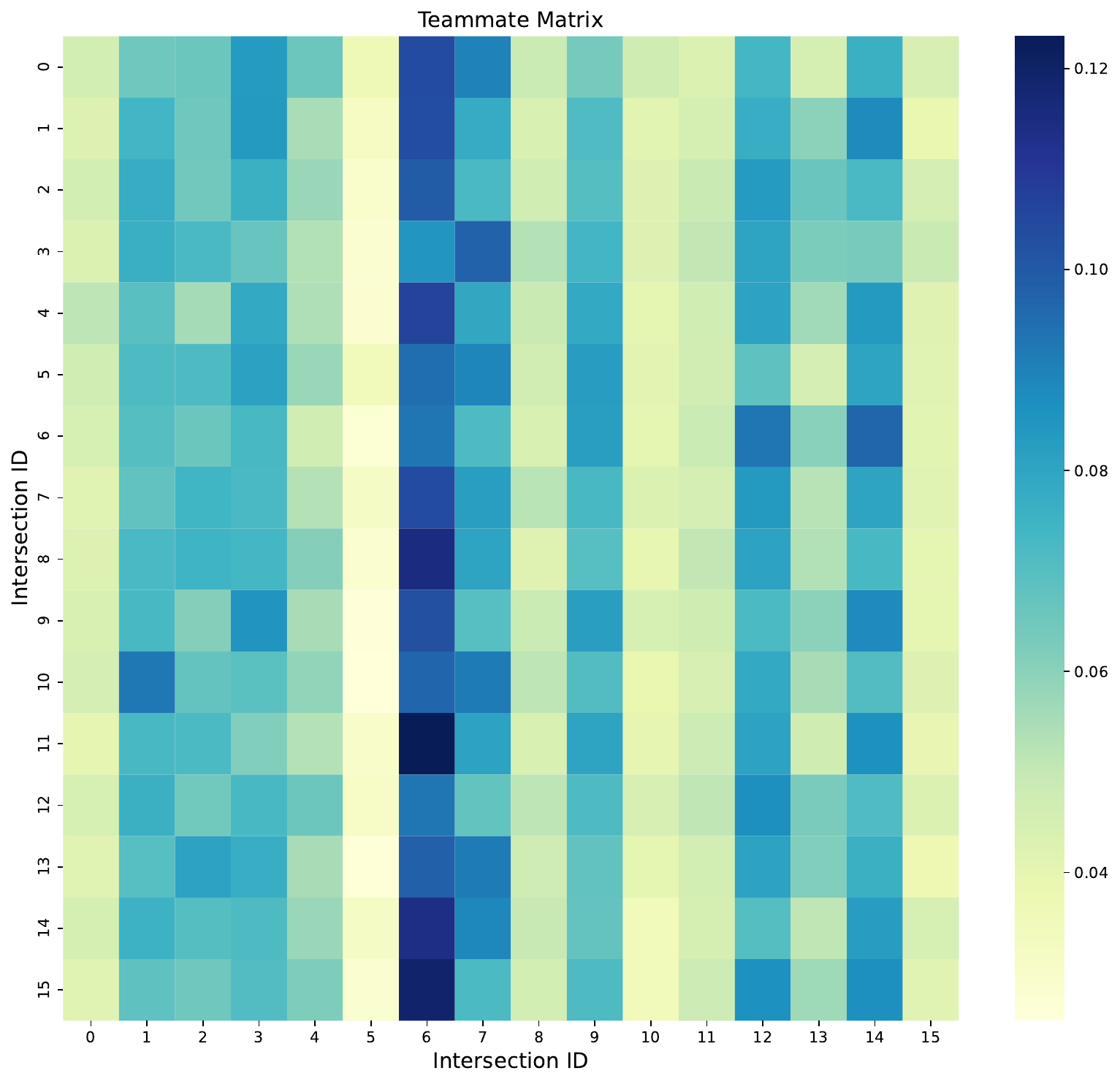}
		\end{minipage}
	}
	\subfigure[Ending in Avenue 4 $\times$ 4]{
		\begin{minipage}[t]{0.21\textwidth}
			\centering
			\includegraphics[width=\textwidth]{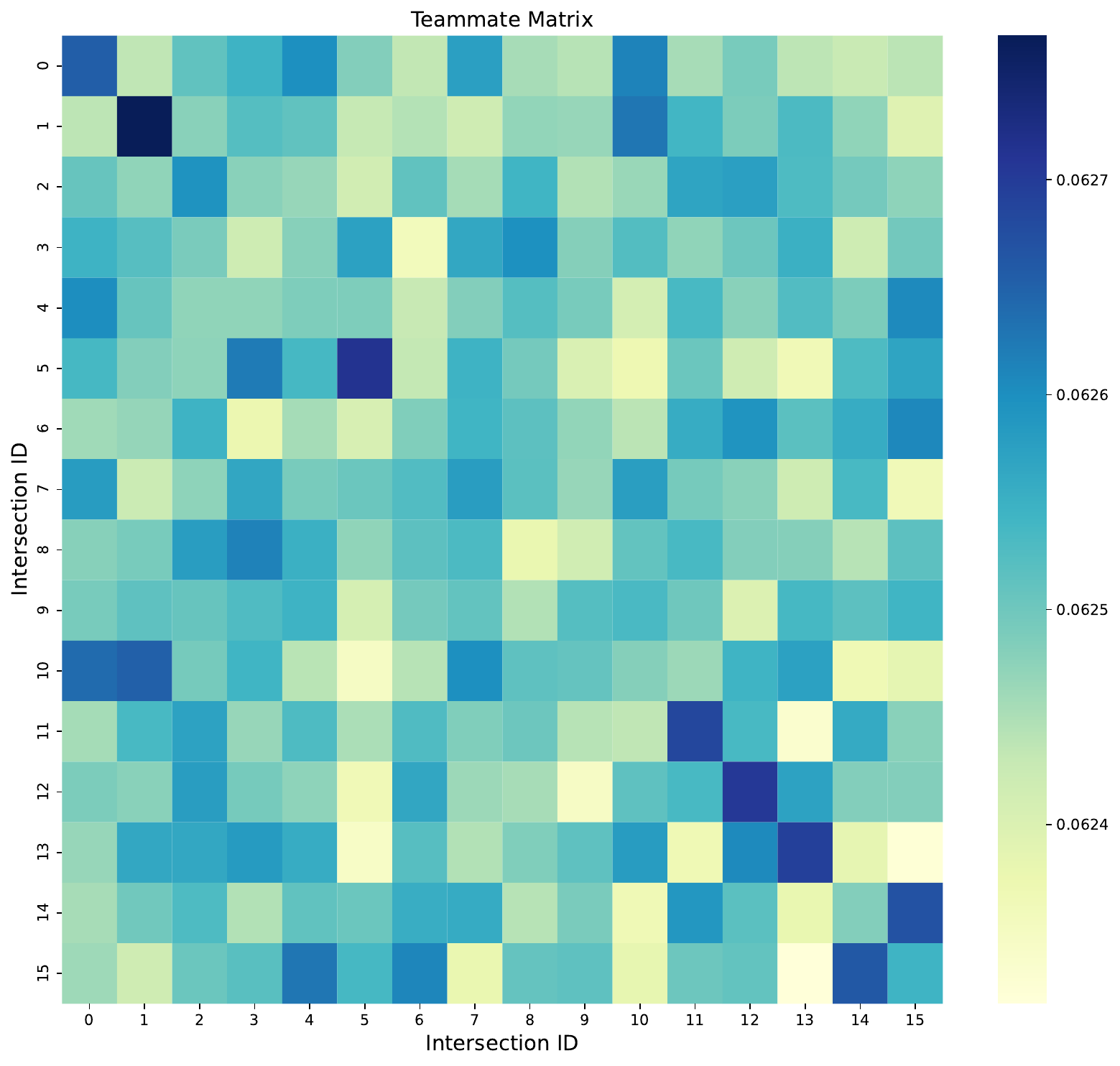}
		\end{minipage}
	}
 
	\subfigure[Beginning in Grid 5 $\times$ 5]{
		\begin{minipage}[t]{0.21\textwidth}
			\centering
			\includegraphics[width=\textwidth]{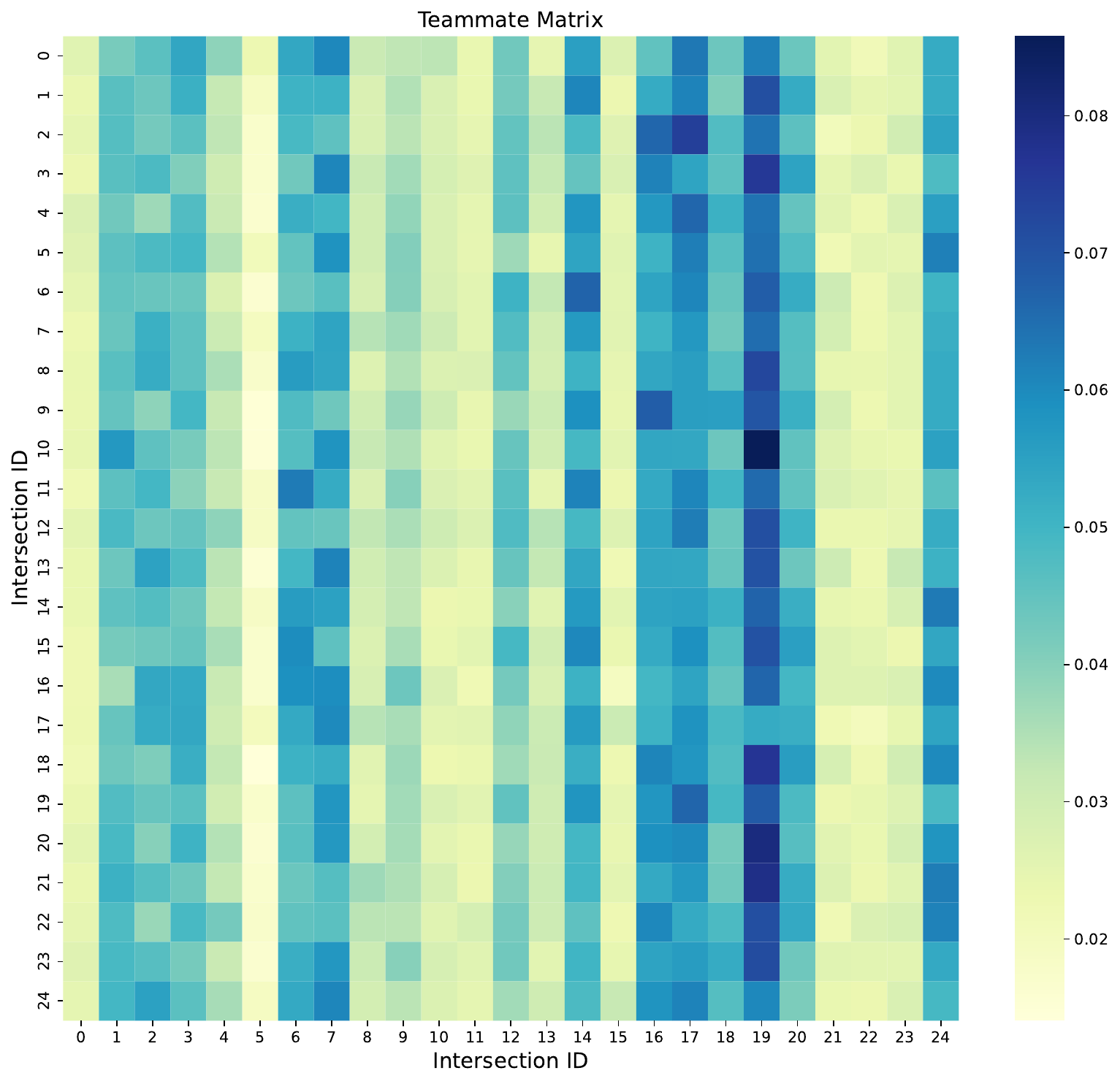}
		\end{minipage}
	}
	\subfigure[Ending in Grid 5 $\times$ 5]{
		\begin{minipage}[t]{0.21\textwidth}
			\centering
			\includegraphics[width=\textwidth]{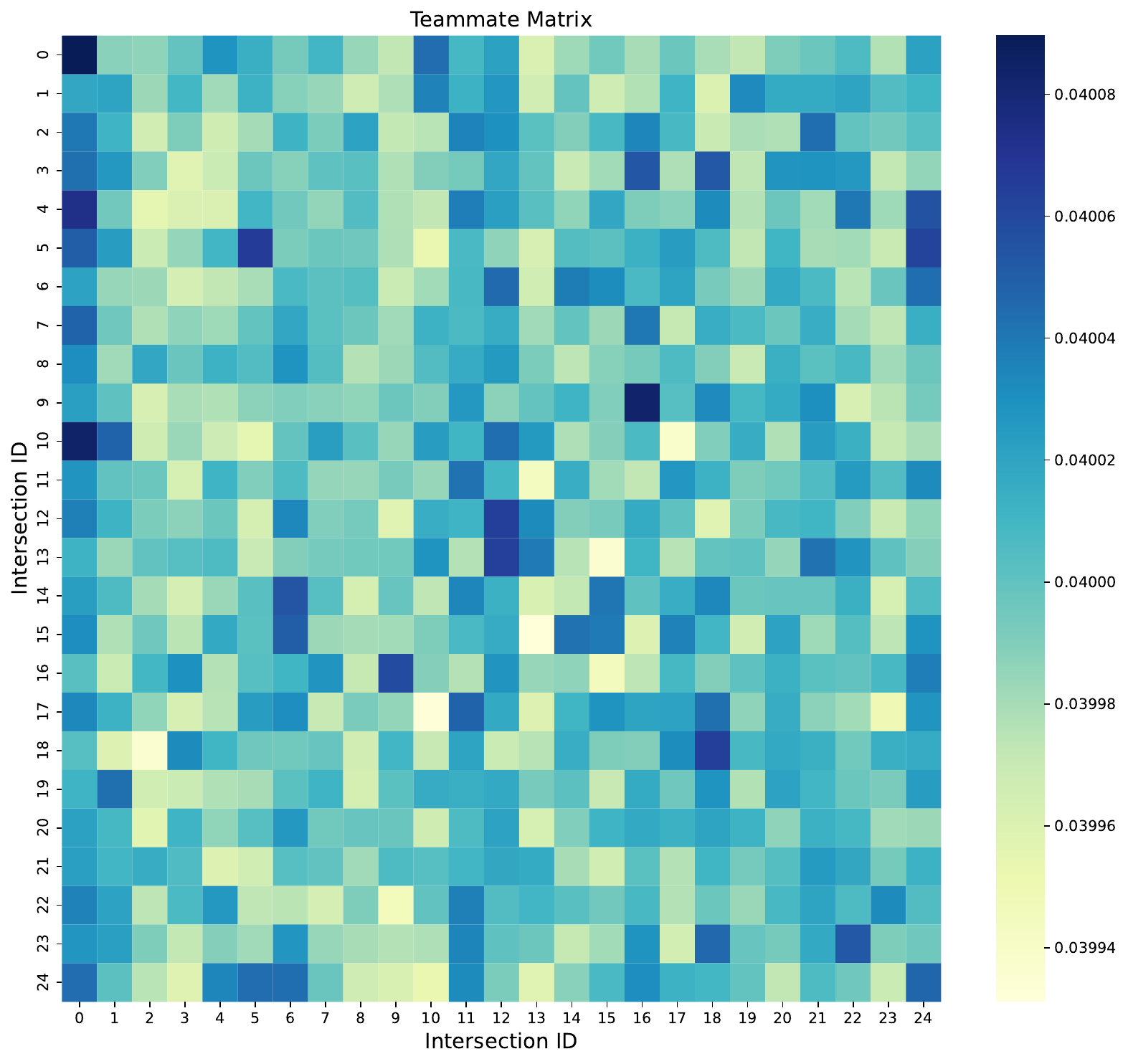}
		\end{minipage}
	}
	\subfigure[Beginning in Cologne8]{
		\begin{minipage}[t]{0.21\textwidth}
			\centering
			\includegraphics[width=\textwidth]{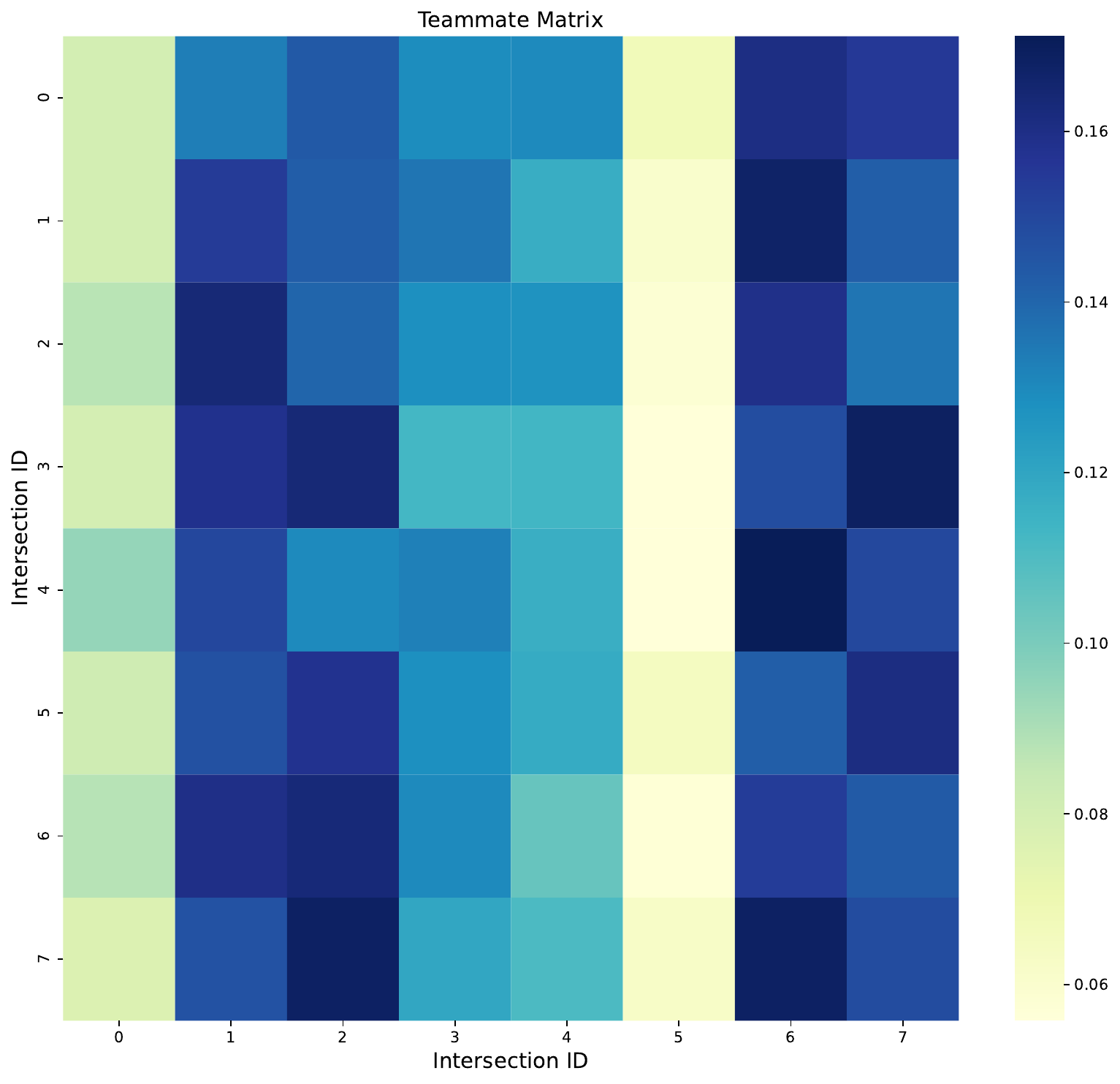}
		\end{minipage}
	}
	\subfigure[Ending in Cologne8]{
		\begin{minipage}[t]{0.21\textwidth}
			\centering
			\includegraphics[width=\textwidth]{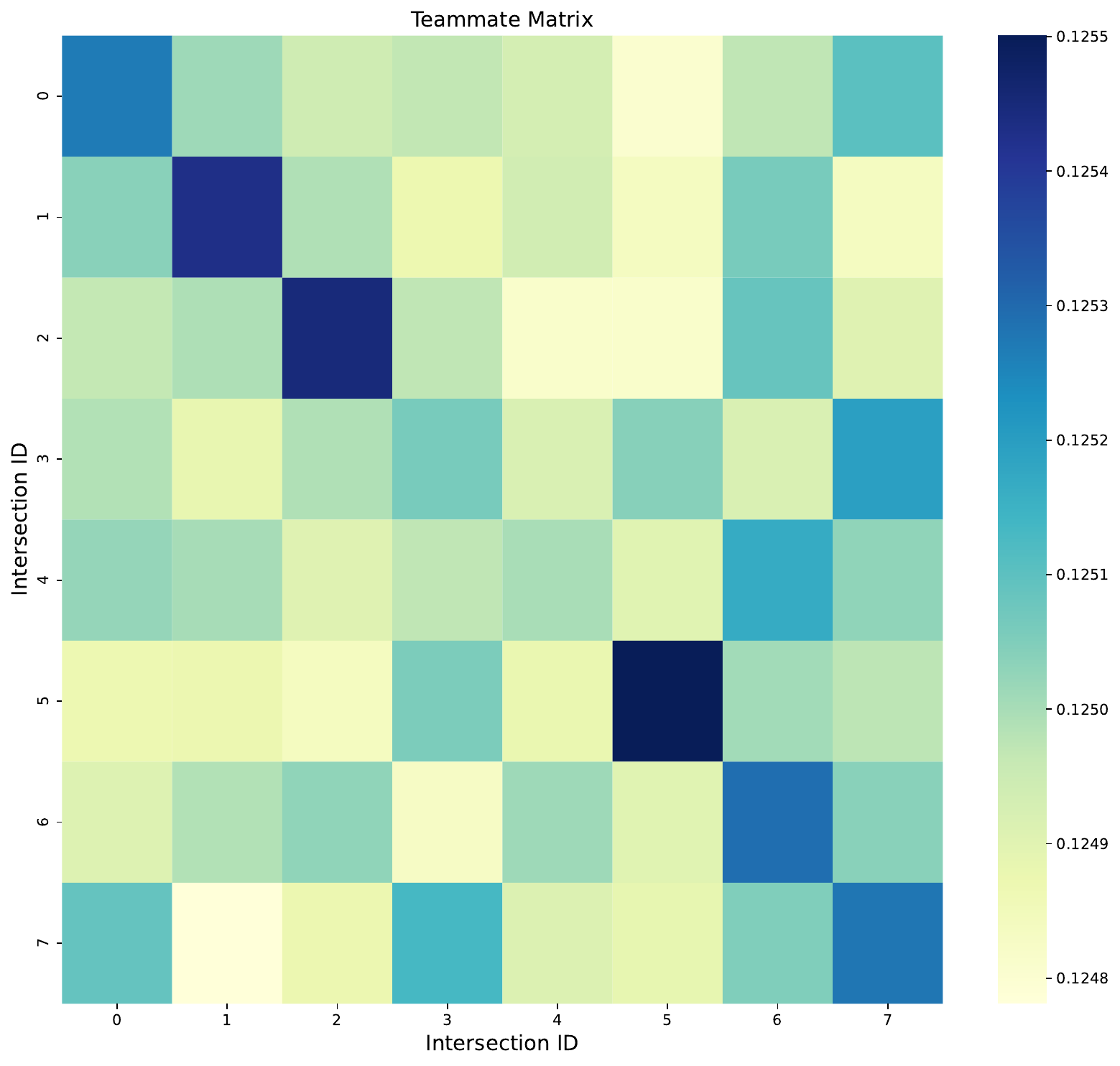}
		\end{minipage}
	}

	\subfigure[Beginning in Nanshan]{
		\begin{minipage}[t]{0.21\textwidth}
			\centering
			\includegraphics[width=\textwidth]{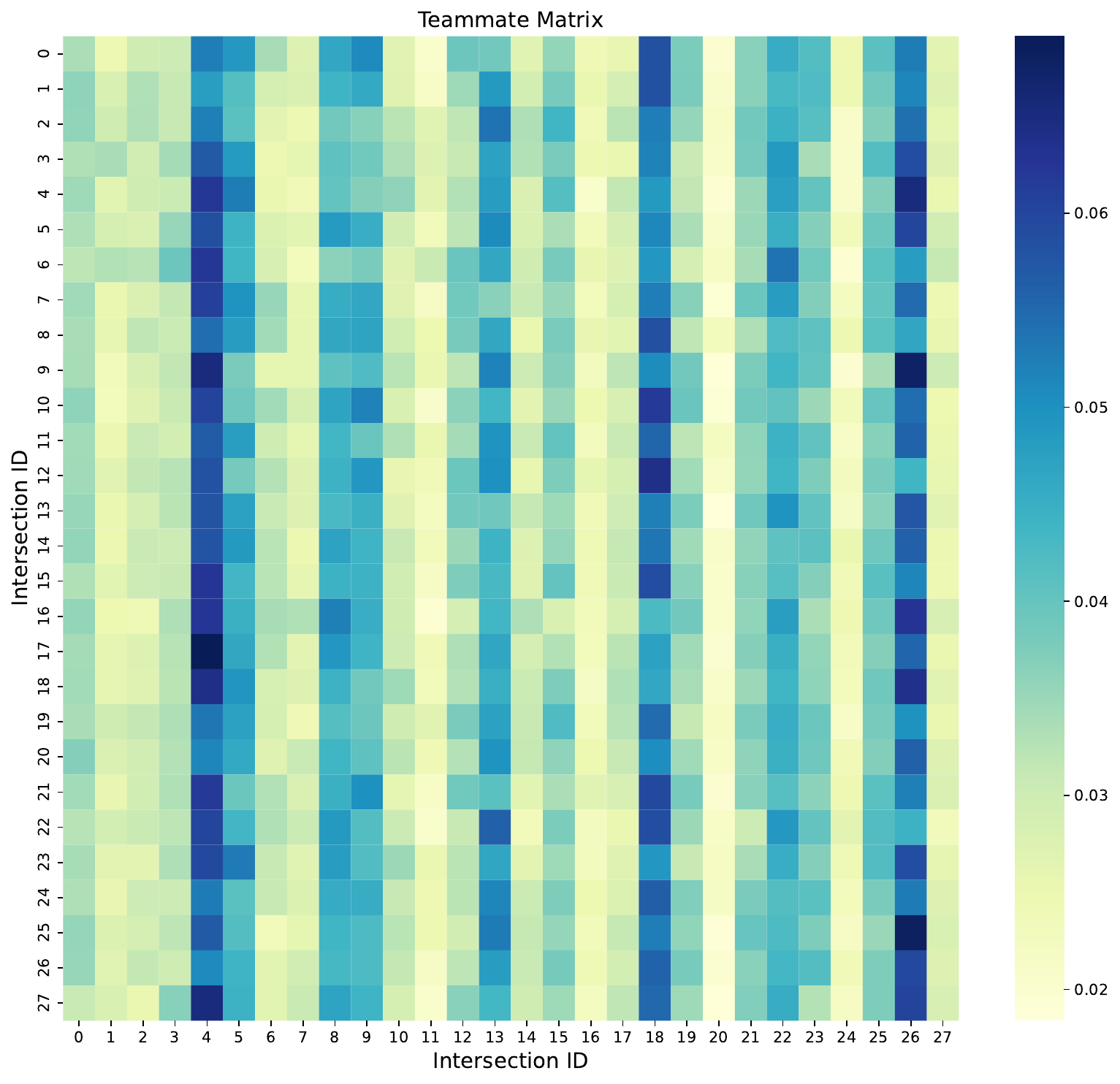}
		\end{minipage}
	}
	\subfigure[Ending in Nanshan]{
		\begin{minipage}[t]{0.21\textwidth}
			\centering
			\includegraphics[width=\textwidth]{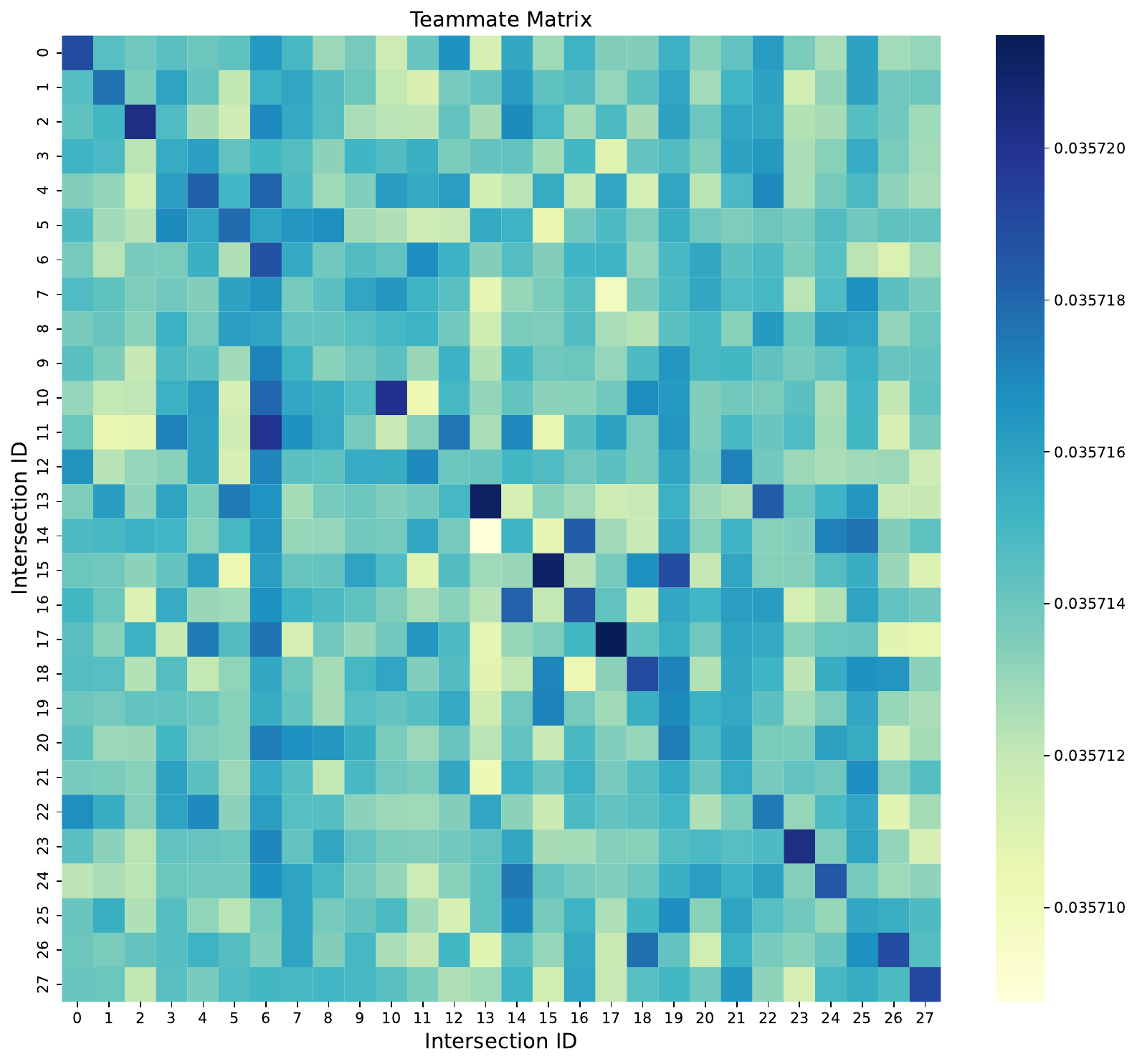}
		\end{minipage}
	}
	\caption{Additional visualization of team matrix. The deeper the color, the stronger the correlation.}
	\label{exp:app_vis_team_matrix}\vspace{-10pt}
\end{figure*}

In this section, we present an extended visualization analysis, incorporating heat maps across all scenarios in Figure~\ref{exp:app_vis_team_matrix} to provide a more comprehensive understanding of the learning dynamics. This additional material affirms the intriguing pattern that we have already discussed: intersections are not only learning to concentrate on their own states, but they also develop an awareness of their peers. This reveals a collaborative learning process among intersections.

Each intersection manages to strike a balance between self-attention and peer attention, learning to recognize when collaboration could bring about performance improvements. In this way, we see that our algorithm has fostered a learned mutual collaboration among intersections. 
These observations offer a broader validation of the effectiveness of our approach in developing a collaborative environment among intersections, which ultimately leads to performance enhancement across different scenarios.

\section{Additional Visualization Analysis of Dual-feature}
\label{app:vis_dual_feas}

\begin{figure}[ht!]
    \setcounter{subfigure}{0}
	\centering
 \subfigure[Grid 4 $\times$ 4]{
		\begin{minipage}[t]{0.22\textwidth}
			\centering
			\includegraphics[width=\textwidth]{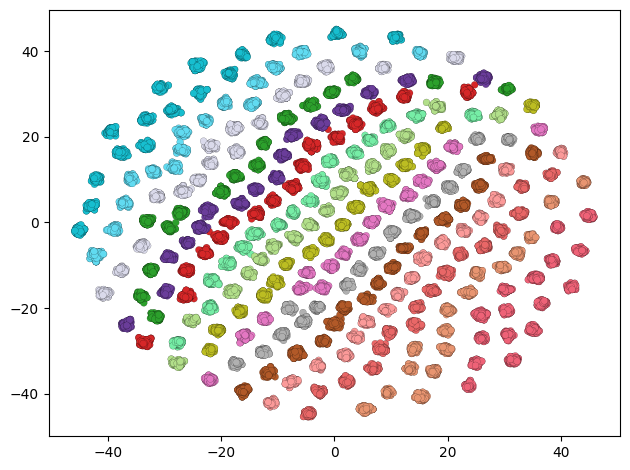}
		\end{minipage}
	}
 \subfigure[Avenue 4 $\times$ 4]{
		\begin{minipage}[t]{0.22\textwidth}
			\centering
			\includegraphics[width=\textwidth]{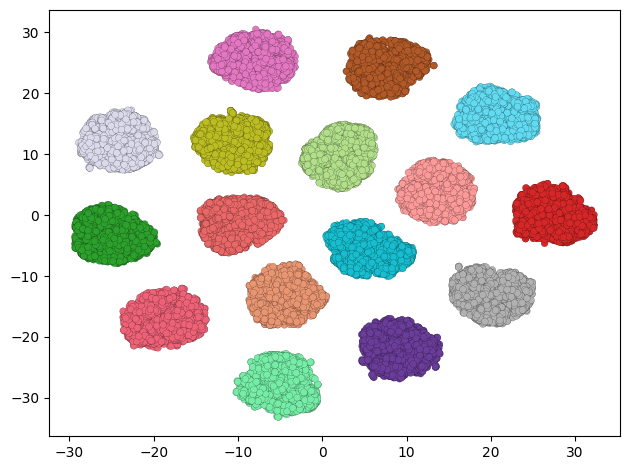}
		\end{minipage}
	}
 \subfigure[Grid 5 $\times$ 5]{
		\begin{minipage}[t]{0.22\textwidth}
			\centering
			\includegraphics[width=\textwidth]{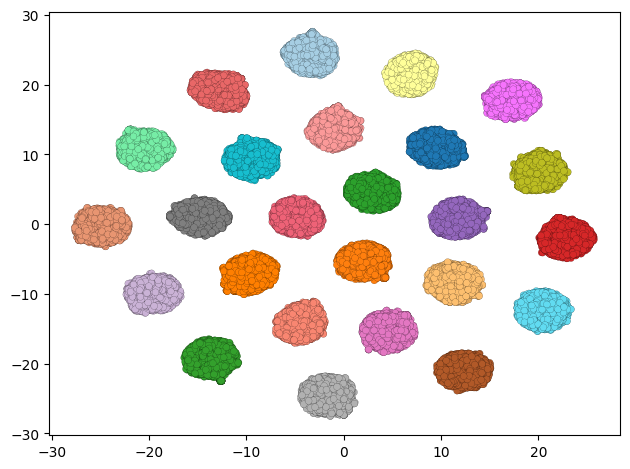}
		\end{minipage}
	}
 
	\subfigure[Cologne8]{
		\begin{minipage}[t]{0.22\textwidth}
			\centering
			\includegraphics[width=\textwidth]{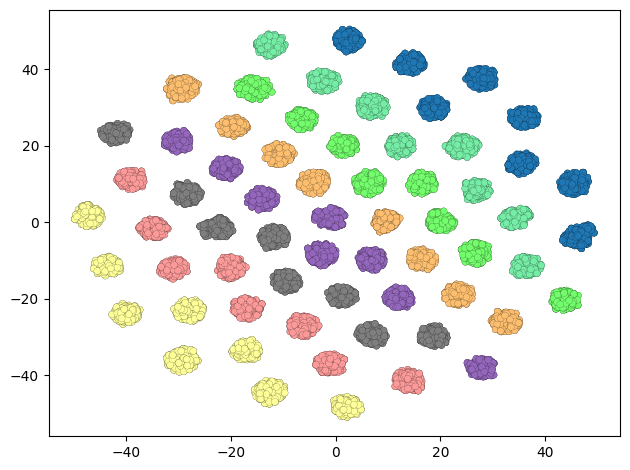}
		\end{minipage}
	}
	\subfigure[Nanshan]{
		\begin{minipage}[t]{0.22\textwidth}
			\centering
			\includegraphics[width=\textwidth]{fig/dual_fea/nanshan_actor_fea1_eps_0.png}
		\end{minipage}
	}
	
 \caption{2-D TSNE Visualization of Dual-feature. Each color represents a specific intersection.}
	\label{fig:app_vis_dual_fea}\vspace{-10pt}
\end{figure}

\begin{figure}[ht!]
    \setcounter{subfigure}{0}
	\centering
 \subfigure[Grid 4 $\times$ 4]{
		\begin{minipage}[t]{0.22\textwidth}
			\centering
			\includegraphics[width=\textwidth]{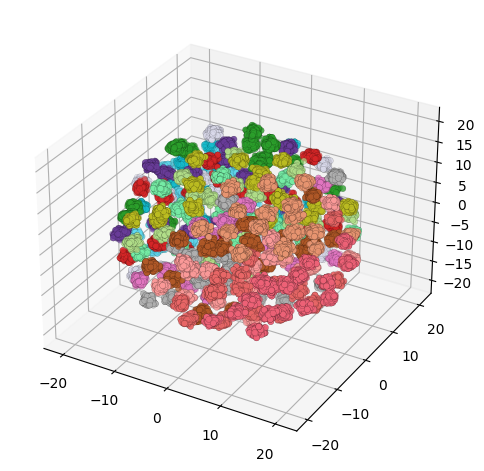}
		\end{minipage}
	}
 \subfigure[Avenue 4 $\times$ 4]{
		\begin{minipage}[t]{0.22\textwidth}
			\centering
			\includegraphics[width=\textwidth]{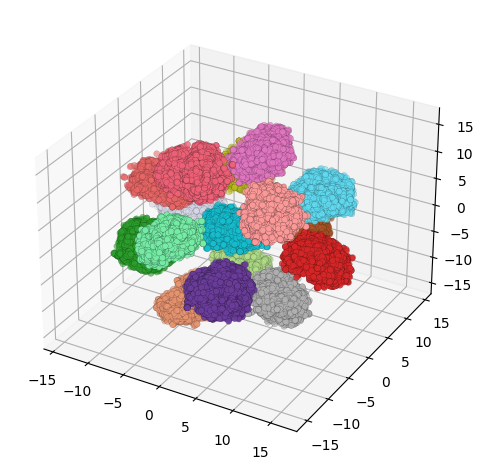}
		\end{minipage}
	}
 \subfigure[Grid 5 $\times$ 5]{
		\begin{minipage}[t]{0.22\textwidth}
			\centering
			\includegraphics[width=\textwidth]{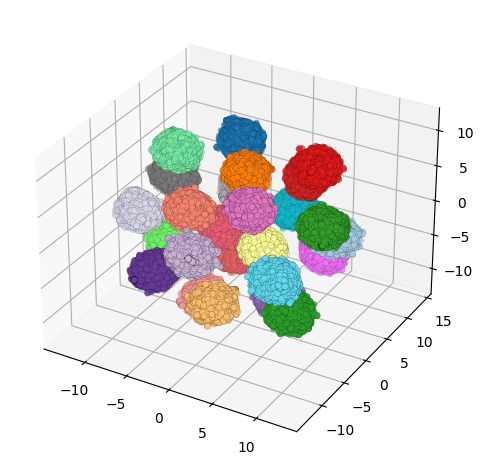}
		\end{minipage}
	}
 
	\subfigure[Cologne8]{
		\begin{minipage}[t]{0.22\textwidth}
			\centering
			\includegraphics[width=\textwidth]{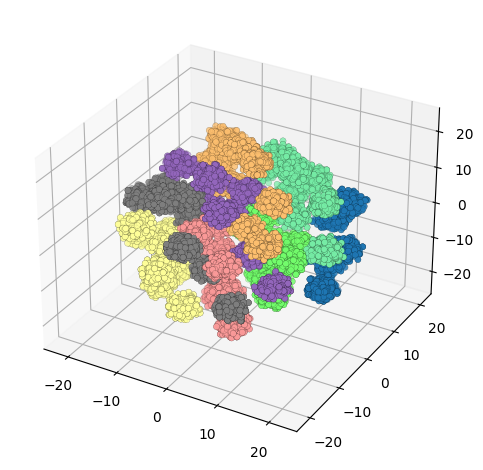}
		\end{minipage}
	}
	\subfigure[Nanshan]{
		\begin{minipage}[t]{0.22\textwidth}
			\centering
			\includegraphics[width=\textwidth]{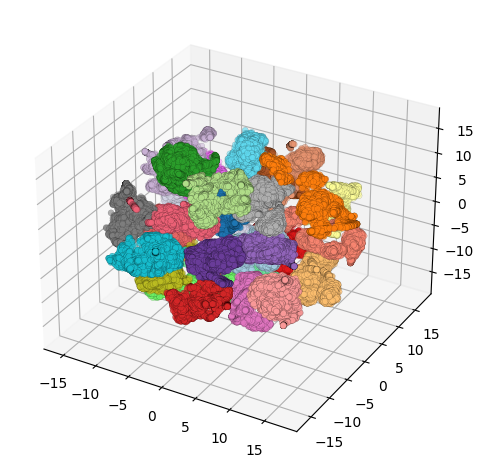}
		\end{minipage}
	}
	
 \caption{3-D TSNE Visualization of Dual-feature. Each color represents a specific intersection.}
	\label{fig:app_vis_dual_fea_3d}\vspace{-10pt}
\end{figure}

In this section, we offer 2D and 3D TSNE visualization results for each scenario, shown in Figure~\ref{fig:app_vis_dual_fea} and~\ref{fig:app_vis_dual_fea_3d}, respectively. These results consistently demonstrate a distinctive clustering characteristic. 
While the embeddings for each intersection may differ due to varying traffic conditions and intricacies inherent to the location, the clustering pattern indicates that the model successfully interprets and distills these unique characteristics into its decision-making process. This ability to learn and adapt to the unique features of each intersection is a key strength of our approach and provides a significant contribution to performance improvement.

\clearpage

\section{Additional Visualization Analysis of Collaborator Assignment}
\label{app:vis_teammate_assign}

In this section, we provide additional results for the visualization of the collaborator assignment. Specifically, we conducted tests on the Cologne8 and Avenue 4$\times$4 scenarios, with collaborator counts of 1, 2, 3, 4, and 5. We refrained from testing scenarios with a larger number of intersections and collaborators due to the diversity of bar colors, which could complicate observation and prevent meaningful conclusions.

\begin{figure}[ht!]
    \setcounter{subfigure}{0}
	\centering
 \subfigure[K=1]{
		\begin{minipage}[t]{0.3\textwidth}
			\centering
			\includegraphics[width=\textwidth]{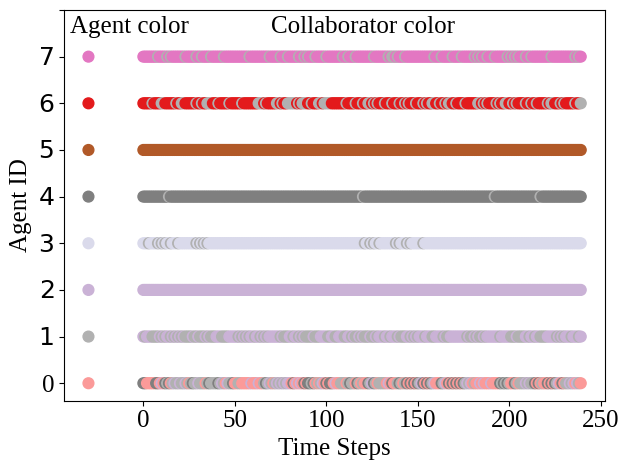}
		\end{minipage}
	}
 \subfigure[K=2]{
		\begin{minipage}[t]{0.3\textwidth}
			\centering
			\includegraphics[width=\textwidth]{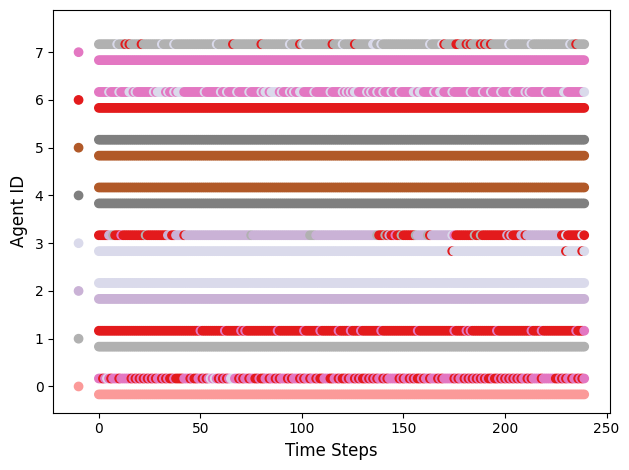}
		\end{minipage}
	}
 \subfigure[K=3]{
		\begin{minipage}[t]{0.3\textwidth}
			\centering
			\includegraphics[width=\textwidth]{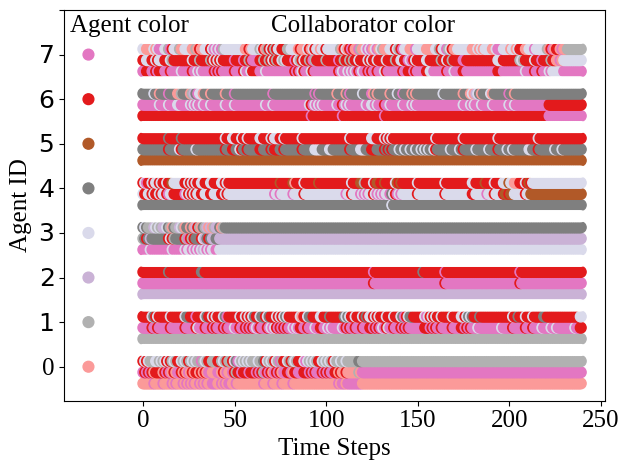}
		\end{minipage}
	}
 
	\subfigure[K=4]{
		\begin{minipage}[t]{0.3\textwidth}
			\centering
			\includegraphics[width=\textwidth]{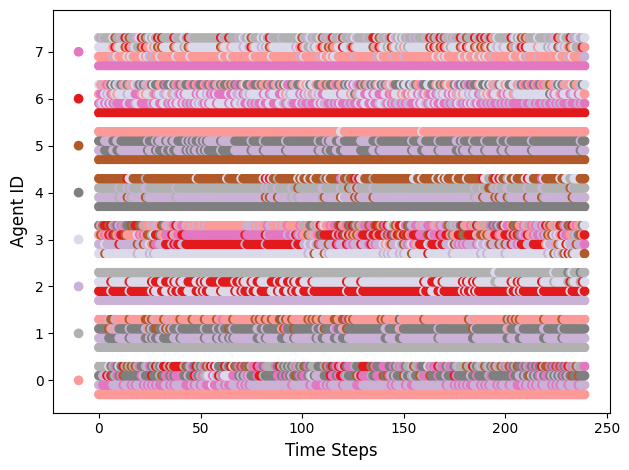}
		\end{minipage}
	}
	\subfigure[K=5]{
		\begin{minipage}[t]{0.3\textwidth}
			\centering
			\includegraphics[width=\textwidth]{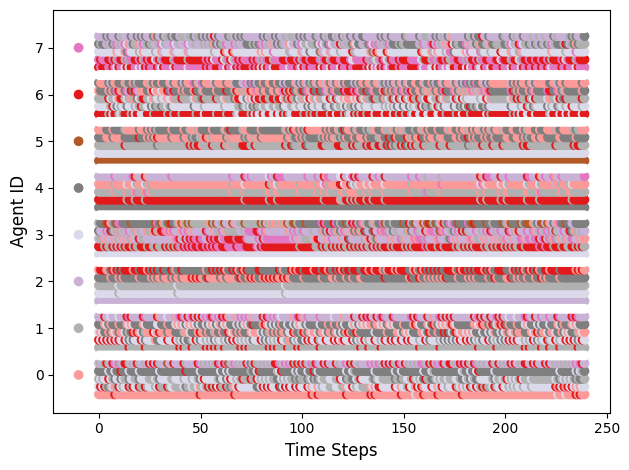}
		\end{minipage}
	}
	
 \caption{Visualization of the collaborator assignment on Cologne8.}
	\label{fig:app_teammate_assign_colo8}\vspace{-10pt}
\end{figure}

In Figure~\ref{fig:app_teammate_assign_colo8}, we present the evolution of the collaborator assignment on the Cologne8 dataset. 
\begin{itemize}
    \item We have previously analyzed scenarios when the collaborator number K equals 1 and 3 in the main manuscript.
    \item For K=2, we observe an intriguing mutual selection trend among agents 2 and 3, as well as agents 4 and 5. Through the reconstruction of the original traffic flow, we identified traffic correlations between these agent pairs, further validating the efficacy of our adaptive method in selecting the most suitable collaborators.
    \item When K is increased to 4, agent 4 primarily chooses agents 4, 2, 1, and 5 as collaborators, agent 2 predominantly selects agents 2, 3, 1, and 6, while agent 1 mainly selects agents 1, 2, 4, and 0. This collaborative behavior suggests that our approach can manage more complex scenarios where multiple agents are involved. 
    \item Upon further increasing K to 5, we observe an emergence of cross-collaboration within the virtual subareas, which solidifies our earlier observations that our algorithm can effectively handle cross-topology collaborator selection.
\end{itemize}
 This cross-collaboration phenomenon also demonstrates the flexibility of our method in creating dynamic, adaptive collaborations, even in complex environments with multiple agents.
In summary, these results strongly indicate that our learning-based strategy can adaptively select the optimal collaborators based on the intrinsic dynamics and the evolving nature of the traffic flow. This adaptive approach to collaborator selection provides a more responsive and effective mechanism to manage the complexities of real-world traffic scenarios.

\begin{figure}[ht!]
    \setcounter{subfigure}{0}
	\centering
 \subfigure[K=1]{
		\begin{minipage}[t]{0.3\textwidth}
			\centering
			\includegraphics[width=\textwidth]{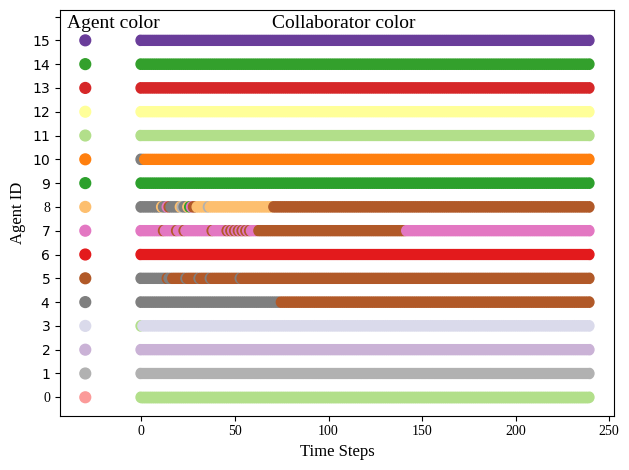}
		\end{minipage}
	}
 \subfigure[K=2]{
		\begin{minipage}[t]{0.3\textwidth}
			\centering
			\includegraphics[width=\textwidth]{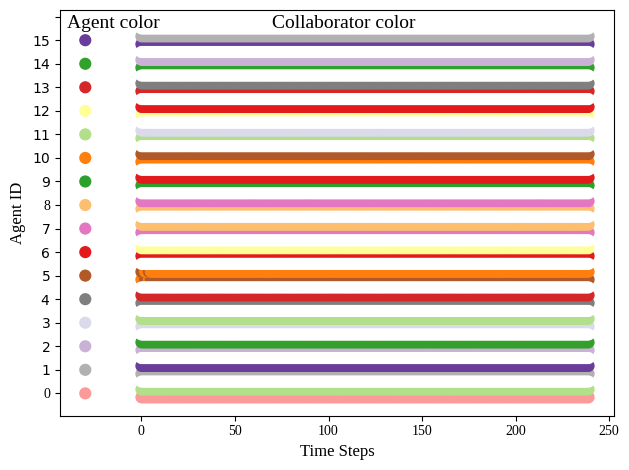}
		\end{minipage}
	}
 \subfigure[K=3]{
		\begin{minipage}[t]{0.3\textwidth}
			\centering
			\includegraphics[width=\textwidth]{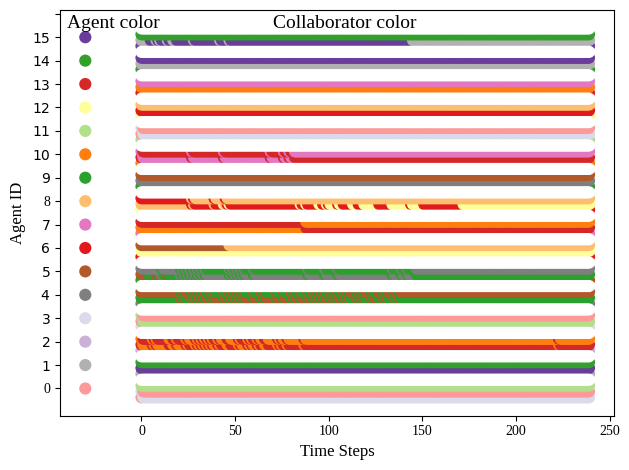}
		\end{minipage}
	}
 
	\subfigure[K=4]{
		\begin{minipage}[t]{0.3\textwidth}
			\centering
			\includegraphics[width=\textwidth]{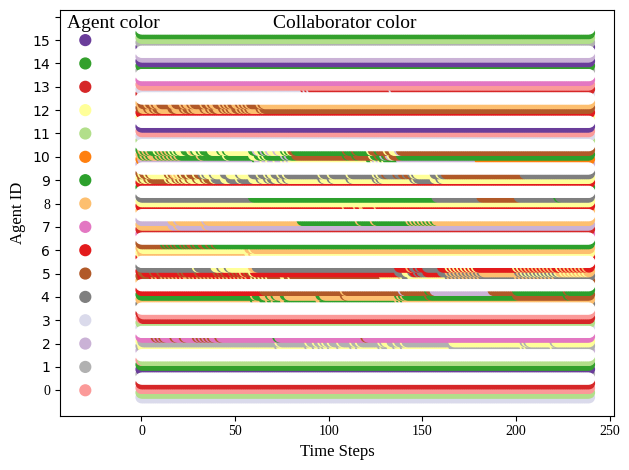}
		\end{minipage}
	}
	\subfigure[K=5]{
		\begin{minipage}[t]{0.3\textwidth}
			\centering
			\includegraphics[width=\textwidth]{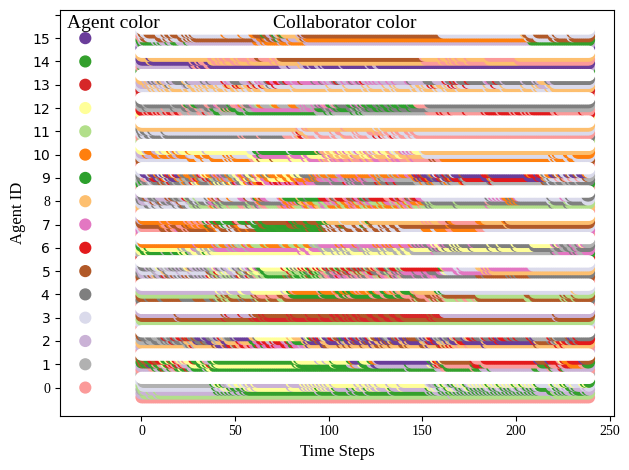}
		\end{minipage}
	}
	
 \caption{Visualization of the collaborator assignment on Avenue 4$\times$4.}
	\label{fig:app_teammate_assign_art4}\vspace{-10pt}
\end{figure}

Subsequently, in Figure~\ref{fig:app_teammate_assign_art4}, we present an examination of collaborator assignment on the Avenue 4 $\times$4 dataset to further substantiate the effectiveness of our approach.
\begin{itemize}
    \item At K=1, the vast majority of agents are capable of selecting themselves as collaborators. Interestingly, agent 0 consistently chooses agent 11 as its collaborator. Through the reconstruction of the original traffic network, we observed that agents 0 and 11 are situated adjacent to each other on the network, with highly correlated traffic flow. This finding indirectly confirms our method's ability to learn inherent traffic patterns. Other instances of mutual selection are occasionally observed, for example, between agents 4 and 5.
    \item When K is incremented to 2, agent 0 continues to select agent 11 and adds itself to the list of collaborators. A mutual selection trend is also observed among agent pairs 5 and 10, 1 and 11, as well as 4 and 13.
    \item For K=3, agent 0 includes a new collaborator, agent 3, in addition to itself and agent 11. Correspondingly, agent 3 tends to choose agents 3, 0, and 11 as its collaborators. Other agents also display mutual selection within their collaborative subareas.
    \item As K expands to 4, agent 0 adds another new collaborator, agent 6, to its previous collaborators (0, 11, 3). This pattern of incrementally adding new collaborators as the number of options increases demonstrates our algorithm's capacity to recognize the intrinsic traffic dynamics and interrelation among multiple intersections.
    \item With K=5, upon careful analysis, the aforementioned phenomena and supporting conclusions continue to hold.
\end{itemize}
In conclusion, these results convincingly demonstrate that our approach is effective in selecting collaborators adaptively in different scenarios. Not only does it capture the intrinsic dynamics of the traffic but also discerns the inherent relations between multiple intersections, thus establishing the value of our method in dealing with complex traffic scenarios.